\documentclass[a4paper,11pt]{article}
\pdfoutput=1 

\usepackage{jheppub} 

\usepackage[enableskew]{youngtab}
\usepackage{hyperref}
\usepackage{fancybox}
\usepackage{float}
\usepackage{amsfonts}
\usepackage{amsmath}
\usepackage{amsbsy}
\usepackage{graphicx}
\usepackage{amssymb}
\usepackage{amsthm}
\usepackage{bm}
\usepackage{comment}
\usepackage{epsfig}
\usepackage{latexsym}
\usepackage{pdflscape}
\usepackage{color}
\usepackage{slashed}

\makeatletter
\def\@fpheader{\relax}
\makeatother

\renewcommand{\b}{\bar}

\DeclareSymbolFont{AMSa}{U}{msa}{m}{n}
\DeclareSymbolFont{AMSb}{U}{msb}{m}{n}
\DeclareMathSymbol{\fieldR}{\mathalpha}{AMSb}{"52}

\DeclareMathOperator\Tr{tr}
\DeclareMathOperator{\tr}{tr}

\newcommand{\beq}{\begin{eqnarray}}
\newcommand{\eeq}{\end{eqnarray}}
\newcommand{\bea}{\begin{eqnarray}}
\newcommand{\eea}{\end{eqnarray}}
\newcommand{\be}{\begin{equation}}
\newcommand{\ee}{\end{equation}}
\newcommand{\bq}{\begin{equation}}
\newcommand{\eq}{\end{equation}}

\newcommand{\half}{\frac{1}{2}}
\newcommand{\nn}{\nonumber}

\def\Xint#1{\mathchoice
   {\XXint\displaystyle\textstyle{#1}}%
   {\XXint\textstyle\scriptstyle{#1}}%
   {\XXint\scriptstyle\scriptscriptstyle{#1}}%
   {\XXint\scriptscriptstyle\scriptscriptstyle{#1}}%
   \!\int}
\def\XXint#1#2#3{{\setbox0=\hbox{$#1{#2#3}{\int}$}
     \vcenter{\hbox{$#2#3$}}\kern-.5\wd0}}

\def\dashint{\Xint-}

\def\l{\lambda}

\def\z{\zeta}

\def\e{\epsilon}

\def\half{\frac12}

\def\6{\partial}

\def\a{\alpha}
\def\b{\beta}

\def\tr{{\rm Tr}}

\def\6{\partial}

\def\normord{ {\scriptstyle {{\bullet}\atop{\bullet}}} }

\def\hri#1#2{\href{http://arxiv.org/abs/#1}{[ArXiv:#1]#2}}
\def\hre#1#2{\href{http://arxiv.org/abs/#2/#1}{[ArXiv:#1/#2]}}

\def\hrj#1#2{\href{https://doi.org/#1}{#2}}

\title{\boldmath Microstates of a $2d$ Black Hole in string theory}

\author[a]{Panos Betzios,}
\affiliation[a]{\href{https://phas.ubc.ca/}{Department of Physics and Astronomy} , University of British Columbia, \\
6224 Agricultural Road, Vancouver, B.C. V6T 1Z1, Canada}

\author[b]{Olga Papadoulaki}
\affiliation[b]{\href{https://perimeterinstitute.ca/}{Perimeter Institute for Theoretical Physics} , Waterloo, \\
 Ontario N2L 2Y5, Canada}

\emailAdd{pbetzios@phas.ubc.ca}
\emailAdd{opapadoulaki@perimeterinstitute.ca}

\abstract{We analyse models of Matrix Quantum Mechanics in the double scaling limit that contain non-singlet states. The finite temperature partition function of such systems contains non-trivial winding modes (vortices) and is expressed in terms of a group theoretic sum over representations. We then focus in the case when the first winding mode is dominant (model of Kazakov-Kostov-Kutasov). In the limit of large representations (continuous Young diagrams), and depending on the values of the parameters of the model such as the compactification radius and the string coupling, the dual geometric background corresponds to that of a long string (winding mode) condensate or a $2d$ (non-supersymmetric) Black Hole. In the matrix model we can tune these parameters and explore various phases and regimes. Our construction allows us to identify the origin of the microstates of these backgrounds, arising from non trivial representations, and paves the way for computing various observables on them. 
}


\begin{document} 
\maketitle
\flushbottom

\section{Introduction}

One of the most interesting physical systems, whose complete description requires a deep understanding of the merging of quantum mechanics and gravity are Black Holes. The pertinent questions one would like to address are related to the spacetime physics of horizons and singularities as well as to the possibility of acquiring a microscopic description of their macroscopic thermodynamic properties such as their entropy. Traditional (worldsheet) string theory can be used to describe only a few aspects of them, since it is mainly based on perturbation theory (and is well developed around very symmetric backgrounds). Non perturbative objects, such as D-branes greatly expanded the tools and prospects that string theory has to attack this problem from a microscopic perspective. This eventually led to some remarkable results in the study of supersymmetric (extremal) black holes, for which non-renormalization theorems guarantee an explicit counting of their microstates using the dual D-brane system~\cite{Strominger:1996sh}. Even though after the advent of the $AdS/CFT$ correspondence, we are in a rare position to have an in principle non-perturbative description of quantum gravity in asymptotically $AdS$ spaces, it is still fair to say that there are various aspects of black holes\footnote{Most of them have to do with the structure and properties of their interior and singularities.}, even in $AdS$, that still defy a complete understanding.

The situation of course is even worse in the case of non-supersymmetric black holes for which even a method to count their microstates is not available (unless one can invoke the Cardy formula in special examples), let alone when one wishes to describe realistic examples in asymptotically flat spacetimes, such as the common Schwarzschild black hole. The main motivation behind our work is to find a (matrix) model that is solvable (or at least analysable via saddle point methods at large $N$), that can describe microscopically a (non-supersymmetric) black hole in string theory\footnote{The matrix model is there to provide an in principle non-perturbative description of the black hole.}. Of course this is a tall order and the price we have to pay is that such a model can be explicitly constructed in non-critical lower dimensional versions of string theory, the most prominent example based on the duality between $c=1$ Liouville theory and Matrix Quantum Mechanics (MQM) in a double scaling limit. With such a model at hand one would hope to address the deepest questions related to black holes, such as the spacetime physics behind the horizon and the nature of the black hole singularity, if the lessons to be learned exhibit any form of universality across dimensions (see the discussion section~\ref{Discussion} for more details).

Most of the past works on MQM (some excellent reviews are~\cite{Klebanov:1991qa,Ginsparg:1993is,Nakayama:2004vk,Martinec:2004td}) have focused in its singlet sector using an $SU(N)$ gauged version of the model, that reduces to the dynamics of (non-relativistic) free fermions in an inverted oscillator potential. The culmination of these works resulted in matching numerous physical observables with the dual $c=1$ Liouville string theory on a linear dilaton background as well as making (exact) predictions for observables that are not yet computable from the string theory side. In contrast, all the efforts to uncover some aspect of black hole physics in high energy scattering using the singlet sector of MQM~\cite{Martinec:2004qt,Friess:2004tq,Karczmarek:2004bw} proved fruitless, demonstrating thus that states resembling black holes can only (potentially) exist in the non-singlet sector of the theory\footnote{An exponential degeneracy of states can be found more generally in models for which the aforementioned fermions carry additional indices, such as those arising from dimensional reduction on extra compact space dimensions, see for example~\cite{Betzios:2016yaq}.}. This result is consistent with the infinite $\mathcal{W}_{1+\infty}$ symmetry of the quantum inverted oscillator that is in conflict with the expected thermalising and ``chaotic" properties of systems that are holographic duals of black holes\footnote{Recent works have argued that black holes should exhibit a fast scrambling and chaotic behaviour (which can ascertained by the analysis of four point OTOC's)~\cite{Maldacena:2015waa}.}. At this point we should emphasize that it is not entirely certain which properties of higher dimensional black holes should still persist in the two-dimensional case. Some expected properties characteristic of black holes, are the appropriate scaling of the entropy and the mass of the black hole with the string coupling, and the quasi-normal mode behaviour for the retarded two point functions of localised probes. Other expectations include the presence of an effective non-zero absorption, and an enhanced particle production when scattering highly energetic particles (strings) that can form a black hole\footnote{For example the $2 \rightarrow N$ amplitude should peak at high values of $N$ with the production of many ``soft particles" that would resemble Hawking radiation.}. 

Turning on to the non-singlet sectors, there have been two main proposals for models that are dual to a two dimensional black hole in the literature~\cite{Kazakov:2000pm,Betzios:2017yms}. They are closely related, since they both involve the liberation of vortices on the string worldsheet~\cite{Sathiapalan:1986db,Kogan:1987jd,Gross:1990md} and the presence of non-trivial winding modes around the target space thermal circle~\cite{Atick:1988si}. In the model of~\cite{Kazakov:2000pm}, it was demonstrated  that the entropy scales in a consistent way with the expectations coming from the Euclidean black hole background and its dual \emph{Sine-Liouville theory}~\cite{Gibbons:1992rh,Nappi:1992as,Kazakov:2001pj}. On the other hand in their model,~\cite{Kazakov:2000pm} did not clearly identify the origin of the microstates accounting for this entropy and neither provided a real time description of the physics. Some subsequent developments related the analytic continuation of Euclidean winding modes, with long-strings that stretch and scatter on the linear dilaton background~\cite{Maldacena:2005hi,Gaiotto:2005gd}, but they focused in the regime when they do not backreact on the geometry.

In relation to these works, in~\cite{Betzios:2017yms} it was found that even though the model of~\cite{Kazakov:2000pm} does not have an a priori obvious analytic continuation into Lorentzian signature, it nevertheless appears in a particular scaling limit of a more general class of MQM models which do have a real time description. These models contain in addition to the original $N \times N$ MQM matrix $M_{ij}$ (and non dynamical gauge field $A_{i j}$), bi-fundamental fields $\chi_{\a i}$ transforming under an $SU(N_f) \times SU(N)$ symmetry\footnote{The $SU(N_f)$ symmetry in these models is a global symmetry. This comes from the open strings ending on $N_f$ FZZT branes. We construct in section~\ref{D0D1} a new model for which both symmetries are gauged.} (that source the MQM $SU(N)$ non-singlets that were originally projected out). Going to the matrix eigenvalue basis, they can be equivalently described in terms of dynamical $SU(N_f)$ spin-Calogero models~\cite{Polychronakos:1991bx,Betzios:2017yms} in an inverted oscillator potential. The models of~\cite{Betzios:2017yms} are well defined both in Euclidean and Lorentzian signature, and their Euclidean partition function constitutes a generalisation of that appearing in~\cite{Kazakov:2000pm}, obeying a discrete (Hirota-Miwa) soliton equation instead of the simpler Toda differential equation. They are also related to the ungauged version of MQM, but contain additional parameters (fugacities for vortices/winding modes) as the model of~\cite{Kazakov:2000pm} does. In addition due to the relation of the models of~\cite{Betzios:2017yms} with FZZT branes, these parameters have a natural interpretation from the Liouville side: The masses of the bifundamentals are related to the boundary cosmological constant via: $2 m = \sigma \, ,$  $\mu_B = \sqrt{\mu} \cosh \pi \sigma$, $N_f$ to the number of FZZT branes and so forth, see~\cite{Betzios:2017yms} for more details.

At this point we should mention a known difficulty relating the model of~\cite{Kazakov:2000pm} with an object that behaves like an actual target space black hole. As we review in section~\ref{introduction}, this matrix model is most directly related to \emph{Sine-Liouville} theory, which in turn is related to the $SL(2,R)_k/U(1)$ WZW coset description of the $2d$ black hole~\cite{Elitzur:1990ubs,Witten:1991yr,Dijkgraaf:1991ba}, by a form of strong/weak duality (FZZ duality~\cite{Fateev}). The issue is that the coset is an actual CFT only for a certain compactification radius $R^2 = k=9/4$ that is of string scale (in units where $\alpha'=1$). This brings us to another important topic, that of the \emph{black hole - string transition}~\cite{Susskind:1993ws,Horowitz:1996nw,Sen:1995in,Damour:1999aw}. In short the basic idea is that once black holes become small and reach the string scale (the so called correspondence point), the gravitational semi-classical black hole geometry ceases to be a good description and is replaced by a condensate of (long) strings. For the particular case of the two-dimensional black hole, it can be shown that the winding mode becomes non-normalisable for radii $R^2 = k<3$. This means that it is explicitly sourced and below $k=3$ the bosonic black hole ceases to be a normalisable state in the theory ~\cite{Giveon:2005mi,Kutasov:2005rr}, leaving only a long string condensate to survive, for more details see the end of section~\ref{introduction}. Of course this signals a possible trouble for the interpretation of the WZW coset as a black hole for such a string scale radius, when it is actually a CFT. Nevertheless, even though we do not currently have access to an exact worldsheet description, generally one does anticipate the existence of black hole solutions for a wide range of radii, perhaps with additional fields turned on in the background such as the Tachyon. It is natural then to view this shortcoming of the WZW coset as being merely a technical one, without fundamental importance.

Due to this technical difficulty on the string theory side, from our perspective we elevate the role of the matrix model and postulate that it provides the accurate (UV complete) description of the dual string theory physics across a wide range of parameter space. This is a quite reasonable expectation, since on the one hand it reproduces the string theory results in the regime where we can actually compare them and on the other hand much like to what happens in higher dimensional examples of modern holography (AdS/CFT), we do believe that the CFT (matrix model) provides a UV complete description of the associated bulk string theory, even though in most cases we cannot explicitly construct the later (for example around non-trivial backgrounds). In the end what is most important are the physics contained in the model and whether it exhibits the aforementioned properties we expect of systems dual to black holes. Finally by computing various non-trivial observables, it is perhaps possible to actually reconstruct the effective bulk geometry that probes will register, as we briefly discuss in section~\ref{Discussion}. 

We shall now proceed and summarize the contents and main results of our work. In the rest we shall use units in which $\alpha'=1$, unless otherwise indicated.

\paragraph{Plan of the paper and results}

\begin{itemize}

\item In section~\ref{introduction}, we provide a review of the $2d$ black hole solution and its string theory descriptions either in terms of an $SL(2,R)_k/U(1)$ WZW coset model or its dual \emph{Sine-Liouville} theory. We also present some novel ideas related to the role of the two possible dressings of the winding modes and how they could be physically relevant if one wishes to describe both sides of a \emph{black hole - string transition (or crossover)} as one changes the compactification radius $R$ of the system. This proposal is corroborated further from some explicit matrix model results of section~\ref{Saddles} and~\ref{thermo}.

\item In section~\ref{windingmatrixliouville}, we briefly describe the matrix model dual to Sine-Liouville, its generalisations and the relation of their (grand-canonical) partition functions to $\tau$-functions of the Toda-hierarchy according to the works of~\cite{Kazakov:2000pm,Betzios:2017yms}.

\item In section~\ref{MQMnonsinglets}, we describe a representation theoretic expansion of such $\tau$-functions and the relevant \emph{Schur-measures} for the physically interesting non-singlet models of~\cite{Kazakov:2000pm,Betzios:2017yms}. This description makes manifest the microstates that the system is composed of, since the partition function is a discrete sum of integers labelling the irreducible representations/partitions. We also introduce the notion of a \emph{leading (continuous) Young diagram} in the limit of large representations that is the dominant saddle of the aforementioned sum. This saddle is composed out of many microstates (irreps) which have similar coarse grained characteristics (they are indistinguishable) in the thermodynamic limit of large partitions. A fixed irreducible representation does not carry any form of classical entropy in the thermodynamic limit. Nevertheless a classical entropy is generated for the non-singlet models by the presence of the
appropriate measures weighting the representations/partitions. 

\item The main new results are in section~\ref{Saddles}, where we derive an effective action in the space of highest weights of representations/partitions (using a Frobenius description for them). We solve the resulting saddle point equations in the limit of continuous Young diagrams and derive the resulting \emph{density of boxes} and the shape of the leading Young diagram for the model of~\cite{Kazakov:2000pm}\footnote{That can also be obtained from the more general dynamical models of~\cite{Betzios:2017yms} in a certain limit as we alluded to above.}. For radii $R<2$, we find two main types of leading saddles that correspond to two different phases depending on the strength of the winding mode/Sine-Liouville deformation parameter $\xi$ and the compactification radius $R$. In one of them there is a region of ``saturation" for the density of boxes, and the corresponding free energy matches that found in~\cite{Kazakov:2000pm}, in the sense that it exhibits the same leading behaviour $\mathcal{F}\sim (2-R)^2 \xi^{4/(2-R)} \sim 1/g_{s}^2$, with $\xi$ corresponding to the Sine-Liouville coupling and $g_s$ the effective string coupling (their relation can be determined via KPZ/DDK scaling arguments as we explain in section~\ref{windingmatrixliouville}). This is the saddle that exists in the parametric regime that is continuously connected to the undeformed case $\xi = 0$ and can be reached via conformal perturbation theory. For larger values of $\xi$ there is a (continuous) transition to an ``unsaturated" saddle. Well above the transition the scaling of the free energy becomes more complicated and intricate, but asymptotes to a D-brane scaling law $\mathcal{F}\sim 1/ g_{s}$ at smaller string coupling.

The radius $R_{KT} = 2$ has a special significance because it corresponds to the Kosterlitz-Thouless temperature above which worldsheet vortices get liberated and proliferate~\cite{Kogan:1987jd,Sathiapalan:1986db,Gross:1990md}. 
Remarkably the matrix model can still be defined for compactification radii above $R>2 = R_{KT}$ (lower temperatures), which were thought not to exist (for non-zero $\xi$) according to the analysis of~\cite{Kazakov:2000pm} (this was based on the fact that the Sine-Liouville action is not well defined in this regime). In this case, one of the two previous saddles becomes completely pathological (negative density of boxes), but the ``saturated" one still exists for some parameter range. The difference with $R<2$, is that now this saddle becomes metastable (the effective potential in the space of representations is unbounded from below for $R>2$). Depending on the values of the parameters we find two significant types of behaviour for the free energy, either a logarithmic scaling behaviour in $\xi$ (which has the interpretation of a non singlet free energy with a gap with respect to the singlet~\cite{Gross:1990md,Klebanov:1991qa}), or a $\mathcal{F} \sim \xi^{4/(R+2)}$ behaviour, revealing the presence of an opposite type of Liouville dressing for the winding mode (see section.~\ref{windingmatrixliouville}). This dressing makes the winding mode normalisable and relevant (having support in the strongly coupled region of Liouville). 

Based on recent literature~\cite{Attali:2018goq,Yogendran:2018ikf,Giveon:2019gfk,Jafferis:2021ywg} relating the presence of long strings in the strong coupling region of Liouville with black holes (\emph{stringy ``ER-EPR"}), one might expect that this alternative dressing would be crucial for describing a state of the system resembling a semi-classical black hole for radii $R> R_{KT} = 2$. Nevertheless, as we shall describe in more detail, an analysis of additional observables is needed to make this proposal more solid.

\item In section~\ref{thermo} we discuss the physical consequences of our results from section~\ref{Saddles} for the thermodynamic properties of the non-singlet MQM models and compare them with the existing literature. These results and the phase structure of the non-singlet models when the first winding mode is dominant are summarised in fig.~\ref{fig:Phasediagram}.

\item In section~\ref{D0D1} we introduce a new, completely gauged model of a system of 
$D0$-$D1$ (ZZ-FZZT) branes, that takes into account all the possible types of open strings that can be stretched between them. We briefly describe its properties and partition function - its most important feature is that its effective potential in the space of representations can be bounded from below even for $R>R_{KT}$ -, but leave a more thorough analysis for the future.

\item Finally in the discussion section~\ref{Discussion}, we close with some general remarks and list some problems that can be addressed in the near future.

\end{itemize}

We also include various detailed appendices~(\ref{Partitionsreps}-\ref{Ungauged}), proving several formulae that are used in the main text.

\paragraph{A note added:} While finishing this work we became aware of related work by~\cite{Ahmadain:2022gfw}. We therefore decided to coordinate our submissions. Our results to the extend they can be compared are in agreement, but our approaches elucidate different aspects of the problem and should be thought of as complementary.

\section{The $2d$ Euclidean black hole and the FZZ duality}\label{introduction}

The low energy (leading in $\alpha'$) target space effective action of string theory in two dimensions is~\cite{Callan:1985ia}\footnote{It is also related to the CGHS action with no matter fields~\cite{Callan:1992rs}. We momentarily reintroduce $\alpha'$ for clarity.}
\be
S_{eff} = \half \int d^2 X \sqrt{-G} e^{- 2 \Phi} \left[ \frac{16}{\alpha'} + R + 4 (\nabla \Phi)^2 - (\nabla T)^2 + \frac{4}{\alpha'} T^2 \right] \, ,
\ee
with $\Phi$ the dilaton and $T$ the tachyon field. This action is known to admit a black hole solution with non trivial metric and dilaton~\cite{Witten:1991yr,Elitzur:1990ubs,Mandal:1991tz}. Its Euclidean version that we shall be interested in is
\be\label{I1}
ds^2 =  \left( (1 - e^{2 Q \phi}) d \tau^2 + \frac{1}{1 - e^{2 Q \phi}} d \phi^2  \right)\, , \quad \Phi = \Phi_0 + Q \phi  \, , \quad Q^2 = \frac{4}{\alpha'} \, ,
\ee
where $- \infty < \phi < 0$ and for $\phi = 0 $ we reach the tip of the cigar. $\Phi_0$ is the only integration constant that fixes the value of the dilaton at the tip of the cigar. One should notice that asymptotic radius of $\tau$ (and hence the temperature) is fixed in terms of $Q$, if we wish to have a classical cigar geometry that is smooth at the tip.

In fact there exists an exact worldsheet CFT for the Euclidean black hole, in the form of a WZW coset model~\cite{Witten:1991yr}, the target space coset being
\be\label{I0}
H_3^+/U(1) \, , \qquad H_3^+ = \frac{SL(2,C)}{SU(2)} \, .
\ee
This coset in the case of a compact $U(1)$ gives rise to a manifold with the topology of a cigar and hence describes a Euclidean black hole\footnote{When taking the quotient with respect to a non-compact $U(1)$ one obtains the Lorentzian black hole geometry.}.
The exact coset solution to all orders in $\alpha'$ was described in~\cite{Dijkgraaf:1991ba} using algebraic CFT techniques (see also~\cite{Kazakov:2001pj} for a description of the thermodynamic properties of the exact solution). The background and the relation of variables is
\bea
ds^2 = \frac{\tanh^2 Q r}{1 - p \tanh^2 Q r } d x^2 \, + \, d r^2 \, , \quad \Phi = \Phi_0 - \log \cosh Q r - \frac{1}{4} \log (1 - p \tanh^2 Q r) \, , \nn \\
  p \, = \, \frac{2 \alpha' Q^2}{1+ 2 \alpha' Q^2} \,  = \,  \frac{2}{k} \, . \qquad \qquad \qquad \qquad
\eea
After a coordinate redefinition, and expanding to leading order in $\alpha'$, this background can be mapped to the background of eqn.~\eqref{I1}. Remarkably it is also possible to map the two backgrounds exactly, via a field redefinition that mixes the metric and the dilaton, and this is a manifestation of the fact that the expression for the background extracted from a gauged WZW model is scheme dependent~\cite{Kazakov:2001pj}. The absence of conical singularity at the tip ($r=0$), fixes the horizon and asymptotic inverse temperatures
\be
\beta_{hor} = \frac{2 \pi}{Q}  \,  , \qquad \beta_{asym} = 2 \pi R = \frac{2 \pi}{Q \sqrt{1-p}} \, \Rightarrow \quad \frac{R^2}{\alpha'} = k \, .
\ee
This means that the level of the associated WZW model $k$ determines the asymptotic radius of the cigar that should be $R/\sqrt{\alpha'} = \sqrt{k} = 3/2$ when $Q^2 = 4/\alpha'$. The $SL(2,R)_k$ algebra dictates the central charge and the spectrum of primaries for the coset CFT. In particular one finds for the central charge
\be\label{I2}
c_{\text{cigar}} = \frac{3 k}{k-2} - 1 \, .
\ee
We notice that conformal invariance of the worldsheet theory requires $k = 9/4$, that is the same condition as the smoothness at the tip. It is possible though, to append to this theory extra degrees of freedom of an ``internal" CFT. This later possibility would allow to change the radius of the cigar\footnote{The MQM models of~\cite{Kazakov:2000pm,Betzios:2017yms} seem to realise such a possibility.}. Additionally one expects the existence of marginal deformations of this CFT and the presence of a more general class of solutions for which the tachyon field also has a non-trivial profile. Unfortunately we do not know an exact CFT desription of this more general class of $2d$ black holes for which additional fields are turned on (see also~\cite{Mukherji:1991kz} for generalisations of the $2d$ black hole using string field theory).
 Moreover, since for $k = 9/4$ the geometry is of string scale, the gravity description could be misleading. As an example, the string coset black hole does not exhibit the same asymptotic conditions for all the modes as in the linear dilaton background without the black hole. The discrepancy comes from string winding modes (and not due to the local metric and dilaton fields). The winding modes even though decrease as we approach the asymptotic weakly coupled region, they do not do so fast enough as in higher dimensions and fail to be a normalisable deformation\footnote{See also eqn. \eqref{I10}, for the definition of the winding modes in the Sine-Liouville description.}. This means that we are deforming the linear dilaton background by adding a source at infinity for the winding string\footnote{We expect analogous types of deformations with sources, to give rise to $2d$ Euclidean (target space) wormhole backgrounds in Liouville theory~\cite{Betzios:2021fnm}. It would be interesting to explicitly construct such \emph{target space wormhole} solutions in string theory.}.
 
The thermodynamics of both~\ref{I1} and the exact coset background of~\cite{Dijkgraaf:1991ba} have been studied in the works of~\cite{Gibbons:1992rh,Nappi:1992as,Kazakov:2001pj}, but it has proven difficult to extract quantitative results unambiguously, precisely due to the scheme dependence of the WZW model~\cite{Kazakov:2001pj} and the lack of an effective target space action for which the exact background is a solution of its equations of motion. Different subtraction schemes give very different quantitative results for the various thermodynamic quantities such as the free energy of the black hole. 
This is also to be expected due to the fact that the background is a string scale background and one should be very careful when defining thermodynamic quantities in fully fledged string theory.  Nevertheless one can safely make some qualitative estimates that are scheme independent, for example the entropy should scale roughly as the mass of the black hole as can be verified using either of the two backgrounds
\be
S \sim M \sim e^{- 2 \Phi_0} \sim 1/g^2_{tip} \, ,
\ee
with $g_{tip}$ the value of the string coupling at the tip of the cigar geometry\footnote{Since there is no area of a horizon in $2d$, the entropy is related to the value of the dilaton at the tip of the cigar geometry.}. 

The conformal primaries $V_{j ; m, \bar{m}}$ of the coset CFT have the following conformal dimensions
\be\label{I3}
\Delta_{j; m , \bar{m}} = - \frac{j(j+1)}{k-2} + \frac{m^2}{k} \, , \quad \bar{\Delta}_{j; m , \bar{m}} = - \frac{j(j+1)}{k-2} + \frac{\bar{m}^2}{k}
\ee
with
\be\label{I4}
m = \half (n_1 + n_2 k) \, , \quad \bar{m} = - \half (n_1 - n_2 k) \, , \qquad n_{1,2} \in \mathbb{Z} \, .
\ee
There exist various indications, including computations of two point and three point correlation functions of such primaries~\cite{Fateev} in the bosonic case and an explicit result based on mirror symmetry for the supersymmetric case~\cite{Hori:2001ax}, that the coset CFT is dual to the so called \emph{Sine-Liouville theory}. The dual Sine-Liouville Lagrangian was defined in the work of~\cite{Fateev} as\footnote{In our conventions the strongly coupled region is for $\phi \rightarrow \infty$. These conventions are opposite to those of~\cite{Kazakov:2000pm}. We also set $\alpha' = 1$ for simplicity in the rest of the formulae. We finally replace the parameter $\lambda$ used in the work of~\cite{Kazakov:2000pm}, by the variable $\xi$, since $\lambda$ in our work is reserved to label partitions/representations.}
\be\label{I7}
L = \frac{1}{4 \pi} \left((\partial x)^2 + (\partial \phi)^2 + Q \hat{R} \phi + \xi e^{b \phi} \cos R \left(x_L - x_R \right) \right)
\ee
where in order to match it with the coset, we set the radius of $x$ to be $R = \sqrt{k}$ and
\be\label{I8}
c_{\text{cigar}} = c_{SL} = 2 + 6 Q^2 \, , \quad Q^2 = \frac{1}{k-2}  \, , \quad b =  \sqrt{k-2} = \frac{1}{Q}
\ee
We find again that conformal invariance sets $k=9/4$.
In this parametrisation, the asymptotic weakly coupled region is $\phi \rightarrow -\infty$ while the strongly coupled region is for $\phi \rightarrow \infty$ near the potential wall. The winding modes are not conserved, since in the original cigar the asymptotic circle shrinks to zero size at the tip, whereas in the Sine-Liouville description the symmetry is broken by the SL potential. The duality between the two models is a strong-weak duality in the sense that the cigar CFT is weakly coupled and semi-classical in the regime $k \rightarrow \infty \, , Q \rightarrow 0$, for which the wavefunction of the SL potential blows up near $\phi \rightarrow  \infty$. On the other hand for $k \rightarrow 2, \, Q \rightarrow \infty$, the SL theory is weakly coupled with a wavefunction supported far from the potential wall, while the coset description becomes highly curved and therefore strongly coupled. This can be given as evidence that for small radii, the black hole is better described in terms of a condensate of winding modes (SL-theory), while for large radii the black hole description is the simplest. A transition between these two behaviours is usually termed \emph{the black-hole string transition}~\cite{Susskind:1993ws,Horowitz:1996nw,Sen:1995in,Damour:1999aw}. 

In SL theory we can define the following winding operators
\be\label{I10}
\mathcal{T}^-_{\pm R} = e^{\pm i R(X_L - X_R) } e^{\left(Q - |Q - 1/Q| \right) \phi} \, , \quad \mathcal{T}^+_{\pm R} = e^{\pm i R(X_L - X_R) } e^{\left(Q + |Q - 1/Q| \right) \phi}
\ee
These are all operators of dimension $(1,1)$ and hence marginal. The upperscript sign refers to the two possibilities of SL-dressing. The $(-)$ case corresponds to a non-normalisable operator whose wavefunction grows at weak coupling $\phi \rightarrow - \infty$ and creates a local-disturbance on the worldsheet~\footnote{To compute the corresponding wavefunction of the operators in a linear dilaton background one needs to multiply them with a factor $\sim e^{- \Phi_0} e^{-Q \phi}\sim g^{-1}_s$.}. The $(+)$ case is a normalisable operator that creates a macroscopic loop/hole on the worldsheet (supported at strong coupling $\phi \rightarrow + \infty$). Another operator that one can consider is the black-hole operator (found by expanding the black hole background around the linear dilaton)
\be\label{I11}
B = \left(\partial X \bar{\partial} X - \partial \phi \bar{\partial} \phi \right) e^{2 Q \phi} \, ,
\ee
where the second term is pure gauge in the BRST cohomology. Turning on this operator deforms the CFT which then flows to the CFT of the Euclidean black hole. It was observed in~\cite{Mukherjee:2006zv} that the OPE between two SL operators~\ref{I10} with opposite dressing gives the black hole operator~\ref{I11} together with BRST trivial terms (and this continues to hold for higher windings). This indicates that in order to ensure exact marginality under both perturbations, one effectively includes the black hole operator into the Lagrangian. The dressing chosen in the original Sine-Liouville Lagrangian~\ref{I7}, was such that for $Q>1$ it corresponds to the non-normalisable operator of~\ref{I10} with a reasonable semiclassical limit for $Q \rightarrow \infty $ (small radius - strongly coupled coset). On the other hand, the structure of the OPE indicates that a more complete description of the black hole at different radii could involve a Sine-Liouville worldsheet theory in which both dressings are used, their importance being inverted with respect to the black hole string transition point, should such a point exist, see~\cite{Mukherjee:2006zv,Zamolodchikov:1995aa} for somewhat related ideas. Our motivation to adopt this perspective comes from sections~\ref{Saddles} and~\ref{thermo}, where we observe that the matrix model somehow seems to take into account both types of dressing, and one is able to change at will both the string coupling $g_s$ and the compactification radius $R$, allowing to describe different phases of the bulk string theory (with and without explicit winding sources from the perspective of the dual worldsheet string theory).

\section{The matrix model with winding (vortex) perturbations and its dual}\label{windingmatrixliouville}

The original Lagrangian of the $c=1$ Liouville string that is dual to gauged matrix quantum mechanics is\footnote{We should notice that the grand canonical ensemble in the matrix model corresponds to the presence of this Liouville potential, while the canonical ensemble to the presence of the term $\sim \phi e^{2 \phi}$. This gives rise to a change in sign for the genus zero term, between the two ensembles~\cite{Klebanov:1994pv,Kazakov:2000pm}.}
\be\label{I12}
L = \frac{1}{4 \pi} \left((\partial x)^2 + (\partial \phi)^2 + 2 \hat{R} \phi + \mu  e^{ 2 \phi}  \right) \, ,
\ee
and describes the linear dilaton/exponential tachyon background that shields the strongly coupled region $\phi \rightarrow + \infty$. Notice that now the value of $Q$ is fixed to $Q=2$ regardless of the compactification radius $R$, the physical reason being the presence of the non trivial tachyon. The a priori physical $(1,1)$ operators in this case are the vertex/vortex operators that take the form\footnote{There exist also additional discrete states, the black hole operator being among them, see~\cite{Mukherjee:2006zv} and references therein.}
\bea\label{I13}
{T}^-_{\pm n/ R} &=& e^{\pm i \frac{n}{R}(X_L + X_R) } e^{\left(2 - \frac{n}{R} \right) \phi} \, , \quad {T}^+_{\pm n/ R} = e^{\pm i \frac{n}{R}(X_L + X_R) } e^{\left(2 + \frac{n}{R} \right) \phi} \, , \nn \\
\mathcal{T}^-_{\pm n R} &=& e^{\pm i n R(X_L - X_R) } e^{\left(2 - n R \right) \phi} \, , \quad \mathcal{T}^+_{\pm n R} = e^{\pm i n R(X_L - X_R) } e^{\left(2 + n R \right) \phi}
\eea
We should now point that as long as $\mu \neq 0$, we recover one linear combination for the possible dressing due to the presence of the tachyon wall that reflects the modes. In particular only the $(-)$ (non-normalisable) modes are considered as the physical asymptotic vertex/vortex operators.
The idea used in~\cite{Kazakov:2000pm} was to deform the $c=1$ model using conformal perturbation theory (so that the resulting theory maintains conformal invariance at the quantum level) via the first winding terms\footnote{We shall replace the parameter $\lambda$ used in the work of~\cite{Kazakov:2000pm}, by the variable $\xi$, since $\lambda$ in our work is reserved to label partitions/representations.}
\be\label{I14}
L_{w} = \frac{1}{4 \pi} \xi e^{(2-R) \phi} \cos R \left(x_L - x_R \right) \, \sim  \, \mathcal{T}^-_{+ R} + \mathcal{T}^-_{- R} \, ,
\ee
and try to approach the $\xi \rightarrow \infty , \, \mu \rightarrow 0$ region of parameters (SL-point). This perturbation agrees with the SL term in~\ref{I7} for $R=3/2$, when $Q=2$, so one is describing the same system at this point. More generally, eqn.~\eqref{I14} makes sense as long as $R < 2$, so that the perturbation of the Lagrangian does not blow up in the asymptotic weakly coupled region and is a relevant deformation\footnote{Notice that similarly to what happens with the Sine-Liouville operators, the operators with the $(+)$ dressing in eqn. \eqref{I13} are always relevant and grow in the strongly coupled region $\phi \rightarrow \infty$.}.  It is also revealing to perform a KPZ-DDK scaling analysis of the action/free energy that indicates that the two independent scaling ratios are
$g_{s}/\mu$ and $g_{s}/\xi^{2/(2-R)}$. The relative strength is dictated by the parameter $z = \mu^{R-2} \xi^{2}$, that governs the relative dimensionless strength of the vortex perturbation (large $z$ corresponds to a strong perturbation). For the scaling of the $n$-th winding mode one has to replace $R \rightarrow n R$ in these formulae. Additionally we should emphasize that all these scalings depend crucially on the functional relation of $\xi$ with $\mu, R$ (it was assumed to be independent of them).

These last observations, led to consider a deformed matrix model that incorporates winding (vortex) perturbations~\cite{Kazakov:2000pm}. These winding modes correspond to the inclusion of Wilson loops to the original gauged matrix model. The mapping is $\mathcal{T}_{n R} \leftrightarrow  \tr P \left[ e^{i \oint A} \right]^n = \tr U^n $ between Liouville theory and matrix model operators (up to possible leg-pole factors). For the leg-pole factors, the recent analysis of~\cite{Balthazar:2017mxh} would indicate that the Liouville theory result will match the matrix model result once the operators on the Liouville side of the duality are properly normalised so that there is no need to introduce further leg-pole factors in the matrix model computations, but this should also be verified a posteriori by matching the computations on the two sides of the duality\footnote{This is because the analysis of~\cite{Balthazar:2017mxh} was performed for scattering modes and not winding modes, but we would expect that something similar should hold for the later as well.}. 

A basic quantity that one can describe in both sides of the MQM/Liouville duality is the general vortex perturbed free energy
\be\label{mainmatch}
F(t; \mu, R) = \left\langle  e^{ \sum_n t_n \left( \mathcal{T}_{n R}+ \mathcal{T}_{- n R}  \right) }   \right\rangle_c  = \log \mathcal{Z}_{MQM}(\tilde t ; \mu, R )
\ee
where the average contains only the connected contributions of the vortex correlators.
In this expression we did not include a superscript in the vortex operators in the Liouville side of the duality. The reason is that we expect the presence of different phases for our models in which the Liouville dressing of the vortex operators could potentially differ.

The matrix model grand canonical free energy is computed from the canonical one by
\be
\mathcal{Z}_{MQM}(\tilde{t}; \mu, R) = \sum_{N=0}^\infty e^{\beta \mu N} \langle e^{ \sum_n \tilde{t}_n \left( \tr U^n + \tr {U^\dagger}^n  \right) } \rangle_U \, , \qquad \beta = 2 \pi R \, .
\ee
The relation between the Liouville couplings $t_n$ and the matrix model couplings $\tilde{t}_n$ is\footnote{One can view this relation as an explicit winding mode leg pole factor between the two descriptions.}
\be 
t_n  = \frac{\tilde{t}_n}{2 i \sin n \pi R} \, .
\ee
This relation is derived upon realising the matrix model grand canonical partition function as a $\tau$-function~\cite{Kazakov:2000pm,Betzios:2017yms} of the general Toda hierarchy 
\be
\mathcal{Z}_{MQM}( \tilde t ; \mu + i k, R ) = e^{\sum_n n t_n t_{-n}} \tau_k( t ; \mu, R )  
\ee
We observe an additional possible parameter $k$ corresponding to the overall vacuum $U(1)$ ``charge" of the $\tau$-function. This parameter descends from the inclusion of a Chern-Simons term $k \int d \tau \tr A$ in the original matrix model action, but has not yet been given a very precise Liouville theory interpretation apart from the fact that it makes the string coupling complex~\cite{Betzios:2017yms} and seems to introduce some form of flux in the background. For a general $\tau$-function the individual couplings to vortices/anti-vortices could differ $\tilde{t}_{n} \neq \tilde{t}_{-n}$, and we shall label them by $\tilde t^+_n$ and $\tilde t^-_n$ with $n > 0$. The ``zero time" $\tilde{t}_0$ can be thought of as a conjugate variable to the parameter $k$, see~\cite{Betzios:2017yms}, for more details.

Since in the matrix model we are free to tune the radius of compactification $R$, it is natural to ponder what happens when $R >2 =R_{KT}$ and in particular for large $R$ where the geometry becomes semiclassical and string corrections are suppressed. On the other hand as we discussed, the deformation in the $c=1$ Liouville Lagrangian proposed by~\cite{Kazakov:2000pm}, see eqn. \eqref{I14}, makes sense only for $R<2$ and it was argued that for $R > 2$ the model simply flows to the undeformed $c=1$ Liouville theory (thermal linear dilaton background). In contrast we find that the class of matrix models with non singlets or winding perturbations~\cite{Kazakov:2000pm,Betzios:2017yms}, can somehow incorporate correctly both types of dressing for the operators existing in eqn. \eqref{I13}. In section~\ref{Saddles} we find a physically reasonable saddle (albeit metastable) to the effective matrix model equations even for radii $R>2$ and the appearance of the opposite ($+$) type of Liouville dressing for the winding modes, whose coupling should scale as $\xi^{2/(R+2)} \sim 1/g_s$ via the KZP/DDK analysis. If this saddle corresponds to a black hole in this parameter regime, this would mean that it cannot be described just by the coset $SL(2,R)_k/U(1)$ WZW model that is conformal only for $R^2 = k =9/4$ and neither by Sine-Liouville if we do not include the second $(+)$ type of dressing for the winding modes. We therefore have to resort to the analysis of the matrix model and try to infer from it the properties of the dual geometric background in this regime.

\section{Partition function and representations}\label{MQMnonsinglets}

\subsection{Expanding the $\tau$-function in terms of representations}

As alluded to above the partition function including arbitrary winding modes, parametrised by a collection of Miwa ``time"variables $t_+, t_-$, can generally be expressed as a $\tau$-function
\be
\mathcal{Z}(\tilde{t}_+, \tilde{t}_- ;\mu + i k, R) \equiv \tau_k( t_+ , t_- ;\mu ,R) \, .
\ee
The $\tau$-function is also described as an expectation value
\be\label{C32}
\tau_k( t_+ , t_- ;\mu, R ) = \langle k | e^{J_+(t_+)} {\bf G} e^{- J_-(t_-)}   | k \rangle
\ee
In this expression $J_{\pm}(t_\pm)$ are the currents generating the ``time" flows, see appendix~\ref{Fermionic} for more details. The $GL(\infty)$ element/operator ${\bf G} = {\bf G}(\mu, R)$ can be expressed as a bilinear of free fermion operators. The state $|k \rangle$ describes an overall $U(1)$  charge $k$ that was found to be tunable
by introducing a Chern-Simons type of term in MQM~\cite{Betzios:2017yms}. The model of~\cite{Kazakov:2000pm} and the ones we study in this work have $k=0$. In appendix~\ref{hierarchies} we provide more details about free fermions, $\tau$-functions and the specific realisation of the $GL(\infty)$ element for our MQM system.

Under T-duality the $\tau$ function is also a generating function for the reflection amplitudes as shown in~\cite{Dijkgraaf:1992hk}. In this case the $GL(\infty)$ element has the interpretation of an S-matrix element or reflection amplitude ${\bf G}(\mu, R) \equiv {\bf S}(\mu, R)$. Even though one usually defines the S-matrix in Lorentzian time,  in this equation we kept the Euclidean time compactified and considered the scattering/reflection of modes on a Euclidean cylinder of radius $R$, see appendices~\ref{reflctau} and~\ref{fixedrepreflect} for more details on this notion of scattering. T-duality also allows us to acquire the winding mode partition function by computing instead the $\tau$ function containing the reflection amplitude, and then performing a T-duality operation in the result, by sending $\mu \rightarrow \mu R \, , \, R \rightarrow 1/R$ to relate them. Remarkably, we can verify that the operation of T-duality in the matrix model can be performed without the need of including additional leg-pole factors between the two descriptions, see appendix~\ref{MQMnonsingletsapp}.

There exists a very useful representation theoretic expansion for the $\tau$ function that we shall use in the rest and which will allow us to give a meaning to the microstates comprising the long string condensate/black hole background. Using the formulae in section~\ref{Tau} and in particular \eqref{C17}
\bea
e^{J_-(t_-)} | k \rangle &=& \sum_\lambda (-1)^{b(\lambda)} s_\lambda(t_-) | \lambda ; k \rangle \nn \\
\langle k | e^{J_+(t_+)}   &=& \sum_\lambda (-1)^{b(\lambda)} s_\lambda(t_+) \langle \lambda ; k | \, ,
\eea
where the state $| \lambda ; k \rangle$ corresponds to a representation/partition $\lambda$ created by acting with fermions on the vacuum of charge $k$ (see appendices~\ref{Partitionsreps} and~\ref{Frobeniusreps} for more details on partitions and the definition of these states), we can expand the $\tau$ function as a statistical sum in terms of transition amplitudes between different representations as follows:
\bea\label{C34}
\tau_k( t_+ , t_- ;\mu, R) = \sum_{\lambda , \nu} {\bf G}^{(k)}_{\lambda \nu}(\mu + i k , R) s_\lambda(t_+) s_\nu( t_-)\, , \nn \\
{\bf G}^{(k)}_{\lambda \nu}(\mu + i k, R) = (-1)^{b(\lambda)} (-1)^{b(\nu)} (-1)^{|\nu|} \langle \lambda ; k | {\bf G}(\mu , R) | \nu ; k \rangle \, .
\eea
In these expressions the summations over $\lambda, \nu$ are summations over all possible Young diagrams that describe the different partitions/representations and $s_\lambda(t_\pm)$ are Schur polynomials/characters in a Miwa time notation using the time variables $t_\pm$, see appendix~\ref{Schur} for details. The sign factors $(-1)^{b(\lambda)}$ can be written explicitly (see~\cite{Alexandrov:2012tr} and the appendix), but will not play any role since it is possible to show that MQM dynamics is diagonal in the representation basis and only ${\bf G}^{(k)}_{\lambda \lambda}$ are non trivial, see appendices~\ref{reflctau} and~\ref{windtau} for a proof of this statement and equations~\eqref{PartitionRep} and~\eqref{C36} for the explicit $GL(\infty)$ element.

\subsection{Measures in the space of representations}\label{Measures}

Using the formula for the $\tau$ function~\eqref{C34} (and taking into account that the $GL(\infty)$ element is diagonal for the double scaled MQM and that the vacuum charge $k$ is set to zero)
\be\label{taumain}
\mathcal{Z} = \tau_k( t_+ , t_- ;\mu) = \sum_{\lambda } s_\lambda(t_+) s_\lambda(t_-) (-1)^{|\lambda|} \langle \lambda ; 0 | {\bf G} | \lambda ; 0 \rangle \, ,
\ee
we find the presence of a general measure that weights the representations/partitions $\lambda$, the so-called Schur-measure~\cite{Okounkov1}
\be\label{measuredefinition}
\mathfrak{M}_\lambda(t) =  \frac{1}{Z_0} s_\lambda(t_+) s_\lambda( t_-) \, , \qquad Z_0 = \sum_\lambda s_\lambda(t_+) s_\lambda( t_-) = e^{ \sum_{k>0} k t_k^+ t_k^-} \, .
\ee
In this expression we have parametrised the Schur polynomials $s_\lambda(x)$ using the Miwa variables $t_k = \sum_i x_i^k/k$, see appendix~\ref{Tau} for some explicit formulae. The expected average size of the partition with respect to the Schur measure is
\be\label{partsize}
\langle |\lambda| \rangle_\mathfrak{M} = \sum_k k^2 t_k^+ t_k^- \, ,
\ee
and hence becomes very large, for large deformations/time variables. If we wish to work with Young diagrams
of fixed area (fixed total number of boxes) we can rewrite the sum over partitions as follows
\be
\sum_\lambda = \sum_{n=0}^\infty \sum_{\lbrace \lambda_i \rbrace} \delta \left( \sum_{i=1}^{\ell} \lambda_i  - n \right) \, ,
\ee
the zeroth term capturing the trivial representation (no boxes).

We shall now describe in detail some important specialisations of the Schur measure that appear also in the actual physical problems we are interested in,~\cite{Borodin1,Vershik1,Kerov} provide a mathematical perspective on these cases\footnote{Appendix~\ref{Schur} contains some additional information about these specialisations of Schur polynomials.}.

\begin{itemize}

\item Turn on only the first winding mode/time parameter. If one sets $t^+ =  t^- = (\xi, 0 , 0 ...)$, one obtains the ``Poissonised Plancherel" measure in the space of partitions $\lambda$
\be\label{poissPlancherel}
\mathfrak{M}_\lambda(\xi) = e^{- \xi^2} \xi^{2 |\lambda|} \left(\frac{\dim \lambda}{|\lambda|!}\right)^2 \, = \,  e^{- \xi^2} \sum_{n=0}^\infty \frac{\xi^{2 n}}{n!} M_n(\lambda) \delta(|\lambda| - n)\,,
\ee
where the quantity $M_n(\lambda)$ in eqn. \eqref{poissPlancherel} is called the ``Plancherel measure" on the partitions of $n$ 
\be\label{Planch}
M_n(\lambda) = \frac{(\dim \lambda)^2}{n!} \, , \quad |\lambda|=n\,.
\ee
The relevant measure in the model of Kazakov-Kostov-Kutasov~\cite{Kazakov:2000pm} is precisely this ``Poissonised Plancherel" measure\footnote{We shall replace the parameter $\lambda$ used in the work of~\cite{Kazakov:2000pm}, by the variable $\xi$, since $\lambda$ in our work is reserved to label partitions/representations.}.

In the shifted highest weight coordinates $h_i = \lambda_i - i + N$ ($\ell(\lambda) = N$), the ratio involving the dimension of the partition is~\cite{Macdonald}
\be
\frac{\dim \lambda}{|\lambda|!} = \frac{\prod_{i<j \leq N}(h_i - h_j)}{\prod_{i\leq N} h_i !} \, .
\ee
The same ratio in the half-integer Frobenius coordinates $p_i,q_j \in \mathbb{Z}^{+} + \half$, defined in appendices~\ref{Partitionsreps} and \ref{Frobeniusreps}, can be re-expressed as
\be\label{dimensionFrob}
\frac{\dim \lambda}{|\lambda|!} = \frac{1}{\prod_{i=1}^d (p_i-\half)! (q_i-\half)!} \det_{1 \leq i,j \leq d} \left[ \frac{1}{p_i + q_j} \right] \, .
\ee
These formulae are useful in order to explicitly express the partition function as a sum over the Frobenius half integers, see appendix~\ref{determinantal} and section~\ref{Saddles}.

As the size of the partitions goes to infinity $n \rightarrow \infty$, the Plancherel measure \eqref{Planch} exhibits a Cardy-like growth
\be
\lim_{n \rightarrow \infty} \, M_n(\lambda) \,  \sim \,  \exp \left(2 \sqrt{n} \right) \, ,
\ee
and concentrates to a universal limiting shape for the Young diagrams\footnote{See appendix~\ref{ContRep} for more details on how to appropriately define the limit of continuous representations and the $y$ coordinates.}, the \emph{Vershik-Kerov-Logan-Shepp limiting shape}~\cite{Vershik1,Logan}
\be
\mathbf{\Omega}(y) = \begin{cases} \frac{2}{\pi} \left( y \arcsin (y/2) + \sqrt{4 - y^2} \right) \, , \quad &|y|<2  \, , \\
|y|\, ,\quad &|y|>2  \, ,
\end{cases}
\ee
that we depict in fig.~\ref{fig:Plancherel}.

\begin{figure}
\begin{center}
\includegraphics[width=0.45\textwidth]{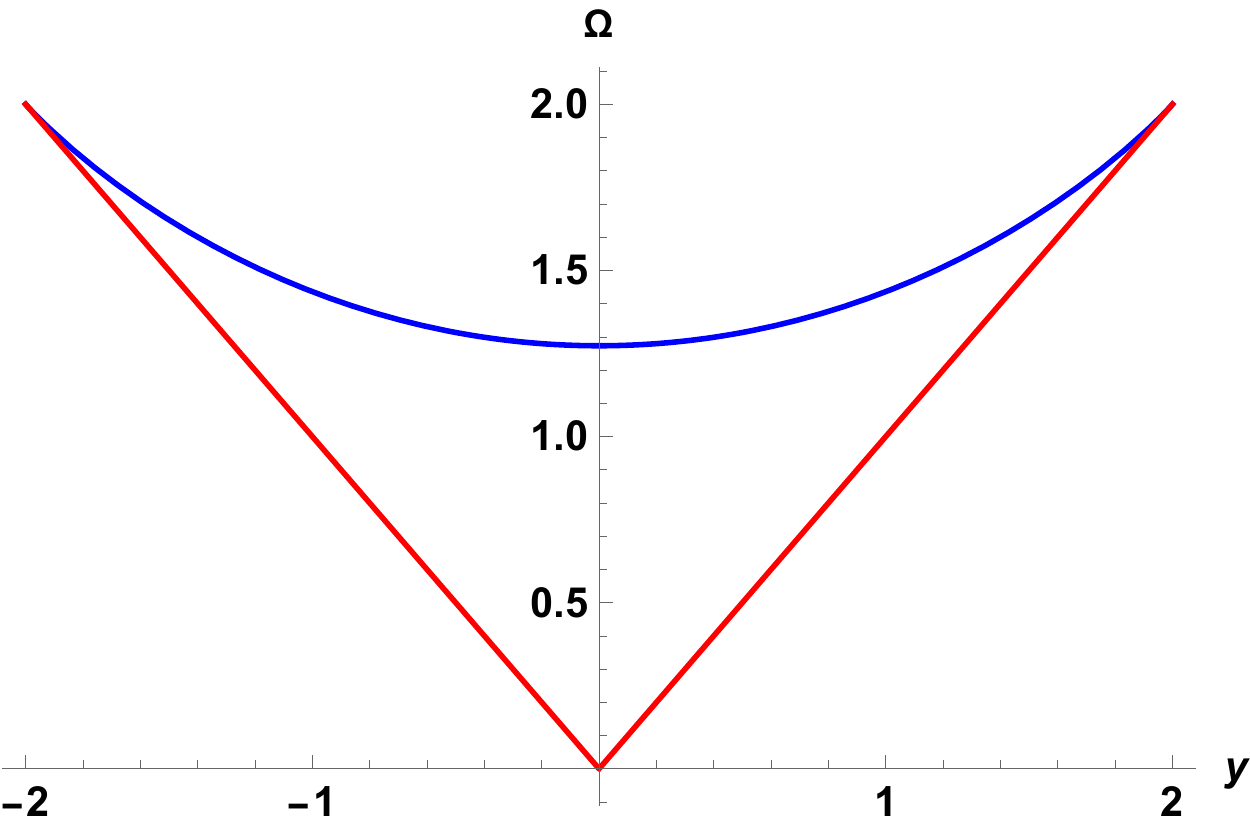} \hspace{0.1em} \hspace{0.1em} \includegraphics[width=0.45\textwidth]{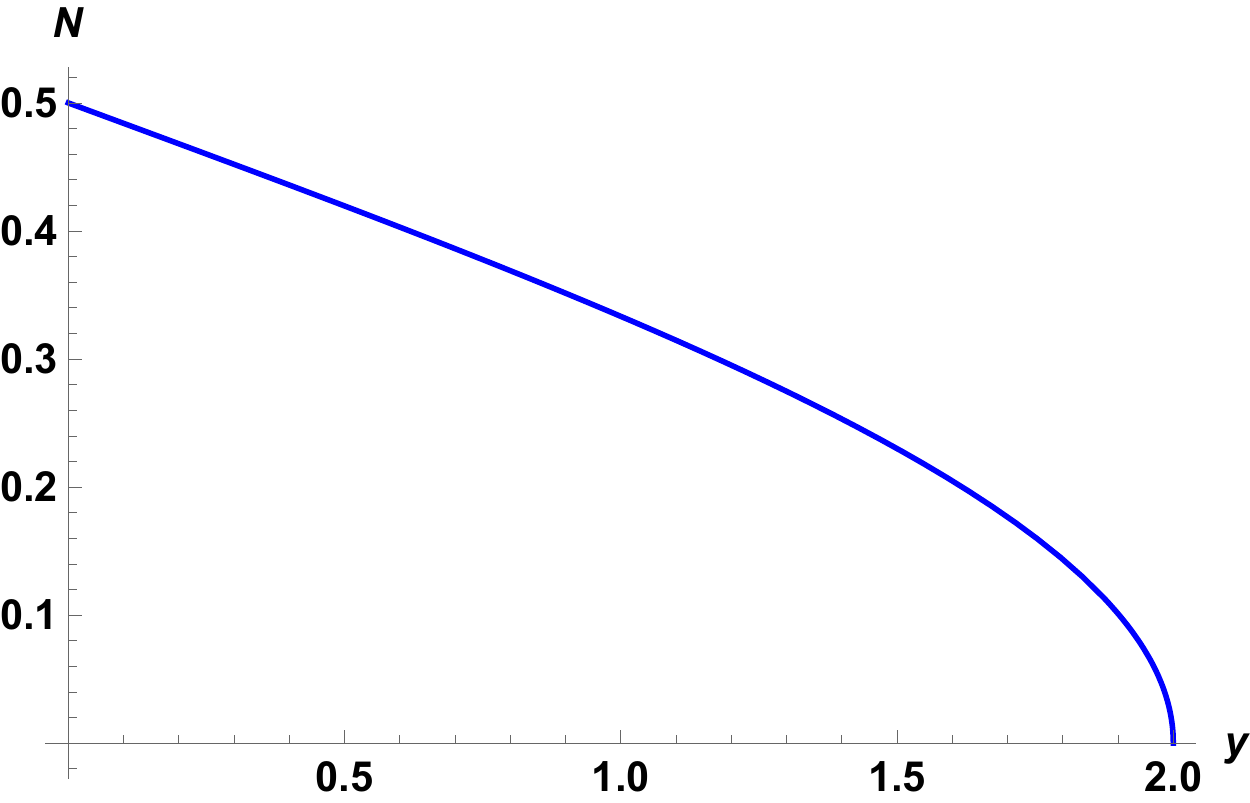}
\end{center}
\caption{The limiting Young diagram shape $\mathbf{\Omega}(y)$ and density of boxes $\mathcal{N}(y)$  for the Plancherel measure.}
\label{fig:Plancherel}
\end{figure}

The density of boxes in this limit is found by 
\be
\mathcal{N}(y) = \frac{1 - \mathbf{\Omega}'(y)}{2} = \frac{1}{\pi} \arccos(y/2) \, , \quad \text{for} \quad 0 < y < 2 \, ,
\ee
An alternative physical method to determine the density of boxes and diagram shape using a coherent state formalism is presented in appendix~\ref{Coherent}.

\item A more general and fundamental case is the ``Poissonised z-measure"~\cite{Borodin2,Kerov,Okounkov2} (relevant for the models of the type studied in~\cite{Betzios:2017yms} that do admit a Hamiltonian description). In this case the specialisation of the time variables is the following
\be
t^+_k = z \frac{\xi^{k}}{k} \, , \qquad  t^-_k = z' \frac{\xi^{k}}{k} \, ,
\ee
and the measure reads
\bea
\mathfrak{M}_{z, z'}(\lambda) &=& \left(1 - \xi^2 \right)^{z z'} \xi^{2 |\lambda|} s_\lambda \underbrace{(1,1,1...)}_{z-times} s_\lambda \underbrace{(1,1,1...)}_{z'-times} \, \nn \\
&=& \left(1 - \xi^2 \right)^{z z'} \xi^{2 |\lambda|} (z)_\lambda (z')_\lambda \left(\frac{\dim \lambda}{|\lambda|!}\right)^2 \, ,
\eea
with $(a)_k = a(a+1) ... (a+k-1)$ the rising Pochhammer symbol. For the model of~\cite{Betzios:2017yms}, the physical parameters are $z= z'= N_f$ counting the number of (anti)-fundamentals (FZZT branes), and $\xi = e^{-\beta m} = e^{- \pi \sigma R}$ is related to their masses (or the boundary cosmological constant $\mu_B$ of the open strings ending on the FZZT branes via the relation $\mu_B = \sqrt{\mu} \cosh \pi \sigma  $). This parameter plays again the role of a fugacity for the winding modes.

The Poissonised Plancherel measure~\eqref{poissPlancherel} is obtained in a limit ($N_f, \, m  \rightarrow \infty$ with $N_f e^{-\beta m} $ fixed) of this more general measure in the space of all partitions~\footnote{This limit was first mentioned in~\cite{Maldacena:2005hi} and studied in detail in~\cite{Betzios:2017yms} from a physical perspective. It corresponds to a large number of heavy ``quenched" open strings (FZZT branes) that can be replaced by the insertion of an exponential of the Wilson loop operator in the partition function.}. 
Using the negative binomial distribution
\be
\pi_{z z', \xi}(n) = (1-\xi^2)^{z z'} \frac{(z z')_n}{n!} \xi^{2n} \, , \quad \xi \in (0,1) \, ,
\ee
we then find the relation
\be
\mathfrak{M}_{z, z'}(\lambda) = \sum_{n=0}^\infty  \, \pi_{z z', \xi}(n) \, M^{(n)}_{z, z'}(\lambda) \, \delta(|\lambda| - n) \, , 
\ee
in terms of the de-Poissonised z-measure, which in Frobenius coordinates reads
\be\label{z-measure}
M^{(n)}_{z, z'}(\lambda) = \frac{n! (z z')^d }{(z z')_n} \prod_{i=1}^d \frac{(z+1)_{p_i-\half} (z'+1)_{p_i-\half} (-z+1)_{q_i-\half} (-z'+1)_{q_i-\half}}{(p_i- 1/2)! (p_i- 1/2)! (q_i- 1/2)! (q_i- 1/2)!} \det \left[\frac{1}{p_i + q_j} \right]^2
\ee
where $n = |\lambda|$. An explicit formula for the limiting form of this measure when $n \rightarrow \infty$ does not appear to exist in the literature. Nevertheless we manage to obtain this limiting form using the coherent state method in appendix~\ref{Coherent}.







\end{itemize}

We notice that in general the Poissonised measures, weigh the partition function in each fixed irrep~\ref{taumain} and contain extra parameters in the form of chemical potentials/fugacities. One needs to de-Poissonise such measures in order to pass to associated measures in the space of representations/partitions of fixed size $n = |\lambda|$. This can be achieved via a transformation between the dual variables $\xi \leftrightarrow n$, for example the free energies are related by 
\be\label{fixedsizetransform}
e^{-F(n)} = \oint \frac{d \xi}{2 \pi i \xi^{2 n + 1}} 	\, e^{-\mathcal{F}(\xi)} \, .
\ee
The physical interpretation of $F(n)$, is that of a free energy containing a fixed number $n$ of vortex anti-vortex pairs~\cite{Kazakov:2001pj}. We explain in more detail the thermodynamic interpretation and properties of these two ensembles in section~\ref{thermo}.

\subsection{Microstates and the origin of entropy (thermodynamic limit)}

The previous analysis of the partition function, written as a sum over representations of $GL(\infty)$ (that are labelled by a set of ordered integers - a partition), elucidates the discrete nature of the microstates that the non-singlet models are composed of.  

It is natural then to inquire about the nature of the coarse grained entropy contained in the thermodynamic limit of the non-singlet models. In particular we would like to understand whether in the associated string theory expansion it would correspond to a genus zero ``classical" entropy, scaling as $S \sim 1/{g_{s}^2}$, as expected from a thermodynamic analysis of the two dimensional black hole~\cite{Gibbons:1992rh,Nappi:1992as,Kazakov:2001pj}\footnote{Since there is no area of a horizon in $2d$, the entropy is related to the value of the dilaton at the tip of the cigar geometry.}. Defining the thermodynamic quantities
\be
\log \mathcal{Z} = - \mathcal{F} \, , \qquad \mathcal{F} = \beta M - S \, ,  \qquad S = \left( R \partial_R - 1 \right) \mathcal{F} \, ,
\ee
we observe that terms in the free energy that are linear in $R$, do not contribute in the entropy of the system. In particular we show in appendix~\ref{MQMnonsingletsapp} that for a fixed irreducible representation, the piece of the partition function~\eqref{taumain} related to the scattering amplitude/MQM dynamics $ \langle \lambda ; 0 | {\bf G} | \lambda ; 0 \rangle \equiv {\bf G}_{\lambda \lambda}^{(k=0)}(\mu, R)$ does not contribute to any classical form of entropy (but only to a quantum mechanical form of entropy arising from loops/higher string genera). Any genus zero entropy in the models of~\cite{Kazakov:2000pm,Betzios:2017yms}, can arise from the presence of the Schur measure in the space of partitions and in particular from the Plancherel/z-measures that contain the dimension of the representation/partition $\dim \lambda$. In particular the Cardy like growth of these measures ($ M_n(\lambda) \sim \exp 2 \sqrt{n}$) as we increase the size $n$ of the partitions is consistent with the presence of an object in the dual geometric background resembling a black hole.

In order to clarify this discussion, it is important to understand the precise fashion in which one should take the thermodynamic limit in order to connect with the coarse grained geometric properties of the dual background. We propose that this limit is exactly the limit of continuous representations/partitions. In this limit, we shall find that there exist \emph{leading shapes governing the Young diagrams}, each shape corresponding to a different phase of the model, that is determined dynamically by solving appropriate saddle point equations (see section~\ref{Saddles}). These saddles are also expected to correspond to the different geometric backgrounds on which the dual string is propagating. The entropy is provided by having many different partitions
with similar coarse grained characteristics in the aforementioned thermodynamic limit.
In order to check this proposal, we should compare the thermodynamic quantities on the two sides of the duality between the matrix model and string theory (and any other observables that we can compute). This comparison, along with further comments regarding thermodynamics can be found in section~\ref{thermo}.

\section{The limit of continuous representations}\label{Saddles}

\subsection{Effective action and saddle point equations }

If we wish to understand the limit where the ``time" variables (or the deformations away from the singlet MQM) become large, according to eqn. \eqref{partsize}, we need to understand the limit of large Young diagrams/partitions. If we scale the Young diagram appropriately, this is a limit of continuous representations, where the shape of the diagram acquires a continuous curve (the boxes grow in number but their size shrinks so that we keep the area of the diagram fixed).  The precise fashion in which one can implement this continuum limit is described in appendix~\ref{ContRep} and follows the analysis of~\cite{Douglas:1993iia}.

The partition function (or its T-dual) at fixed representation is most naturally described in Frobenius coordinates (see appendices~\ref{reflctau} and~\ref{windtau}), and we shall use this coordinate system in what follows. In this case, we introduce two positive semi-definite continuous densities
$p(x), q(y) \geq 0$, by defining
\be
y = \frac{j}{d} \, , \quad q(y) = \frac{q_j}{d} \, , \qquad x = \frac{i}{d} \, , \quad p(x) = \frac{p_i}{d} \, ,
\ee
where $d(\lambda)$ is the number of diagonal elements of the partition\footnote{This number scales as a fraction of the square root of the area of the diagram in the thermodynamic limit.} and then taking the $d \rightarrow \infty $ limit. Both densities obey the inequality
\be
p(x_1)\geq p(x_2) \, , \, \, \, \,  x_1 \leq x_2 , \qquad q(y_1)\geq q(y_2) \, , \, \, \, \,  y_1 \leq y_2 .
\ee
The corresponding densities of boxes are
\be
 \mathcal{N}(q) = - \frac{d y}{d q}\,  , \qquad  \widetilde{\mathcal{N}}(p) = - \frac{d x}{d p} \, , 
\ee
and obey the constraint $\mathcal{N}(q), \, \widetilde{\mathcal{N}}(p)  \, \leq 1 $ due to the aforementioned inequalities. We can also normalise them to $\int \mathcal{N}(q) d q = \int  \widetilde{\mathcal{N}}(p) d p = 1/2 $ (if the diagram is symmetric). The replacement rule for sums is $d^{-1} \sum_{i=1}^d \rightarrow \int_0^1 d x$. 

In the explicit summand for the partition function~\eqref{taumain}, one finds the presence of several Gamma functions, either from the specific Schur measures we are interested in (eqns. \eqref{z-measure} and \eqref{dimensionFrob}), or from the $GL(\infty)$ element that is expressed in terms of the reflection amplitude, see eqn.~\eqref{PartitionRep}. In the large-$d$ limit, one can use Stirling's approximation for the Gamma functions to find
\be\label{largedgamma}
\sum_{i=1}^d \log \Gamma \left(p_i + \half \right) \, \rightarrow d^2 \int_0^1 d x \, p(x) ( \log p(x) - 1)  \, .
\ee
For the part in the free energy containing the $GL(\infty)$ element/reflection amplitude (see eqn.~\eqref{PartitionRep} and eqn.~\eqref{Freeenergygeneralrep}), the limit is more involved and in general depends on the non-perturbative definition of the amplitude or scattering phase, see appendix~\ref{fixedrepreflect}. If we use the perturbative expression for the reflection amplitude, it takes the following explicit form
\be\label{BBB}
{\bf G}_{\lambda \lambda}^{(k=0)}(\mu, R) = Z_{singlet}  \prod_{j=1}^d \sqrt{ \frac{\Gamma\left(\half + i \mu R + p_j R \right)}{\Gamma\left(\half - i \mu R - p_j R \right)}} \sqrt{ \frac{\Gamma\left(\half - i \mu R + q_j R \right)}{\Gamma\left(\half + i \mu R - q_j R \right)}} \, .
\ee
Expanding the logarithms of the $\Gamma$ functions, if we do not scale $\mu$, then it drops out and the result becomes independent of it. On the other hand we can choose to keep $\mu/d = \mu_r$ fixed as a renormalised coupling in the limit of large representations. The (perturbative\footnote{It would be very interesting to extend our analysis, capturing non-perturbative effects.}) scattering phase appearing in~\eqref{BBB} contains the following leading contributions in the $1/d$ expansion

\be\label{largedreflection1}
i \sum_{j=1}^d \Phi_{pert.}\left(-i p_j R + \mu R  \right) \, \rightarrow d^2 \int_{0}^{1} dx\, R\left(p(x)+i \mu_r\right)\left(\log\left[R\left(p(x)+ i  \mu_r \right)\right]-1\right)
\ee


\be\label{largedreflection2}
- i \sum_{j=1}^d \Phi_{pert.}\left(i q_j R + \mu R  \right) \, \rightarrow d^2 \int_{0}^{1} dy\, R\left(q(y)-i \mu_r\right)\left(\log\left[R\left(q(y)- i  \mu_r \right)\right]-1\right)
\ee
Using the continuum variables, the partition function and free energy are expressed in the following form\footnote{If we wish to include the contribution of the singlet part of the free energy into $S_{eff}$, one simply has to replace $\mathcal{N} \rightarrow \mathcal{N} - \half$ in the expressions involving the densities, see appendix~\ref{ContRep} for a proof.}
\be\label{pathintegralpartition}
\mathcal{Z}\, = \, e^{- \mathcal{F}} \, = \, e^{- \mathcal{F}_{singlet}} \, \int D p D q e^{- d^2 S_{eff}(p,q)} \, ,
\ee
with the the leading order effective action for the model of~\cite{Kazakov:2000pm} that contains only the first winding modes being (see appendix~\ref{bifundeffective} for the effective action of the more general bi-fundamental model of~\cite{Betzios:2017yms}) 
\bea\label{effectiveaction1}
S_{eff} &=&  - \int_0^1 d x \int_0^1 d y  \, \log \frac{|(p(x) - p(y))(q(x) - q(y))|}{(p(x)+q(y))^2}   \, + \tilde{V}(p) + V(q) \, , \nn \\
\tilde{V}(p) &=& \int_0^1 d x \, 2 p(x) ( \log p(x) - 1 -  \log \xi)  - R\left(p(x)+i \mu_r\right)\left[\log\left[R\left(p(x)+ i  \mu_r \right)\right]-1\right], \, \nn \\
V(q) &=& \int_0^1 d x \,  2 q(x) ( \log q(x) - 1 -  \log \xi)  - R\left(q(x)-i \mu_r\right)\left[\log\left[R\left(q(x)- i  \mu_r \right)\right]-1\right]\, . \nn \\
\eea
This effective action can also be derived from a coupled (normal) two-matrix integral by turning the sum over Frobenius coordinates $(p_i | q_j)$ to a two matrix integral, see appendix~\ref{couplednormal} for the details.

The resulting saddle point equations of this effective action are a coupled system of equations written in terms of the densities of boxes (we shall use $\dashint$ for a shorthand of the principal value integral $\mathcal{P} \int$ and drop the index $r$ in $\mu_r$)
\bea\label{saddlepoint1}
 \log \left(\frac{q^2}{\xi^2} \right) - R \log\left[R\left(q - i \mu \right) \right]\,  = \, 2  \dashint d s \frac{\mathcal{N}(s)}{q - s} \, - 2  \dashint d s \frac{\tilde{\mathcal{N}}(s)}{q + s}\, , \nn \\
 \log \left(\frac{p^2}{\xi^2} \right) - R \log\left[R\left(p + i\mu \right)\right] \,  = \, 2 \dashint d s \frac{\tilde{\mathcal{N}}(s)}{p - s} \, - 2  \dashint d s \frac{{\mathcal{N}}(s)}{p + s}\, .
\eea
By adding and subtracting the equations in~\eqref{saddlepoint1} we find
\bea\label{saddlepoint2}
 2\log \left(\frac{p^2}{\xi^2} \right) - R \log\left[R^2 \left(p^2 + \mu^2 \right) \right]\,  = \, 4  \dashint d s  \frac{s(\mathcal{N}(s)+\tilde{\mathcal{N}}(s))}{p^2 - s^2}\, , \nn \\
 - R \log\left[\frac{p - i\mu}{p + i\mu}\right] \,  = \, 4p \dashint d s \frac{(\mathcal{N}(s)-\tilde{\mathcal{N}}(s))}{p^2 - s^2}
\eea
We observe that any spectral asymmetry in the Frobenius coordinates will lead to terms that are odd in the $1/\mu$ expansion (open string/instanton contributions) and that the densities are complex conjugates to each other $\tilde{\mathcal{N}}(s) = {\mathcal{N}}^*(s)$, so that
the spectral asymmetry will contribute only to the imaginary part of the free energy (decaying states). It is also possible to introduce the variable $u=p^2$ and rewrite the first equation of \eqref{saddlepoint2} in the form
\be\label{saddlerealpart}
 \log \left(\frac{u}{\xi^2} \right) - \frac{R}{2} \log\left[R^2 \left(u + \mu^2 \right) \right]\,  = \, 2  \dashint d v  \frac{\Re \mathcal{N}(u)}{u - v}\, .
\ee
The on shell action using the equations of motion can be written as
\bea\label{onshell1}
S_{eff} \, = \,  \half \int d p \mathcal{N}(p) \tilde{V}(p) + \half \int d q \tilde{\mathcal{N}}(q) V(q) + \half \tilde{C} + \half C \, \nn \\
\tilde{V}(p) - 2 \int d s \tilde{\mathcal{N}}(s) \log |p-s| + 2 \int d s {\mathcal{N}}(s) \log |p+s| = {C} \, \nn \\
V(q) - 2 \int d s \mathcal{N}(s) \log |q-s| + 2 \int d s \tilde{\mathcal{N}}(s) \log |q+s| = \tilde{C} \, .
\eea
We then define the resolvent $\omega(q)$
\be
\omega(q) =  \int_{\text{supp.}} d s \, \frac{\mathcal{N}(q)}{q - s} \, , \quad    \quad \mathcal{N}(q) = - \frac{1}{2 \pi i} \left( \omega(q+ i \epsilon) - \omega(q - i \epsilon) \right)   \, ,
\ee
and similarly $\tilde{\omega}(p)$ for the $p$ distribution. The resolvents also obey
\be
\omega(q) = \slashed{\omega}(q) \mp i \pi \mathcal{N}(q) \, , \qquad \tilde \omega(p) = \tilde{\slashed{\omega}}(p) \mp i \pi \tilde{\mathcal{N}}(p) \, .
\ee
Using these properties of the resolvents one can express the saddle point equations \eqref{saddlepoint1} as follows
\bea\label{saddlepoint3}
 \log \left(\frac{q^2}{\xi^2} \right) - R \log\left[R\left(q - i \mu \right) \right]\,  = \, 2  \slashed{\omega}(q) \, + 2  \tilde{\omega}(-q) = 2 \slashed{\Omega}(z=q)  \, , \quad q-\text{cuts} \, , \nn \\
 \log \left(\frac{p^2}{\xi^2} \right) - R \log\left[R\left(p + i\mu \right)\right] \,  = \, 2  \tilde{\slashed{\omega}}(p) \, + 2  {\omega}(-p) = 2 \slashed{\Omega}(-z = p) \quad p-\text{cuts}  \, ,
\eea
where we also defined the total resolvent $\Omega(z) = \omega(z) + \tilde{\omega}(-z)$. This total resolvent has cuts in both the positive and negative $z$ plane, the first being the $q$-cuts and the second the $p$-cuts, making it possible to rewrite equations \eqref{saddlepoint3} in terms of the total resolvent. The non-singlet susceptibility of the effective action \eqref{effectiveaction1} is computed using
\be\label{nssuscept}
\chi^{n.s.} \, = \, - \partial^2_{\mu} S^{eff} \, = -  R \int_{supp.} d p \, \frac{\tilde{\mathcal{N}}(p)}{p + i \mu}  \, - R \int_{supp.} d q \, \frac{{\mathcal{N}}(q)}{q - i \mu}  \, = \,  R \, \Omega(z =  i \mu) \, .
\ee
Its value is hence related to that of the total resolvent $\Omega(z)$ (analytically continued in a region outside its physical cut). Another useful quantity to determine is the first derivative of the on-shell action with respect to $\xi$
\be\label{firstderxi}
-  \frac{\partial S_{eff.}}{\partial \xi}  = \frac{2}{\xi} \left(\int_{supp.} d p \tilde{\mathcal{N}}(p) p  + \int_{supp.} d q \mathcal{N}(q) q  \right) \, .
\ee
Using eqns.~\eqref{poissPlancherel} and~\eqref{pathintegralpartition}, this is related to the average size of the partition via
\be\label{averagepartitionsize}
\langle |\lambda| = n \rangle =  \half \xi \frac{\partial \log \mathcal{Z}}{\partial \xi} = - \half \xi \langle \frac{\partial S_{eff.}}{\partial \xi} \rangle \, ,
\ee
where the last term is approximated by the on-shell action as in~\eqref{firstderxi}. One can in principle compute also the fluctuations around the dominant Young diagram, by expanding the action around its saddle point.

One natural question then, is to compare the properties of the spectral curve we shall acquire in the limit of large representations, with the genus zero spectral curve (string equation) found in~\cite{Kazakov:2000pm,Kostov:2001wv}
\be\label{spectralKKK}
\mu \, e^{\chi_0/R} \, + \, (R-1) \, \xi^2 \, e^{(2-R)\chi_0/R} \, = \, 1 \, , \qquad \chi_0 = -\partial_\mu^2 \mathcal{F}_0 \, .
\ee
In this formula $\chi_0 = - R \log \mu + \chi_0^{n.s.}$ with $\chi_0^{n.s.}$ the genus zero part of the non singlet susceptibility. This should be captured by eqn. \eqref{nssuscept}.

In addition we would also like to understand the different phases that the model exhibits in the thermodynamic limit of large representations, since the effective  potential we found is complicated enough to allow for this possibility, as we describe in section~\ref{determiningresolvent}. We expect that the result for the free energy in~\cite{Kazakov:2000pm,Kostov:2001wv}, obtained solving the Toda equations (or from the spectral curve \eqref{spectralKKK}), with an initial state that corresponds to the undeformed thermal linear dilaton background, corresponds only to one phase of the model. In particular that approach is not expected to be able to capture the possibility of having phase transitions because the said solution is continuously connected to the $\xi = 0 $ linear dilaton solution. For more details, see  section~\ref{thermo}.

\subsection{Determining the resolvent}\label{determiningresolvent}

Since we would like to describe a leading order closed string background, we can consider only reflection symmetric Young diagrams $\tilde{\mathcal{N}}(s) = {\mathcal{N}}(s)$ with vanishing spectral asymmetry (having only a real contribution to the free energy). Inspecting eqns. \eqref{saddlepoint1}, this approximation becomes exact in the limit $\mu \rightarrow 0$ (\emph{Sine-Liouville limit} of section~\ref{windingmatrixliouville}), when the backreaction of the winding modes on the linear dilaton background is strong and the tachyon potential vanishes. In this case we can simplify the set of equations \eqref{saddlepoint1} or \eqref{saddlepoint3} into a single equation 
\be\label{saddlefinal}
 \log \left(\frac{z^2}{\xi^2} \right) - \frac{R}{2} \log \left( R^2 z^2 \right) \,  = \, 2 \slashed{\Omega}(z)  \, ,
\ee
where the allowed cuts of the total resolvent $\Omega(z)= \omega(z)$ are symmetrically distributed with respect to $z=0$ and belong strictly on the real axis. Equivalently we can use eqn. \eqref{saddlerealpart} for $\mu \rightarrow 0$, remembering that $u=z^2$ and that double cuts in the $z$-plane map to single cuts in the $u$-plane. This equation becomes
\be\label{saddlefinal2}
\frac{\partial V_{eff.}(u)}{\partial u} = \frac{2-R}{2}  \log \left(\frac{u}{\xi_{eff.}^{4/(2-R)}} \right)  \,  = \, 2 \slashed{\Omega}(u) \, , \quad \xi_{eff.} = \xi R^{-R/2}  \, ,
\ee
and has physical cuts for $u \geq 0$. We observe the appearance of the scaling parameter $\xi_{eff.}^{4/(2-R)}$, related to the KPZ/DDK scaling of the Sine-Liouville coupling with the usual ($-$) type of dressing, see section~\ref{windingmatrixliouville}. A brief analysis of the general eqn.~\eqref{saddlerealpart} with arbitrary $\mu$ is presented in appendix~\ref{methodforgenresolvent}.

The effective potential defined through \eqref{saddlefinal2}, determines essentially all the physical features of the solutions. One has first to distinguish two cases that behave very differently: For $R<2$, the effective potential is stable, while for $R>2$ it is unstable and goes to $-\infty$ as $u \rightarrow \infty$. This should translate into a physical statement about the stability of the solutions and the dual background they correspond to. 

\begin{figure}[t]
\begin{center}
\includegraphics[width=0.45\textwidth]{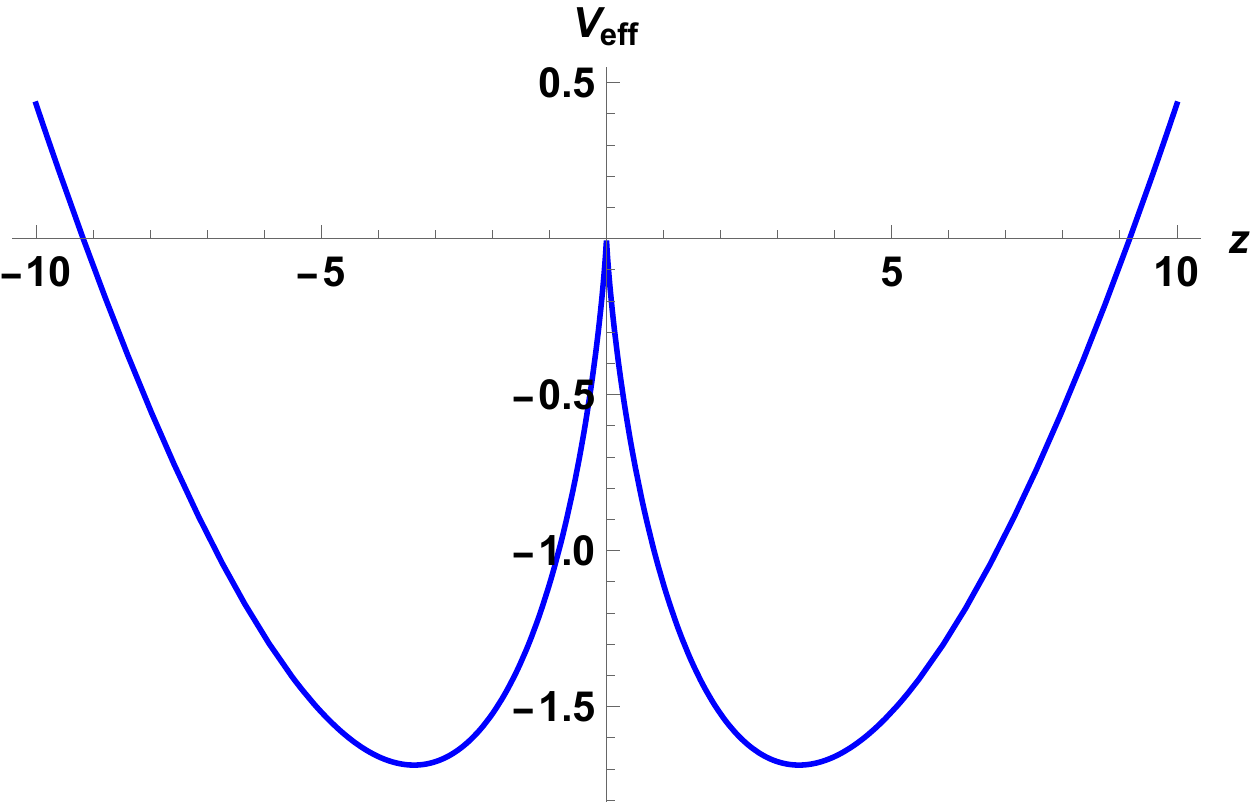} \hspace{0.1em} \hspace{0.1em} \includegraphics[width=0.45\textwidth]{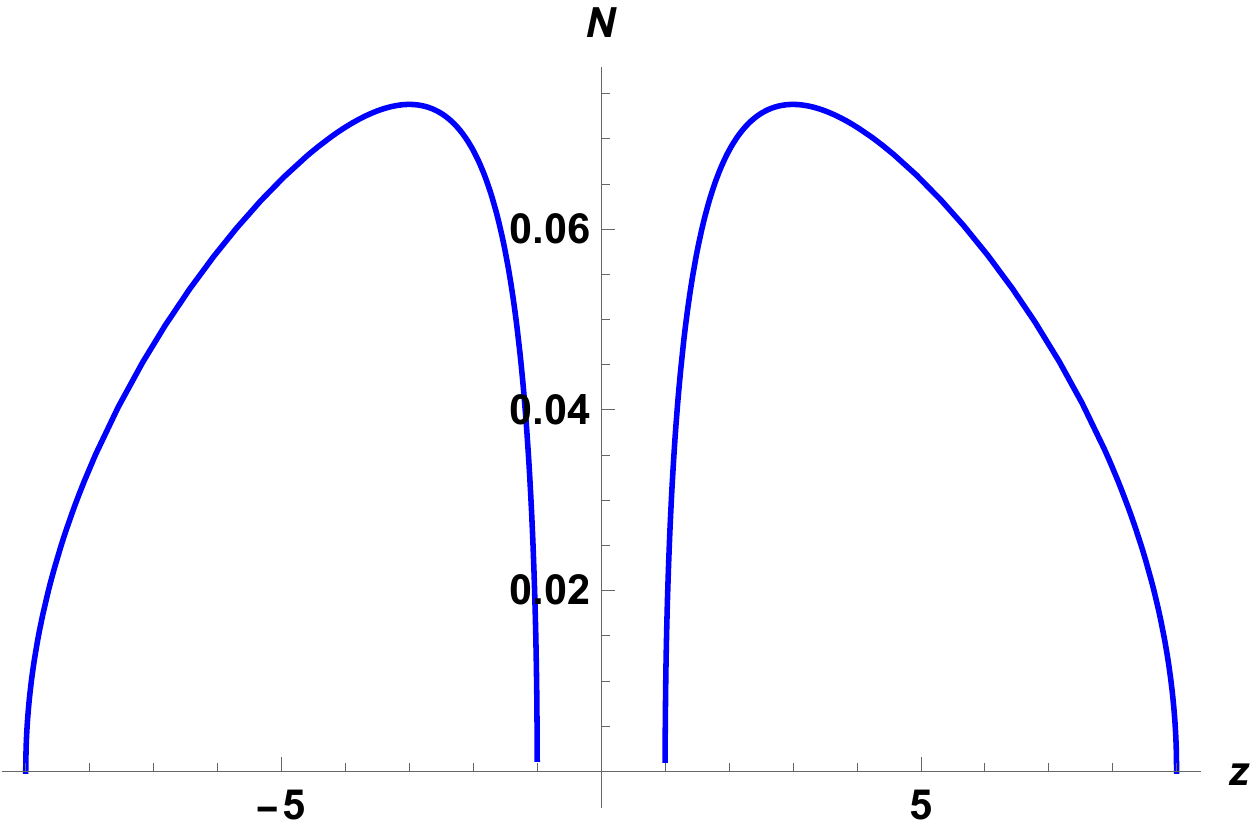}
\end{center}
\caption{In the left plot we depict the typical effective potential for large $\xi_{eff} = \xi R^{-R/2}$ (two deep wells) and in the right one the typical density of boxes that has support on two segments in the $z$ coordinates.  
}
\label{fig:smallRshallow}
\end{figure}

\begin{figure}[t]
\begin{center}
\includegraphics[width=0.45\textwidth]{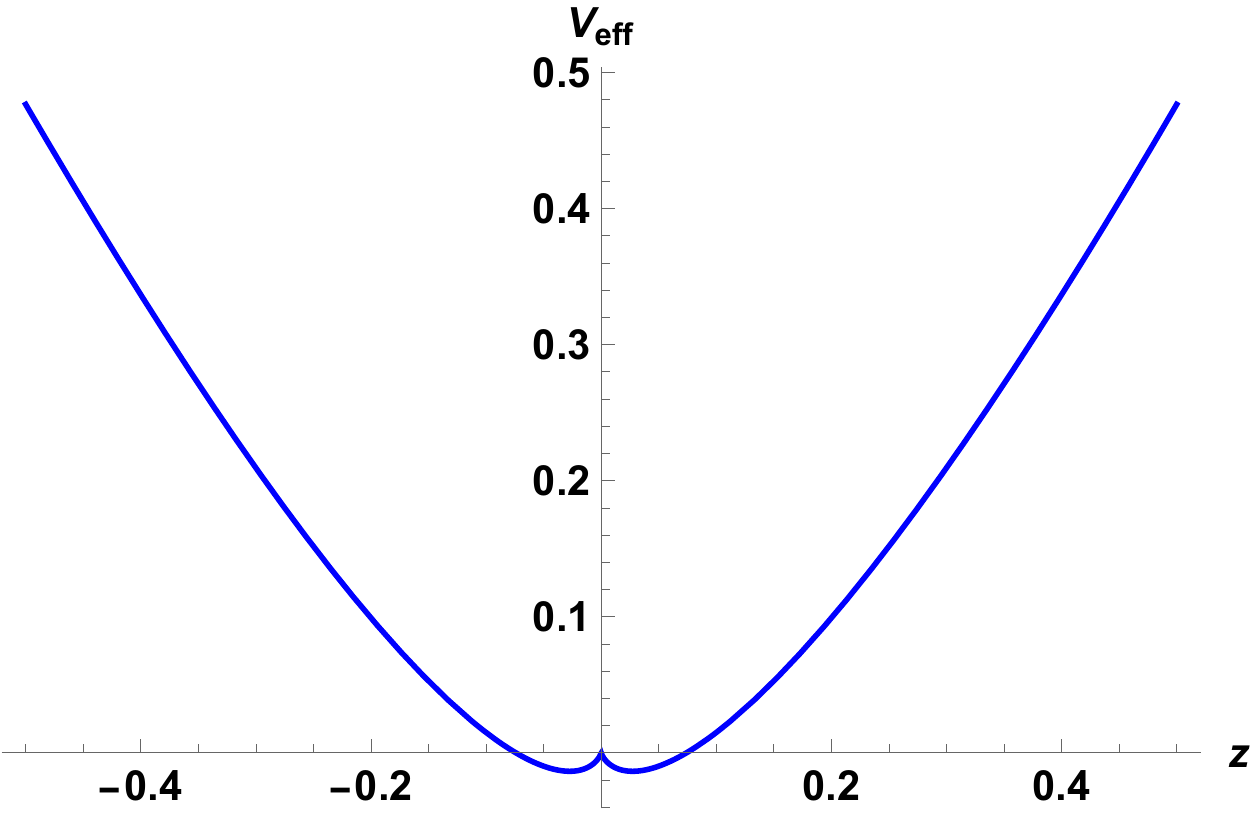} \hspace{0.1em} \hspace{0.1em} \includegraphics[width=0.45\textwidth]{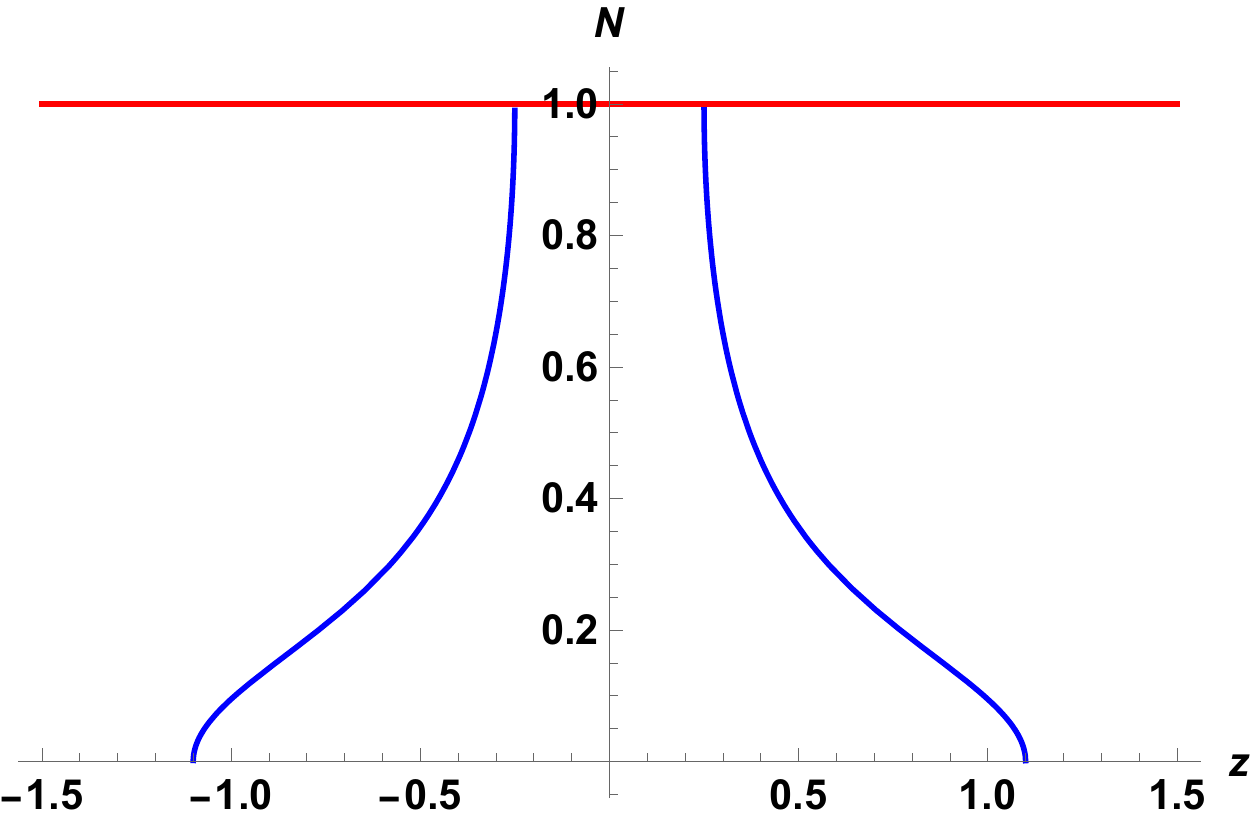}
\end{center}
\caption{In the left plot we depict the typical effective potential for small $\xi_{eff} = \xi R^{-R/2}$ (single well) and in the right one the typical density of boxes that contains a region of saturation near $u = z^2 = 0$.  
}
\label{fig:smallRdeep}
\end{figure}

Let us first analyse the case $R<2$. The single well nature of the effective potential defined through \eqref{saddlefinal2}, indicates the possibility of having two different phases. The first is a phase described by a single cut solution in the $u$-plane (symmetric two cut solution on the real axis of the $z$-plane). This possibility arises only when the parameters make the well deep enough (relatively large $\xi_{eff} = \xi R^{-R/2}$), so that boxes cannot pile up near the origin $u=z^2=0$. The second type of solution contains a saturation region near the $u = z^2 =0$ region, valid when the wells around that region are very shallow (relatively small $\xi_{eff} = \xi R^{-R/2}$).  Since the solutions are $Z_2$ symmetric and double cuts in the $z$ plane map to single cuts in the $u = z^2$ plane, we shall refer to them as single cut solutions with or without saturation, leaving the nomenclature double cut (four cuts in the $z$ plane) for more complicated solutions that can appear when $\mu$ is arbitrary. A typical plot of the potential and the density of boxes in these two cases can be found in figs.~\ref{fig:smallRshallow} and~\ref{fig:smallRdeep}.

\begin{figure}[t]
\begin{center}
\includegraphics[width=0.45\textwidth]{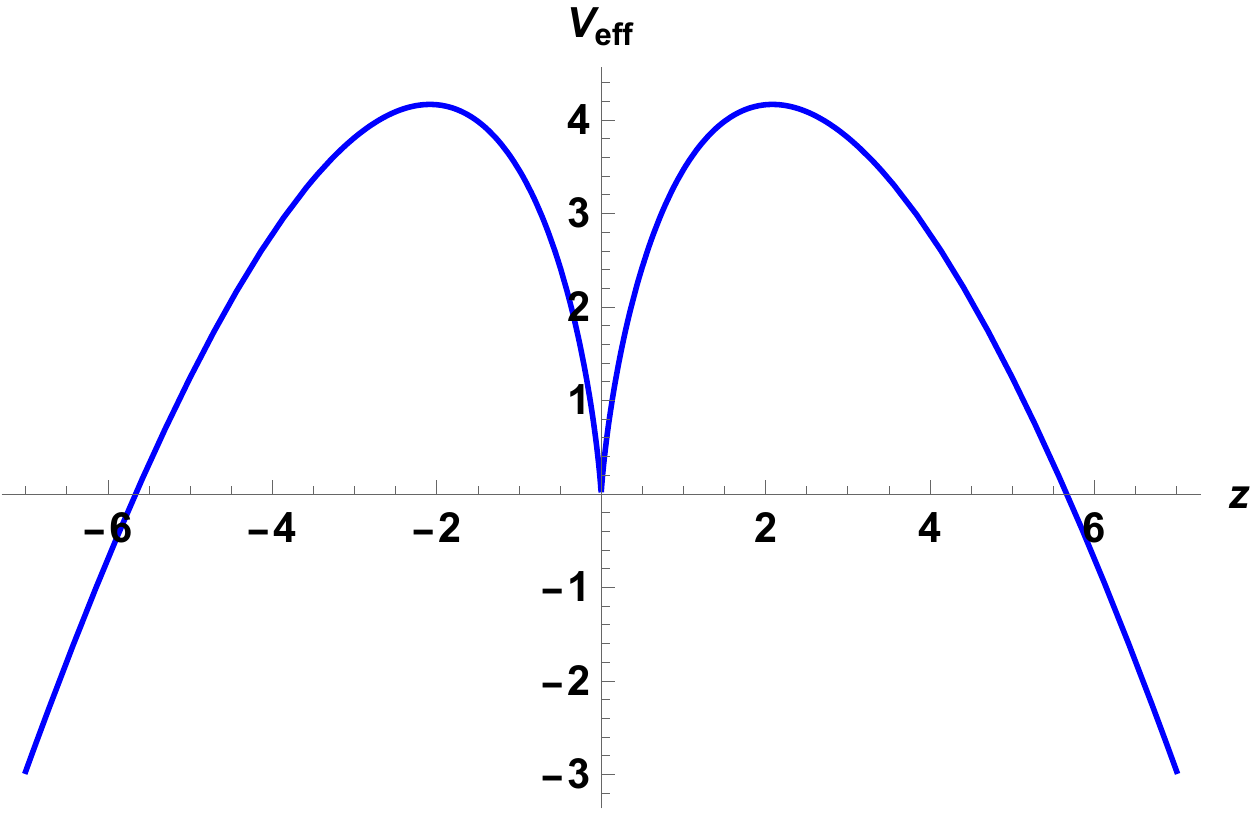} \hspace{0.1em} \hspace{0.1em} \includegraphics[width=0.45\textwidth]{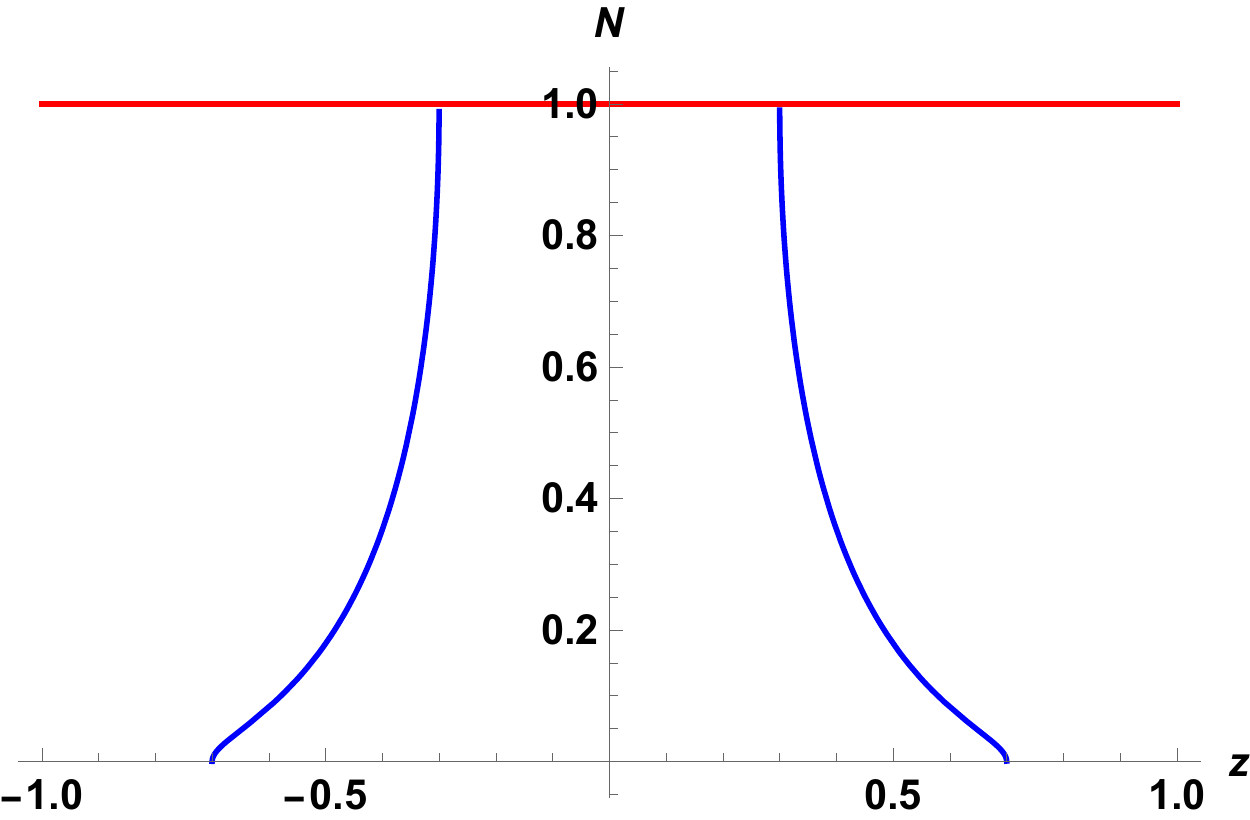}
\end{center}
\caption{In the left plot we depict the typical effective potential when $R>2$ (unstable regime) and in the right one the typical density of boxes that has a saturation region with two adjacent cuts. This solution exists for relatively small $\xi_{eff} = \xi R^{-R/2}$, otherwise the well becomes too shallow to support a metastable solution.
}
\label{fig:bigRdeep}
\end{figure}

The case when $R>2$, is a case of an unstable potential with a single maximum (in the $u$ variables). In such a potential we can define a physical solution only when its maximum $V_{max}(u_{max})$ is high enough and displaced from the origin so that it can support a cut in a region $u \in [0, b)$, with $b < u_{max}$. A typical plot of the potential and the density of boxes in this case can be found in fig.~\ref{fig:bigRdeep}. Otherwise, the only acceptable Young diagram is the trivial one corresponding to the usual $c=1$ linear dilaton background at finite temperature. Notice that this is precisely the regime where one could potentially find a large semi-classical black hole\footnote{The non-perturbative instability of the potential in this regime could perhaps be related to the fact that a large semi-classical black hole is expected to be thermodynamically unstable, because it can radiate away its energy when the space is asymptotically flat.} and in which the usual dressing ($-$) of the winding mode  in~\eqref{I14} becomes irrelevant (singular in the weakly coupled region). At this point we simply notice that the alternative type ($+$) of dressing for the winding mode operator in eqn.~\eqref{I13} is still well defined and relevant in this regime (having support at the strong coupling region), the importance of this will soon become more clear.

\subsection{The stable regime ($R<R_{KT}$)}\label{stableregime}

\subsubsection{A cut with no saturation of the density}\label{cutnosaturation}

\paragraph{Method I -}
There exists a quite simple method to determine the total resolvent for saddle point equations of the type~\eqref{saddlepoint1} and~\eqref{saddlefinal}, see~\cite{Halmagyi:2003ze}. We define the even function
\be
g(z) = e^{\frac{2 \Omega}{2-R}} + \frac{z^2}{\xi_{eff}^{\frac{4}{2-R}}} e^{- \frac{2 \Omega}{2-R}} \, , \qquad \lim_{z \rightarrow \infty} g(z) = \frac{z^2}{\xi_{eff}^{\frac{4}{2-R}}} \, , \qquad \xi_{eff} = R^{-R/2} \xi \, .
\ee
This function due to the saddle point equation \eqref{saddlefinal} does not have any branch cuts in the complex $z$ plane and is regular except at infinity. This means that
\be
g(z) =  \frac{z^2 + B}{\xi_{eff}^{\frac{4}{2-R}} }\, , \qquad g(0) = \frac{ B}{\xi_{eff}^{\frac{4}{2-R}}} \, ,
\ee
hence leading to an algebraic equation for the resolvent and the susceptibility once the parameter $B$ has been determined in terms of the physical parameters $\xi, R$. 

Solving in terms of the resolvent we find
\be\label{ressingle1}
\frac{2 \Omega(z)}{2-R} = \log \left( \frac{1}{2 \xi_{eff}^{\frac{4}{2-R}}} \left(z^2 + B - \sqrt{(z^2 + B)^2 - 4 z^2 \xi_{eff}^{\frac{4}{2-R}}} \right) \right) \, .
\ee
where the quantity under the square root now being $(z^2-a^2)(z^2-b^2)$ for the symmetric cut $|z|\in (a,b)$. Due to this we can identify
\be\label{identifications1}
B = a b \, , \qquad \xi_{eff}^{\frac{4}{2-R}} = \frac{(a+b)^2}{4} \, ,
\ee
so that
\be\label{ressingle2}
\frac{\Omega(z)}{2-R} = \log \left( \frac{\sqrt{(z+a)(z+b)} - \sqrt{(z-a)(z-b)} }{2 \xi_{eff}^{\frac{4}{2-R}}}   \right) \, .
\ee

\paragraph{Method II -}
Another approach is to analyse eqn.~\eqref{saddlefinal2} expressed in terms of the variable $u = z^2$. This equation can be treated as a singular integral equation with physical cut/s only on the positive $u$ axis and its solution is automatically $Z_2$ symmetric in the original $z$ variables. A caveat though is that one should not impose the normalisation condition of the resolvent $\int_{\text{supp.}} d u \mathcal{N}(u) \neq 1$, since the resolvent is normalised in the $z$ and not the $u$ variables.

Fortunately eqn. \eqref{saddlefinal2} has been thoroughly analysed in the works~\cite{Dutta:2007ws,Dutta:2015noa}\footnote{ We extend this analysis for the more general equation~\eqref{saddlerealpart} with non-zero $\mu$ in appendix~\ref{methodforgenresolvent}.}. The solution for the resolvent in the case of a single unsaturated cut $u \in (a^2,b^2)$ takes the form (see the method in appendix~\ref{methodforgenresolvent})
\be\label{ressingle3}
\frac{2 \Omega(u)}{2-R} = \log \left( \frac{ u + a b - \sqrt{(u-a^2)(u-b^2)}  }{(a+b) \xi_{eff}^{\frac{2}{2-R}}}   \right) \, .
\ee
Upon using the identifications \eqref{identifications1} this resolvent  coincides with the one that we obtained in eqn.~\eqref{ressingle1}, verifying the validity of the complementary methods.

\paragraph{Properties of the density of boxes -}

Using the resolvent we find the density of boxes to be
\be\label{dobsinglecut}
\mathcal{N}(z) = \frac{2-R}{2 \pi} \cos^{-1} \frac{z^2+ a b}{|z| (a+b) } \, , \qquad |z| \in \left( a \, ,\, b \right) \, .
\ee
This density acquires its maximum value
\be
\mathcal{N}_{max.} = \frac{2-R}{2 \pi}\cos^{-1} \frac{2 \sqrt{a b}}{(a+b) } \, ,\qquad z = \sqrt{a b} \, .
\ee
and vanishes at the endpoints $z = a, b$. This means that one needs to demand the no 
saturation condition $\mathcal{N}_{max.} \leq 1$ on the solution in order for it to exist. The shape of a typical Young diagram in the unsaturated phase is plotted in fig.~\ref{fig:YoungsmallR} and contrasted with the one appearing in the saturated phase.

\paragraph{Fixing the edges of support -}
Our last task is to fix the edges of support $a,b$ in terms of the physical parameters $\xi, R$. This task is more complicated compared to examples of similar equations in the literature~\cite{Dutta:2007ws,Dutta:2015noa}. We obtained a symmetric two cut solution in the $z$-variables (in terms of which the density is appropriately normalised). Due to this $Z_2$ symmetry, it is not possible to impose the usual normalisation condition $\Omega(z)\sim 1/z$ at $z \rightarrow \infty$. What we have to do instead is to compute $\oint_{\mathcal{C}_i} \Omega(z) dz = \int_{\mathcal{C}_i} \mathcal{N}(z) d z$ around the cuts and use this information to fix the normalisation.

We then explicitly integrate the density of boxes over the positive cut
\be
\frac{1}{2} = \int_{a}^b d z \mathcal{N}(z)  = - \int_a^b d z z \mathcal{N}'(z) =  - \frac{2-R}{2 \pi} \int_a^b d z \frac{a b - z^2}{\sqrt{(b^2-z^2)(z^2-a^2)}}  \, .
\ee
This integral can be performed in terms of elliptic functions. The result is
\be\label{bc1}
\frac{1}{2} = \frac{2-R}{2 \pi} \left(b E\left(1 - \frac{a^2}{b^2} \right) - a K\left(1 - \frac{a^2}{b^2} \right) \right)
\ee
with $K,\, E$ the elliptic integrals of the first and second kind. Equations \eqref{identifications1} and \eqref{bc1} completely fix the edges of support in terms of $\xi,R$ (albeit somewhat implicitly). A simpler result can be given if we assume that the ratio $a/b$ is either close to zero or one (wide vs. narrow cut). In the first case we keep the leading terms in the expansion of the elliptic functions to find
\be
\frac{1}{2} \simeq \frac{2-R}{2 \pi} b  \left(1 + \frac{a}{b} \log \frac{a}{4 b} \right)   \, , \quad 2 \xi_{eff}^{\frac{2}{2-R}}  = a + b \, ,
\ee
provided that $0< a/b \ll 1$ and $\mathcal{N}_{max} \leq 1$
\be
0 < \frac{a}{b} = \frac{ 2 (2-R) \xi_{eff}^{\frac{2}{2-R}}}{ \pi} - 1 \ll 1 \, , \quad \frac{2-R}{2 \pi} \left( \frac{\pi}{2} - 2 \sqrt{\frac{a}{b}}\right) \leq 1 \, .
\ee
These conditions can be simultaneously satisfied in a small locus determined by $(\pi/(4-2R))^{(2-R)/2}  \leq \xi_{eff} \ll ( \pi/(2-R))^{(2-R)/2} $. For $ \xi_{eff} < ( \pi/(4 - 2 R))^{(2-R)/2}  = \xi_{eff}^*(R) $ the solution ceases to exist altogether, since $a$ becomes negative. This is an $O(1)$ number in the regime $1<R<2$. As we shall see later crossing the line $\xi_{eff}^*(R)$ we encounter a phase transition.

In the second limiting case of a narrow cut we find
\be\label{smallcutid}
\frac{8}{2-R} \simeq   b (1-a/b)^2  = b \epsilon^2 \, , \quad 2 \xi_{eff}^{\frac{2}{2-R}}  = a + b = b(2- \epsilon) \, ,
\ee
so that
\be
b - a = \epsilon b \simeq \frac{2}{ (2-R)} \left( 1 + \sqrt{1 + 2 (2-R) \xi_{eff}^{\frac{2}{2-R}}}  \right) \,  .
\ee
In this case the conditions become
\be
\epsilon = \frac{ 4}{1+ \sqrt{1+ 2 \xi_{eff}^{\frac{2}{2-R}} (2-R)}} \ll 1 \, , \quad \frac{2-R}{2 \pi} \frac{\epsilon}{2} \leq 1 \, . 
\ee
These conditions are satisfied as long as
\be
\xi_{eff} \gg  \left( \frac{4}{2 - R} \right)^{(2-R)/2} \, .
\ee
For $R<2$, this condition is generically satisfied far away from the transition line in the limit of very large $\xi_{eff}$,  when the potential develops a deep well making the cut narrow.

\paragraph{The free energy -}
The non-singlet susceptibility is directly related to the resolvent through eqn.~\eqref{nssuscept}. For small $\mu$ we find from~\ref{ressingle3}
\be
\chi^{n.s.} = - \partial_{\mu}^2 \mathcal{F}_{n.s.} \simeq \frac{R(2-R)}{2} \log \frac{\mu^2}{a b} \, .
\ee
Unfortunately it is not convenient to use this expression to determine the free energy, since it contains partial derivatives with respect to $\mu$ and hence if we try to integrate its leading asymptotic expression for small $\mu$ we could potentially miss some $\xi$ dependent terms. It is therefore more natural to use eqn.\eqref{firstderxi} that contains all the dependence in $\xi$. In particular we find that for $u=z^2$ the average size of the partition is
\be
\langle |\lambda| = n \rangle = - \half \xi  \partial_\xi S_{eff.} = \int_{a^2}^{b^2} d u \mathcal{N}(u) \, ,
\ee
that is simply the leading asymptotic coefficient in the resolvent~\eqref{ressingle3} $\Omega(u) \sim c/u$ as $u \rightarrow \infty$. We therefore find
\be
\langle |\lambda| = n \rangle = - \half \xi  \partial_\xi S_{eff.} = \frac{2-R}{16} (b-a)^2  \, ,
\ee
which is positive for $R<2$ as expected.

For a wide cut, we find
\be
\langle |\lambda| = n \rangle \simeq \frac{2-R}{4}  \xi_{eff}^{\frac{4}{2-R}}  \, , 
\ee
and as we proved this holds on a specific locus of $\xi_{eff}$ that defines the near transition region.

The regime of very large $\xi_{eff}$ is related to a narrow cut, when 
\be
\langle |\lambda| = n \rangle \simeq \frac{1}{ 4 (2-R)} \left( 1 + \sqrt{1 + 2 (2-R) \xi_{eff}^{\frac{2}{2-R}}}  \right)^2 \simeq   \half  \xi_{eff}^{\frac{2}{2-R}}  \, .  
\ee
If we integrate these expressions we find the leading contribution to the free energy
(corresponding to the on-shell effective action) 
\be\label{freeunsaturated}
\mathcal{F}_{narrow} \simeq - \frac{(2-R)}{2} \xi_{eff}^{\frac{2}{2-R}} + F(\mu) \, , \quad \mathcal{F}_{wide} \simeq - \frac{(2-R)^2}{8} \xi_{eff}^{\frac{4}{2-R}} + F(\mu) \, ,
\ee
where the second term is subleading in our approximations (and contains the singlet contribution as well). We immediately observe that the scaling of $\xi_{eff}$ in the free energy (and the property that it vanishes for $R=2$), coincide with the results of~\cite{Kazakov:2000pm,Kazakov:2001pj}, when the cut becomes large (near the phase transition region between the single cut and the saturated cut). In the limit of a narrow cut (very large $\xi_{eff}$) the behaviour of the free energy is different and  asymptotically starts to scale as the square root of the free energy near the transition region. In between we have a complicated scaling behaviour dictated by the elliptic functions.



\subsubsection{Saturated cut}\label{solutionsaturatedcut}

\begin{figure}[t]
\begin{center}
\includegraphics[width=0.45\textwidth]{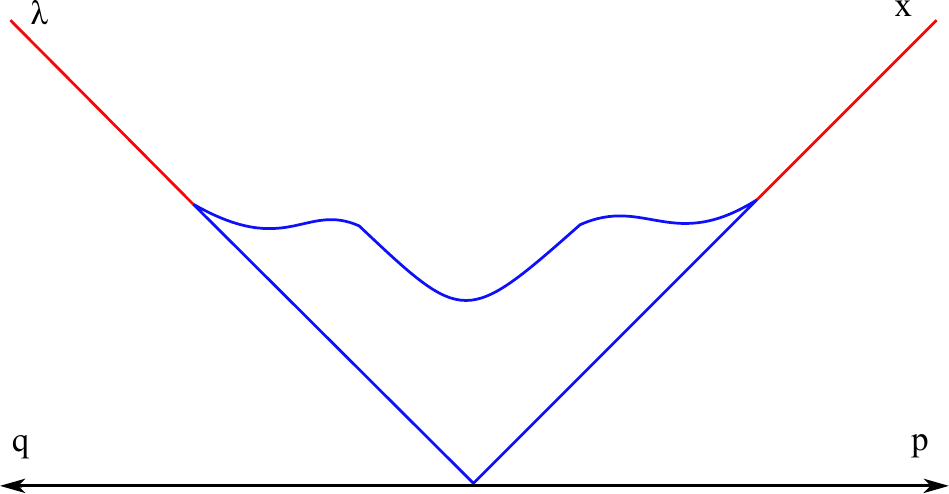} \hspace{0.1em} \hspace{0.1em} \includegraphics[width=0.45\textwidth]{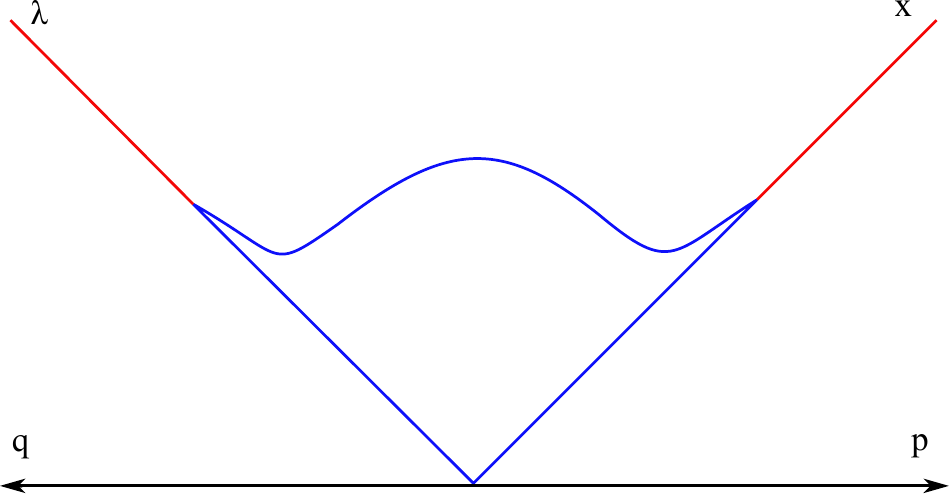}
\end{center}
\caption{In the left figure we depict a typical Young diagram in the unsaturated phase (for larger $\xi_{eff}$) and in the right one a typical Young diagram in the saturated phase (for smaller $\xi_{eff}$). The shape can be understood by considering the slope function $\mathbf{\Omega}'(z) = 1 - 2 \mathcal{N}(z) \, , \, $ for $z \geq 0$ (it becomes one when there is no density and minus one when there is saturation). In the case of exactly zero $\mu$ that we analyse, the shape at $p=q=z =0$ is continuous but has a cusp (that we have smoothed out in order to show the effect of including the singlet contribution in $\mathcal{N}(z)$ for a small non-zero $\mu$). 
}
\label{fig:YoungsmallR}
\end{figure}

Let us now proceed to analyse the case when the potential is such that the density
saturates in a region near $u = z^2 = 0$. We shall therefore take an ansatze for the density of the form
\be
\mathcal{N}(u) = 1 \, , \quad u \in [0,a^2) \, , \qquad \mathcal{N}(u) \neq 0 \, , \quad u \in (a^2,b^2) \, .
\ee
These conditions are satisfied by the following modified integral equation
\bea\label{sat1}
2 \dashint_{a^2}^{b^2} dv \, \frac{\mathcal{N}(v)}{u - v} \, &=& \,  2 \log \frac{u - a^2}{u}  + \frac{2-R}{2} \log \left(\frac{u}{\xi_{eff}^{\frac{4}{2-R}}} \right)  \, \nn \\
\, &=& \, 2 \log(u-a^2) - \frac{R+2}{2} \log \left( \frac{u}{\xi_{eff}^{-\frac{4}{R+2}}} \right)     \, ,
\eea
where remarkably the scaling $\sim \xi_{eff}^{2/(R+2)}$ of the coupling for the ($+$) type of Liouville winding mode appears, see section~\ref{windingmatrixliouville}. This will become important for $R>2$. The resolvent is
\be
\Omega(u) = \log \frac{u}{u - a^2} + \int_{a^2}^{b^2} dv \, \frac{\mathcal{N}(v)}{u - v} \, .
\ee
Solving the modified integral equation with the method we describe in appendix~\ref{methodforgenresolvent}, we find the resolvent
\bea\label{resolventsaturated}
 \Omega(u) -  \log \frac{u}{u - a^2}  =  \int_{a^2}^{b^2} dv \, \frac{\mathcal{N}(v)}{u - v}  = \qquad \nn \\
= 2 \log\left(\frac{u-a^2 - \sqrt{(u-a^2)(u-b^2)}}{\sqrt{b^2-a^2}} \right) - \frac{R+2}{2} \log\left(\frac{u + a b - \sqrt{(u-a^2)(u-b^2)}}{(b+a) \xi_{eff}^{-\frac{2}{R+2}}} \right)  \, . \nn \\
\eea
From the leading asymptotic at $u\rightarrow \infty$ we find the condition
\be\label{satasympt}
(b+a)^{-R} (b - a)^2 = 2^{2-R}  \xi_{eff}^2  \, .
\ee
The associated density of boxes is now given in the $z$-variable by
\be
\mathcal{N}(z) = \begin{cases}  1 \, , \quad |z| \in [0,a] \, , \\ \frac{2}{\pi} \arccos \sqrt{ \frac{z^2-a^2}{b^2-a^2}} - \frac{R+2}{2 \pi} \arccos \frac{z^2 + a b}{|z|(a+b)}    \, , \quad |z| \in (a,b)  \, .
\end{cases}
\ee
The shape of a typical Young diagram in the saturated phase is plotted in fig.~\ref{fig:YoungsmallR} and contrasted with the unsaturated case.

\paragraph{Fixing the edges of support -}
Finally we need to impose the normalisation condition to completely fix the edges of support. Integrating the density of boxes over the positive cut via
\be
\frac{1}{2} - a = \int_{a}^b d z \mathcal{N}(z)  = - \int_a^b d z z \mathcal{N}'(z) \, ,
\ee
we find
\be\label{bc2}
\frac{1}{2} - a = \frac{2}{\pi} b E\left(1 - \frac{a^2}{b^2} \right) - \frac{R+2}{2 \pi} \left(b E\left(1 - \frac{a^2}{b^2} \right) - a K\left(1 - \frac{a^2}{b^2} \right) \right)
\ee
with $K,\, E$ the elliptic integrals of the first and second kind.

If the ratio $a/b$ is close to zero (wide cut) we find the leading result for the boundary conditions
\be\label{smallratio}
\frac{1}{2} - a \simeq \frac{2-R}{2 \pi} b  \, , \quad b \simeq 2  \xi_{eff}^{2/(2-R)} \, ,
\ee
consistent when
\be\label{smallratiocon}
0 < \frac{a}{b} \simeq \frac{1}{4}  \xi_{eff}^{-2/(2-R)} - \frac{2-R}{2 \pi} \ll 1 \,  .
\ee
This condition holds for relatively small $\xi_{eff}$, that is the regime of a shallow potential admitting a solution with a saturation region. For $\xi_{eff}^{*}(R) = ( \pi/(4 - 2 R))^{(2-R)/2}  < \xi_{eff}$ the solution ceases to exist altogether, since $a$ becomes negative (this is an $O(1)$ number in the regime $1<R<2$). This is precisely the opposite inequality compared to the one we found for the single cut unsaturated solution (again in its corresponding wide cut regime), verifying the transition region between the two solutions.

This also means that if we try to construct a saturated narrow cut solution, we have to make the potential narrower by increasing $\xi_{eff}$. But then we inevitably transition to the unsaturated phase. We therefore conclude that there is no consistent narrow cut approximation when $R<2$ for the saturated solution and this phase has only a consistent wide cut approximation.

\paragraph{The free energy -}
The leading term in the non-singlet susceptibility is given once more by combining eqn.\eqref{nssuscept} and eqn. \eqref{resolventsaturated} for $u= - \mu^2$ and expanding near $\mu = 0$ to find
\be
\chi^{n.s.}_0 \simeq \log \frac{\mu^2}{a^2} + \log \frac{a(a+b)}{\sqrt{b^2 - a^2}} - \frac{R+2}{4} \log \frac{\mu^2 (a+b)}{a b \, \xi_{eff}^{-\frac{2}{R+2}}}  \, .
\ee
We can once more determine the average size of the partition and the free energy using eqn.\eqref{firstderxi} and the asymptotic expansion of the resolvent to find
\be
\langle |\lambda| = n \rangle = - \half \xi  \partial_\xi S_{eff.} =  \frac{2(b+a)^2-R (b-a)^2}{16}   \, ,
\ee
that is always positive for $R<2$ (but can still be positive for $R>2$ in some parameter regime, as we describe in the next section). Using the wide cut approximation relevant for this unsaturated phase it becomes
\be
\langle |\lambda| = n \rangle  \simeq \frac{2-R}{4} \xi_{eff}^{\frac{4}{2-R}} \, . 
\ee
If we integrate it we find the free energy (from the on-shell effective action)
\be\label{freesaturated}
\mathcal{F}_{wide} \simeq - \frac{(2-R)^2}{8} \xi_{eff}^{\frac{4}{2-R}} + F(\mu) \,  .
\ee
We observe that in this phase the leading part of the free energy corresponds exactly to that found in~\cite{Kazakov:2000pm,Kazakov:2001pj}. It also coincides with the free energy of the unsaturated phase solution near the transition point (wide-cut), and so do their first derivatives (compare with eqn. \eqref{freeunsaturated}), showing the expected continuous nature of the phase transition.

\subsection{The unstable regime ($R>R_{KT}$)}\label{unstableregime}

When the compactification radius becomes large ($R>2$), the effective potential $V_{eff}(u)$ of eqn.~\eqref{saddlefinal2} becomes unstable. Nevertheless there does exist a regime of parameters in which it has a large positive local maximum, leading to an allowed solution where the density is non trivial in a cut that starts from $u=0$.

For the potential defined via eqn.~\eqref{saddlefinal2}, the maximum is at $u_{max} = \xi_{eff}^{-\frac{4}{R-2}} $ and its value is $V_{max}(u_{max}) = \half (R-2)  u_{max} $, so they both grow in the same fashion. This means that the regime $\xi_{eff} = \xi R^{-R/2} <1 $ is the relevant one for having a large and displaced maximum. Due to this, we then observe that the solution without saturation of section~\ref{cutnosaturation} is pathological for $R>2$, because both the density of boxes~\eqref{dobsinglecut} and the resolvent become negative. 

The only viable (perturbatively stable) possibility thus, is the case with a saturation region near $u=0$ studied in~\ref{solutionsaturatedcut}. For the saturated case the effective potential is instead the one derived from eqn.~\eqref{sat1}. The location of its maximum obeys $(u_{max}-a^2)^2 = u_{max}^{(R+2)/2} \xi_{eff}^2$. We also observe that the natural scaling variable in this equation is $\xi_{eff}^{-\frac{2}{R+2}}$. 

We should then impose the most important additional physical condition when $R>2$, that of the positivity of the density of boxes $\mathcal{N}(z) \geq 0$ (which is trivially satisfied when $R<2$). Considering the density of boxes before fixing the edges
\be
\mathcal{N}(z) = \begin{cases}  1 \, , \quad |z| \in [0,a] \, , \\ \frac{2}{\pi} \arccos \sqrt{ \frac{z^2-a^2}{b^2-a^2}} - \frac{R+2}{2 \pi} \arccos \frac{z^2 + a b}{|z|(a+b)}    \, , \quad |z| \in (a,b)  \, ,
\end{cases}
\ee
we find that
\be
\mathcal{N}'(z) < 0 \, , \quad \text{for} \, \quad z^2 < \frac{a b (2+R)}{R-2} \, , \qquad \mathcal{N}'(z) > 0 \, , \quad \text{for} \, \quad z^2 > \frac{a b (2+R)}{R-2}  \, .
\ee
This means that it is monotonically decreasing until it acquires its minimum value for $z_{min}^2 = \frac{a b (2+R)}{R-2}  $, after which it starts growing. Since the density should be a decreasing function acquiring its minimum zero value at $z = b$, we find that the saturated solution is acceptable as long as
\be\label{positivitycondition}
a < b <  \frac{a (R+2)}{R-2} \quad \Rightarrow \quad r_{c} = \frac{R-2}{R+2} < r = \frac{a}{b}  < 1  \, .
\ee
This means that there exists a new natural approximation that one can make (for the saturated cut) when solving for the boundary conditions determining the edges~\eqref{bc2} when $R>2$. This is the case of a narrow cut $r \simeq 1$. In particular we now find that the boundary condition~\eqref{bc2} together with~\eqref{satasympt} for $1 - a/b = \epsilon \ll 1$ contain the following first terms in the small $\epsilon$ expansion
\be\label{narrowcutasym}
\half  \simeq 2 b - \frac{3 b \epsilon}{2} \, , \quad b^{R-2} \simeq \epsilon^2 \frac{1 }{4 \xi_{eff}^2} \left(1 + \epsilon \frac{R}{8} \right) \, .
\ee
To leading order we find
\be
b \simeq \frac{1}{4} \, , \quad \epsilon \simeq 2^{3-R} \xi_{eff} \ll 1 \, , 
\ee
which holds as long as $\xi_{eff} \ll 2^{R-3} $, which makes this approximation more and more natural for very large radii and bad for radii close to $R=2$.

\begin{figure}[t]
\begin{center}
\includegraphics[width=0.325\textwidth]{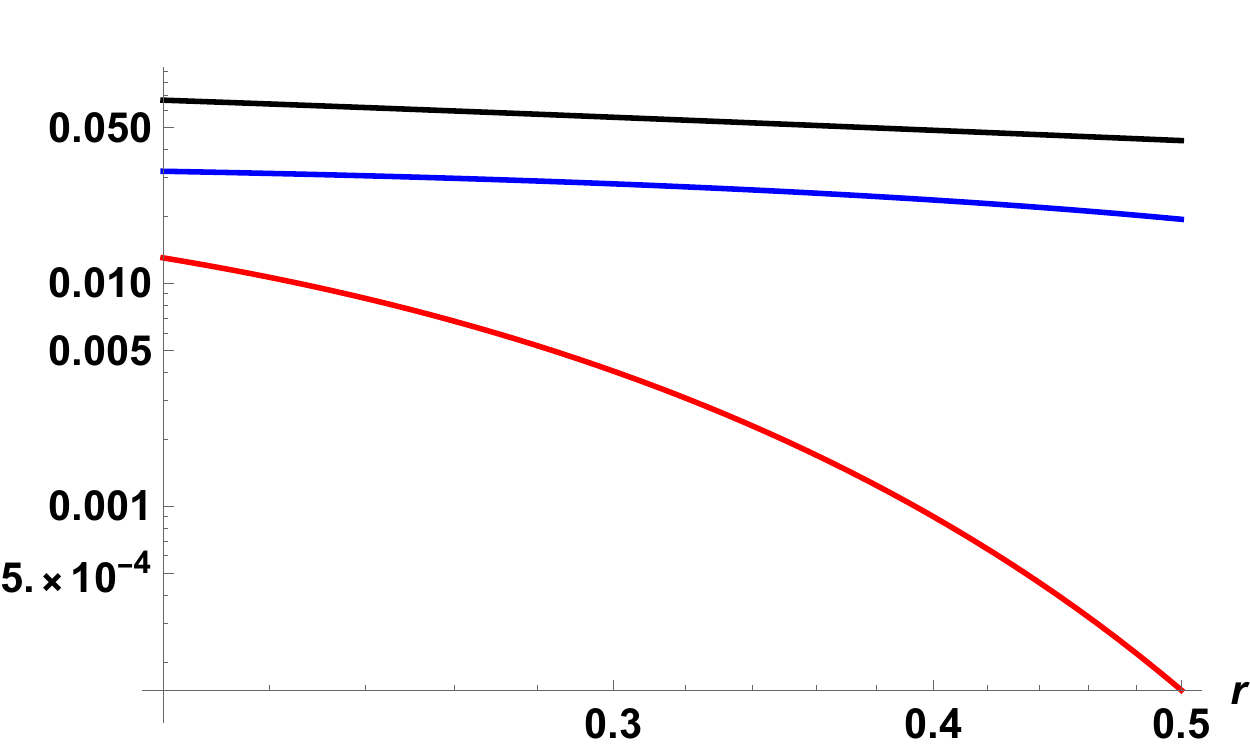}  \includegraphics[width=0.325\textwidth]{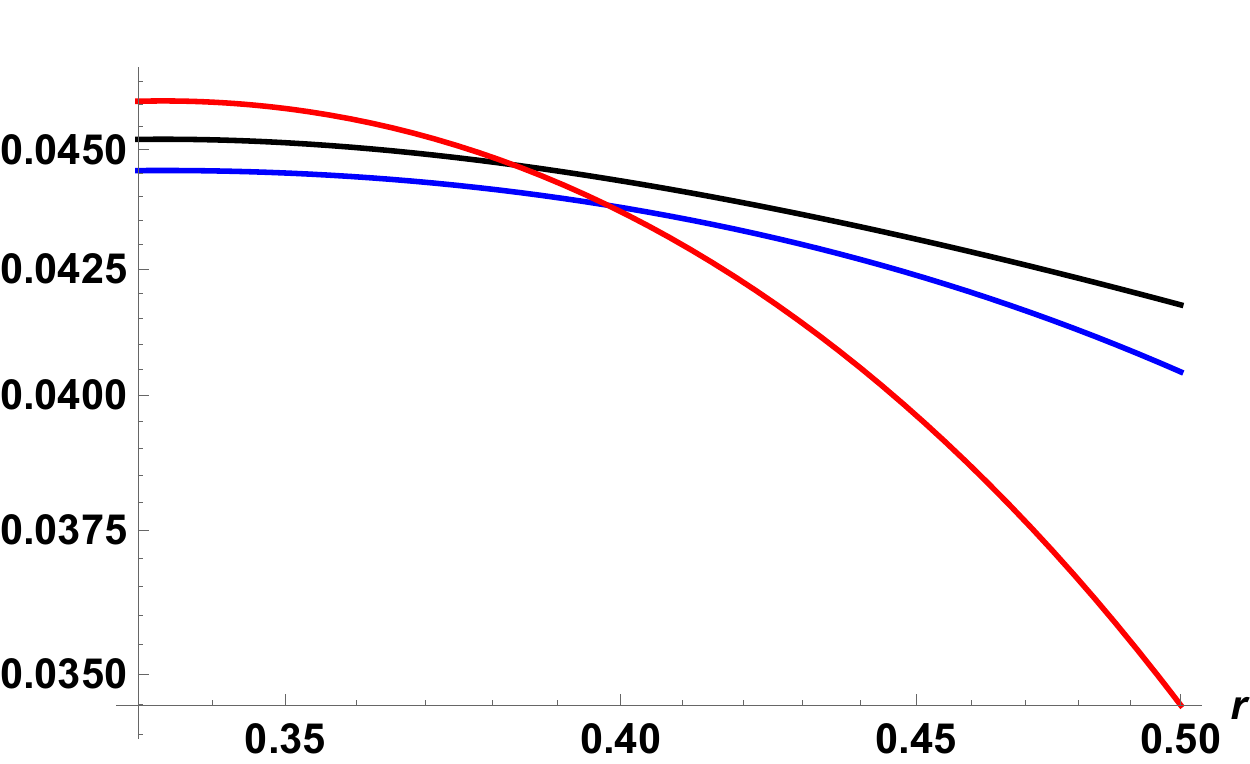} 
\includegraphics[width=0.325\textwidth]{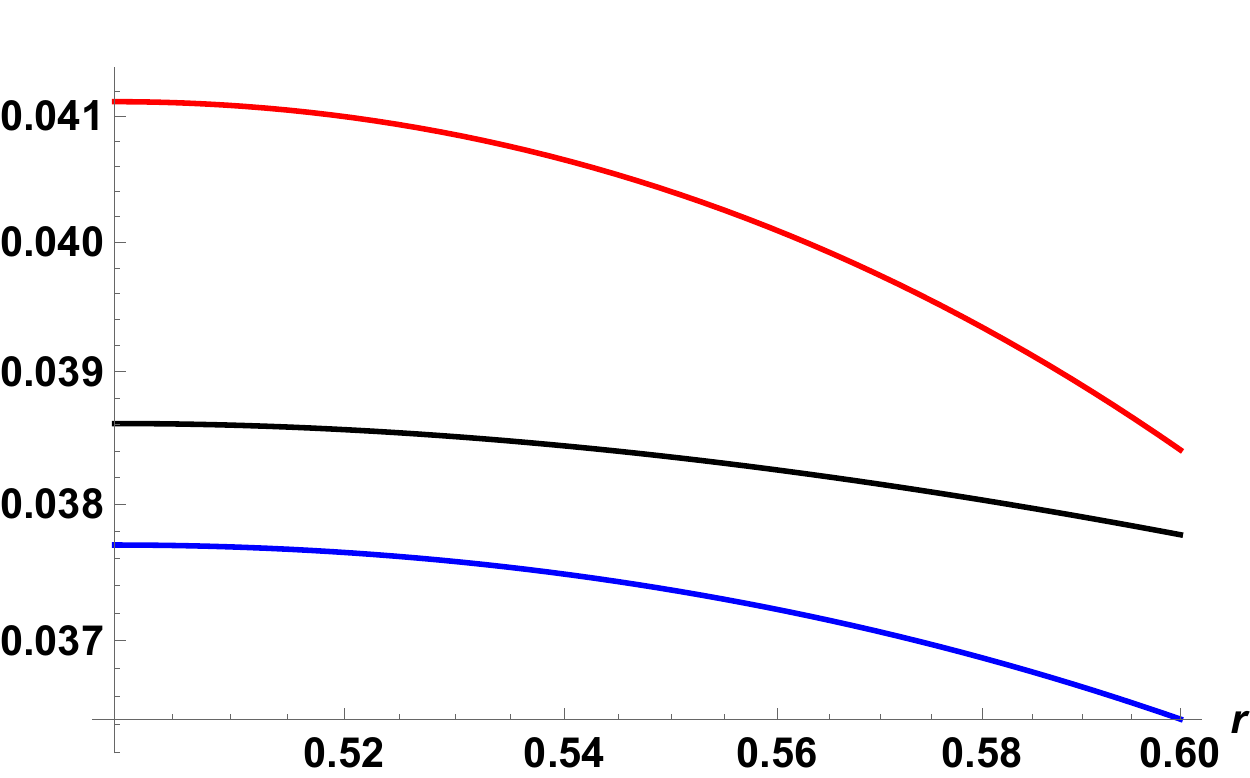}
\end{center}
\caption{Logarithmic plots of the number of boxes $\langle n \rangle$ (black line) for fixed $R= (2.5, 4, 6)$ - from left to right - as a function of $r$. The plots start from its critical value $r_c(R)$. With the blue line we depict the scaling $\xi^{\frac{4}{R+2}}$ and with red line the scaling $\xi^{\frac{4}{R-2}}$. Since the plots are logarithmic
we only care about the slope of the functions (scaling law). We observe that the blue curve captures the correct scaling properties of $\langle n \rangle$, but starts to deviate as we increase $R$, since $r_c(R) \rightarrow 1$, where the correct scaling is given by the narrow cut scaling of eqn.~\eqref{narrowscaling}. The numerical prefactor shifts the plots vertically and can be determined for each $R$, so that the blue curve falls on top of the black one. Similar plots can be given if we keep $r$ fixed and vary $R$.
}
\label{fig:alternatedressing}
\end{figure}

The opposite regime of the narrow cut in the unstable phase, is the limit where the ratio $r=a/b$ approaches the lower critical bound of the positivity condition in eqn.~\eqref{positivitycondition} $r \rightarrow r_{c}$. This is also the critical limit where the eigenvalues fill the metastable effective potential defined by eqn.~\eqref{sat1} as much as possible before spilling on the unstable side. 
We should then consider the edge conditions~\eqref{bc2},~\eqref{satasympt} that can be written as
\bea\label{bc5}
\frac{1}{2 b}  = r +  \frac{2}{\pi}  E\left(1 - r^2 \right) - \frac{R+2}{2 \pi} \left( E\left(1 - r^2 \right) - r K\left(1 - r^2 \right) \right) \, , \nn \\
b^{2-R} (1+r)^{-R} (1 - r)^2 = 2^{2-R}  \xi_{eff}^2   \, ,  
\eea
respectively. If we set $r = r_c$ in these relations and solve them in terms of $\xi_{eff}(R)$, we can define a critical line $\xi_{eff}^c(R)$, below of which a physical solution exists. This is the red dashed line in fig.~\ref{fig:Phasediagram} where the unstable potential becomes critical. We observe that the value of $\xi_{eff}^c(R) \sim O(1)$ as expected. Unfortunately the elliptic functions do not simplify near this line (unless we are very near $R=R_{KT}$). We can nevertheless determine approximately the scaling of the free energy, by numerical methods. The result is summarised in fig.~\ref{fig:alternatedressing}.

\paragraph{The free energy -}
Next, we determine the average size of the partition and the free energy using eqn.\eqref{firstderxi} and the asymptotic expansion of the resolvent to find
\be
\langle |\lambda| = n \rangle = - \half \xi  \partial_\xi S_{eff.} =  \frac{2(b+a)^2-R (b-a)^2}{16}   \, .
\ee
For the narrow cut approximation this becomes
\be\label{narrowscaling}
\langle |\lambda| = n \rangle  \simeq \frac{b^2 (1 - \epsilon)}{2} \simeq \frac{1}{32} ( 1 -  2^{3-R}  \xi_{eff}) \, , 
\ee
and is still positive.
If we integrate it we find the free energy
\be\label{freeRgreat}
\mathcal{F}_{narrow}^{(R>2)} \simeq - \frac{1}{16} \log \xi_{eff} + \frac{1}{16}  2^{3-R}  \xi_{eff}   + F(\mu) \,  .
\ee
We observe that the leading term is a logarithmic contribution,
that could have the interpretation of a ``quantum non-singlet" contribution to the entropy.

The situation is more interesting near the critical curve $\xi^{c}_{eff}(R)$, that is plotted by a red dashed line in fig~\ref{fig:Phasediagram}. In the near critical region we find the scaling law
\be
\langle |\lambda| = n \rangle_{c}  \sim   \xi_{eff}^{\frac{4}{R+2}}  \, , 
\ee
see the logarithmic plots in fig.~\ref{fig:alternatedressing}. This scaling law, makes evident the appearance of the opposite $(+)$ type of Liouville dressing for the winding mode. It is hard to determine the exact prefactor of this expression (it can be done numerically for each fixed $R$), but the sign is positive as expected. The free energy exhibits again the same scaling law, but is negative.

\section{Thermodynamics from continuous representations}\label{thermo}

In section~\ref{MQMnonsinglets} we made manifest the microstates that the MQM system is composed of, in terms of group theoretic representations/partitions of $GL(\infty)$. In the previous section~\ref{Saddles} we introduced the notion of a \emph{leading (continuous) Young diagram} in the limit of large representations that constitutes the dominant saddle of the aforementioned partition sum. This saddle is composed out of many microstates which have similar coarse grained characteristics (they are indistinguishable) in the thermodynamic limit. Here we shall further analyse the thermodynamic properties of the saddles we found in the previous section and the phase transition(s) between them as we vary the parameters of the model. We would also like to compare our results with the literature, since it has proven difficult to extract quantitative results for the black hole thermodynamics unambiguously~\cite{Kazakov:2001pj}. We therefore start by first reviewing some issues on the existing literature and  explain how our analysis overcomes them.

The application of the Toda differential equation to determine the free energy of the system, suffers from two difficulties: On the one hand it is impossible to work directly in the Sine Liouville point of the parameter space ($\mu \rightarrow 0$). The reason is that the string equation eqn.~\eqref{spectralKKK} can be obtained only in the dispersionless limit of the Toda equation, which reduces to a differential equation for the free energy, and this limit requires to consider the regime of large $\mu \rightarrow \infty$. One then obtains this solution and makes an analytic continuation to the opposite regime of small $\mu$.
The argument is that as long as conformal perturbation theory is valid (which is equivalent to saying that we do not encounter any phase transition in the system), 
we can reach the Sine-Liouville point. In~\cite{Kazakov:2000hea} it was argued that this holds as long as the couplings $t_{\pm 1}$ to the first winding modes of~\eqref{I13} remain relatively small, otherwise their scaling dimension could possibly change to the other $(+)$ branch. On the other hand since the solution of the dispersionless Toda equation was obtained using a specific initial condition (that of the undeformed thermal linear dilaton background), it seems impossible for this approach to capture the phase transitions that the system can exhibit (and we explicitly found such transitions in the previous section~\ref{Saddles}). 

\begin{figure}[t]
\begin{center}
\includegraphics[width=0.85\textwidth]{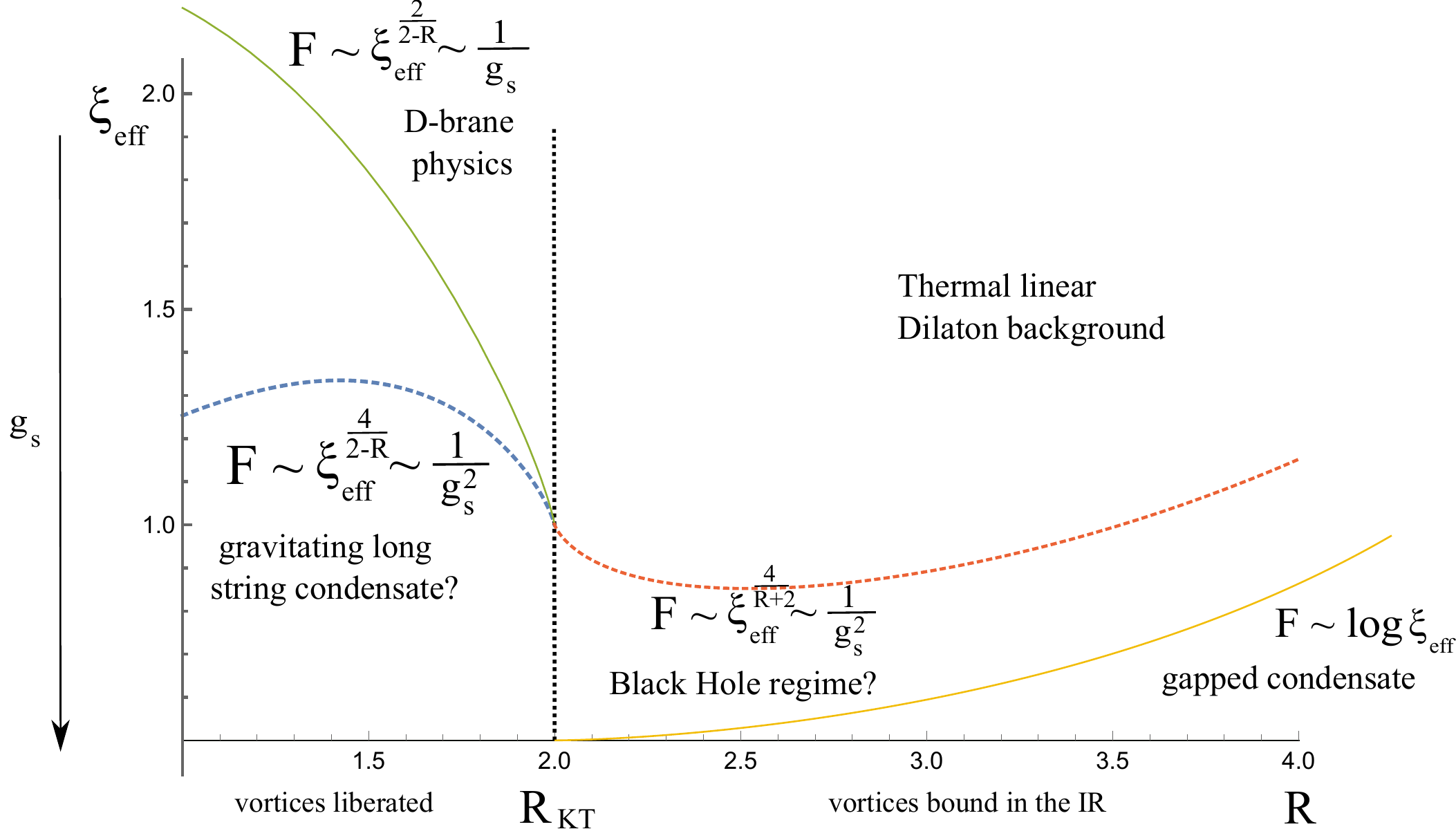}
\end{center}
\caption{The phase diagram. The dashed lines signal phase transitions. $T_{KT} \sim 1/R_{KT}$ is the Kosterlitz-Thouless temperature, above which worldsheet vortices get liberated and proliferate in the IR. The continuous lines do not signal phase transitions, but are regimes of different behaviour (cross-over). The question-mark in the gravitating long string condensate phase, has to do with the fact that one needs additional probes to distinguish the location of the black hole string crossover, since the scaling of the free energy and entropy for $R<R_{KT}$ is consistent with both interpretations. The
scaling of the free energy in the black hole regime for $R>R_{KT}$ reveals an opposite type ($+$) of dressing for the winding modes, that become normalisable and have support in the strongly coupled region of Liouville. This means that a black hole can start existing as an excited state in the spectrum of the theory. This region though, becomes increasingly narrower as $R\rightarrow \infty$ and is replaced by a non-singlet object whose free energy is logarithmically gapped compared to the singlet. The question mark
in the black hole phase for $R>R_{KT}$ has to do with the fact that we would like to understand the fate of the favourable critical region, when we turn on also a non-zero
$\mu$ (a third axis).
}
\label{fig:Phasediagram}
\end{figure}

On the contrary, our approach is direct and does not suffer from any of these complications. In the high temperature regime $1<R<2$, we are able to consider the cases of both small and large $\xi_{eff.} = \xi R^{-R/2}$, and uncover a phase transition between them. The transition line is defined by
\be
\xi_{eff.}^*(R)  = \left( \frac{\pi}{4-2 R} \right)^{(2-R)/2} \, ,
\ee
that we plot with a dashed blue line in fig.~\ref{fig:Phasediagram}. In addition as we show in eqns.~\eqref{freesaturated} and~\eqref{freeunsaturated}, the free energy scales exactly as~\cite{Kazakov:2000pm,Kazakov:2001pj} found, in the regime of $\xi_{eff} < \xi_{eff.}^*$ (in the phase with a saturated cut in the density of boxes), that is
\be\label{freeenergysmallxi}
\mathcal{F} \simeq - \frac{(2-R)^2}{8} \xi_{eff.}^{4/(2-R)} \, , \qquad S = \frac{\xi_{eff.}^{4/(2-R)}}{8} \left( 4 - R^2 - 4 R \log \xi_{eff.} \right) \, ,
\ee
where in our conventions the thermodynamic quantities are defined as
\be
\log \mathcal{Z} = - \mathcal{F} \, , \qquad \mathcal{F} = \beta M - S \, ,  \qquad S = \left(R \partial_R - 1 \right) \mathcal{F}\, ,
\ee
and we assumed that $\xi_{eff.}$ is the parameter that should be held fixed (radius independent) to define the entropy. This leads to an $\sim 1/g_{s}^2$ behaviour for the thermodynamic quantities (see the KPZ/DDK analysis of section~\ref{windingmatrixliouville} that relates $\xi_{eff.}$ with the effective string coupling $g_s$), that is consistent with the presence of a gravitating object in the background such as a black hole or a gravitating long string (winding mode) condensate. The factor $(2-R)^2$ indicates that the free energy vanishes at the Kosterlitz-Thouless point $R_{KT} = 2$, which is the point that worldsheet vortices get liberated in the IR. Moreover, the entropy is positive in the regime of parameters where the formula~\eqref{freeenergysmallxi} is valid. Its first derivative, the specific heat
\be
C_T = T \frac{\partial S}{\partial T} = - R \frac{\partial S}{\partial R} \, ,
\ee
is found to be a small negative number for $1<R<2$ that vanishes at $R=R_{KT}$ (again if we keep $\xi_{eff}$ fixed and radius independent). This indicates that the object is thermodynamically unstable (something that is expected from a non BPS black hole or a gravitating string condensate in asymptotically flat space).

Another important point that has not been emphasised in the literature concerns the sign of the free energy in~\eqref{freeenergysmallxi}, that is negative. This is crucial, since the number of boxes/size of the partition that physically corresponds to the number of wound strings/vortices is 
\be
\langle |\lambda| = n \rangle = - \half \xi  \partial_\xi \mathcal{F} =  \half \xi  \partial_\xi \log \mathcal{Z}   \simeq \frac{2-R}{4}  \xi_{eff}^{\frac{4}{2-R}}  \, , 
\ee
which is a positive number as it should\footnote{ Notice that the approach of~\cite{Kazakov:2000pm,Kazakov:2001pj} was giving a wrong sign in front of the free energy, (our free energy is defined via $\log \mathcal{Z} = - \mathcal{F} $, while~\cite{Kazakov:2000pm,Kazakov:2001pj} used the convention $\log \mathcal{Z} = + \mathcal{F}$ and they were also obtaining a negative sign with their convention). To our knowledge this issue with the sign of~\cite{Kazakov:2000pm,Kazakov:2001pj} was first noticed in~\cite{Maldanotes}. Perhaps the leading term in the singlet free energy that acted as an initial state in the Toda approach, was taken with an incorrect sign convention.}. We can also define a susceptibility of partitions/vortices via
\be
\chi_{vortex.} \, = \, \half \xi  \partial_\xi \langle |\lambda| = n \rangle \, = \, \half \xi_{eff}^{\frac{4}{2-R}} \, ,
\ee
that is also positive. The third derivative of the free energy though, becomes singular
and changes sign at the Kosterlitz-Thouless temperature $R_{KT} = 2$, signalling a third order phase transition (for any \emph{constant} $\xi_{eff} < \xi_{eff}^*$), above which vortices get liberated. This also shows that the order parameter of the transition is the winding mode as expected.

Another perspective for the system, can be obtained if we change ensemble and keep the size of partitions - number of vortices $\langle n \rangle$ fixed (de-Poissonisation). Using eqn.~\eqref{fixedsizetransform} to determine the free energy with fixed number of boxes ${F}(n)$, we find the leading thermodynamic quantities
\be
{F}(n) \simeq - (2-R) n \log \frac{\Lambda}{\sqrt{n}} \, , \qquad M \simeq \frac{1}{2 \pi} n \log \frac{\Lambda}{\sqrt{n}} \, , \qquad S = 4 \pi M \, . 
\ee
In these formulae $\Lambda$ plays the role of a UV cutoff (that also incorporates any non-universal terms) and in order to derive these relations we can either perform a saddle point evaluation of~\eqref{fixedsizetransform}, or equivalently a Legendre transform between the two ensembles. Once again, these thermodynamic quantities are consistent with a gravitating object that is a black hole or long-string condensate, since $S \sim M$.

For $\xi_{eff} > \xi_{eff}^*$ (and still for $R<2$), we enter another phase in a continuous manner. Initially the free energy scales again as in~\eqref{freeenergysmallxi}, but soon after we enter a complicated regime, that is described by Elliptic integrals. For very large $\xi_{eff}$ (which amounts to small effective string coupling $g_{s}$) things simplify, and D-brane (open string) physics govern the system, since the free energy scales as $\mathcal{F}\sim 1/g_{s}$. This is the regime of a very deep effective potential with a narrow cut analysed in section~\ref{cutnosaturation}. These results seem reasonable, since this is the regime of very high temperature and small effective string coupling $g_{s}$, where we do expect D-brane physics to dominate. 

At this point we should also mention the possibility raised in~\cite{Betzios:2017yms}, that $\xi_{eff.}$ is replaced by a parameter descending from a model of FZZT branes. As we mentioned in the introduction and in section~\ref{Measures}, this parameter is related to the boundary cosmological constant (that we called $\sigma$) of the open strings ending on the FZZT branes and the replacement is $\xi_{eff.} \rightarrow \exp (  -\pi R \sigma)$. Then $\xi_{eff.}$ becomes an actual fugacity (depending on the radius $R$) and $\sigma$ is the fixed external chemical potential (that can be both positive or negative). The thermodynamic quantities at fixed $\sigma$, change slightly in detail compared to those at fixed $\xi_{eff.}$, but their qualitative behaviour remains the same, so we do not repeat the analysis here.

This concludes our study of the high temperature $R < R_{KT} = 2$ regime. At the Kosterlitz Thouless line we encounter a phase transition that is again of continuous nature (third order) as long as $\xi_{eff.} < \xi_{eff.}^*$. For very high $\xi_{eff.} \gg \xi_{eff.}^*$, the nature of the transition changes and seems to become second order, as can be seen from the second derivative of eqn.~\ref{freeunsaturated} in the narrow regime. Our reasoning in this case is that the only solution we know for $R>R_{KT}$ and $\xi_{eff.} > \xi_{eff.}^*$ corresponds to the trivial representation - linear dilaton.

The regime $R> R_{KT}$ admits a metastable solution, since the effective potential we found is unstable. Away from $R_{KT}$ the worldsheet vortices are bound in the IR.
In this case we found three physical behaviours as can be seen in fig.~\ref{fig:Phasediagram}. For large $\xi_{eff.}$ there is no non-trivial solution and the physics is dominated by the thermal linear dilaton background. For $\xi_{eff.} < \xi_{eff.}^c$, below the (critical dashed red) line we find a critical regime where the free energy scales as $\mathcal{F} \sim  \xi_{eff.}^{4/(R+2)}$ (and is negative). This is the regime where the ($+$) Liouville dressing of the winding mode takes over, since the $(-)$ dressing is irrelevant for $R>R_{KT}$. From the KPZ/DDK analysis of the $+$ winding mode in section~\ref{windingmatrixliouville}, this means that the free energy scales in terms of the effective string coupling as $\mathcal{F} \sim 1/g_s^2$, which is once more indicative of black hole physics. This black hole behaviour exists for relatively small $R$, roughly up to $R=10$. As we increase $R$ further and further, we find that the behaviour of the free energy changes and scales with the logarithm of $\xi_{eff}$ (narrow cut regime). This corresponds to a non-singlet object whose free energy has a logarithmic gap with respect to the singlet~\cite{Gross:1990md,Klebanov:1991qa}. 

We conclude that our results for the thermodynamics are consistent with the presence of a background that resembles a black hole or gravitating long string condensate for $R<R_{KT} = 2$ (in string units) at sufficiently strong effective string coupling $g_s$. To distinguish between these two possibilities one would have to consider further probes of the (string sized) geometry. For larger $R>R_{KT}$, there is still a favorable critical black hole regime again at relatively strong string coupling. This is perhaps the best location to search for black hole physics, since this region is governed by the opposite type of Liouville dressing that is normalisable making the black hole an allowed excited state of the theory. In order to clarify the physics further, one would have to include the effects of a small non-zero $\mu$ to determine the fate of this favourable regime in the complete $\mu, \xi, R$ phase diagram of the model and study further probes of the effective geometry that can actually distinguish a black hole from a gravitating string condensate.

\section{A completely gauged ZZ/FZZT model}\label{D0D1}

The most complete model from a D-brane perspective is one that contains all possible types of open strings: ZZ-ZZ ($SU(N)$) open strings are known to be described by MQM via the matrix $M_{i j}$~\cite{McGreevy:2003kb,McGreevy:2003ep}. ZZ-FZZT strings were proposed to be described by bi-fundamental fields $\psi_{\alpha i} \, ,  \alpha = 1, ... , N_f$~\cite{Betzios:2017yms}. So far the action has a global $SU(N_f)$ symmetry and it is natural to ponder whether there is some extension of that model for which all the symmetries are gauged.

Related to this, there has been no proposal in the literature so far regarding the dynamics of the purely FZZT-FZZT strings ($SU(N_f)$). One can imagine that since these are $D1$ branes, extending along the Liouville direction, they should be described by a variant of $2d$ YM or generalised BF theory describing their low energy dynamics. In this ``complete" model both $SU(N)$ and $ SU(N_f)$ will be gauged\footnote{ A model of this type with two copies of MQM coupled through a general BF theory was analysed in~\cite{Betzios:2021fnm}, as a model having the potential to describe Euclidean wormholes.}. Let us now describe the various parts that are contained in the action of such a system.

The Hermitian gauged MQM part of model (ZZ-branes) is defined by
\begin{equation}\label{D0}
S_{MQM}=\int dt \, \tr \left(\frac{1}{2}\left(D^A_{t}M\right)^{2} - V(M) \right),
\end{equation}
where the covariant derivative with respect to the gauge group is $D_{t}M={\partial}_{t}M-i\left[A,M\right]\,$ with $V(M)= -\half M^2$ in the relevant case of $c=1$ Liouville theory described by the double scaling limit of MQM that focuses in the unstable inverted harmonic oscillator potential near the tip~\cite{Klebanov:1991qa,Boulatov:1991xz}. The theory has an $SU(N)$ gauge symmetry 
\begin{eqnarray}\label{gauge}
M(t)\rightarrow U(t)M(t)U^{\dagger}(t),\qquad A(t)\rightarrow  U(t)A(t)U^{\dagger}(t)+i  U(t){\partial}_{t}U^{\dagger}(t).
\end{eqnarray}
The conserved current associated to this symmetry is $J= - i [M, \, \dot M]$, note that it is traceless and its diagonal $U(1)$ part is trivial as one can also see from the transformation $M( t)\rightarrow U(t)M(t)U^{\dagger}(t)$.
The ungauged MQM model, see appendix~\ref{Ungauged}, decomposes into different irreducible representations of the $SU(N)$ algebra classified by $J$. More precicely, the allowed representations should be such that they admit a zero weight state and arise in the decomposition of tensor products of the adjoint~\cite{Gross:1990md,Boulatov:1991xz,Karczmarek:2008sc,Betzios:2021fnm}. This selection rule on the possible representations comes from the fact that the diagonal part of $A$ does not appear in the action, see also appendix~\ref{Ungauged}. This is in accord with the fact that J is traceless. 

We proposed that the ($D1$ or FZZT) part of the system is expected to be described by an action of the generalised YM type ($\tilde{A}$ is the one form gauge field, $\tilde{F} = d \tilde{A} +   \tilde{A} \wedge  \tilde{A} $ and $d \mu$ is the volume form)
\be\label{gYMaction}
S_{gYM} = \frac{1}{g_{YM}^2} \int_\Sigma  \Tr B  \tilde{F} \, - \,  \frac{1}{2 g^2_{YM}} \int_\Sigma \Tr \Phi(B) \, d \mu \, .
\ee
In this action $B$ plays the role of an auxiliary zero form, and $\Phi(B)$ is a potential that specifies the gYM theory we are considering. We do not really know the exact version of gYM that is relevant for the dynamics of the FZZT strings, one simple option is that $\Phi(B)=B^2$ when we recover the usual 2d YM theory. At finite temperature, it is natural to take the topology of $\Sigma$ to be the disk or the cylinder. In this case, it can be shown that the dynamics of the action~\ref{gYMaction} is related to a gauged version of an $SU(N_f)$ Unitary matrix quantum mechanics model living on the boundary/ies of the disk/cylinder with the action~\cite{Minahan:1993mv}
\be
S_{u} = - \half \int dt \,  \tr \left[ \left(u^{-1} D^{\tilde{A}}_{t} u \right)^{2} \right] \, , \quad D^{\tilde{A}}_{t} u = \partial_t u + i \left[ \tilde{A}, u \right]
\ee
In order to distinguish the two groups, we will use Latin indices for the $SU(N)$ matrices and Greek for the $SU(N_f)$ and a tilde for the $SU(N_f)$ gauge field.
The ungauged version of this model again decomposes into different sectors and its Hamiltonian is written as
\bea
{H}_{u} = \left[\sum_\alpha^{N_f} \half {p}_\alpha^2 + \half \sum_{\alpha \neq \beta} \frac{Q_{\alpha \beta}^\mathcal{R} Q_{\beta \alpha}^\mathcal{R}}{4 \sin^2(\frac{\theta_\alpha - \theta_\beta}{2})} \right]  \, \nn \\
\hat{H}_{u} \mathcal{R}_{A B}(u) = C_\mathcal{R} \mathcal{R}_{A B}(u) \, ,
\eea
where we wrote both the Classical Sutherland and the quantum version of the Hamiltonian acting on the representation matrices $\mathcal{R}_{A B}(u)$ ($A,B$ label the $SU(N_f)$ generators). The gauging of the model results into considering as admissible states the invariant wavefunctions which in particular are the characters of the representation $\mathcal{R}$, denoted by $\chi_\mathcal{R}(u)$.

We shall now couple these two models while still keeping the total action gauged under $SU(N)\times SU(N_f)$. To achieve this we can add extra fields into the action. In particular we will be interested into adding extra fermionic or bosonic matrices, $\psi_{\alpha i}$ or $V_{\a  i}$ with  $\alpha \in [1, N_f]$, $i \in [1, N]$, transforming as bi-fundamentals. The relevant fermionic action is
\be
S_{f} = \int dt   \left( i \psi_{\a i}^\dagger \left( \delta_{\alpha \beta} \delta_{i j} \partial_t + i \delta_{\alpha \beta} A_{i j} + i \delta_{i j} \tilde{A}_{\alpha \beta} + m \delta_{i j} \delta_{\alpha \beta} \right)  \psi_{\beta j}  \right)  \, ,
\ee
and we shall also need to add a set of oppositely charged anti-fundamentals which we label with $\chi_{\a i}$.
The reason for including them is that in the total model we find the constraints
\be
\frac{\delta S_{total}}{\delta A_{i j}} = 0 \, \Rightarrow \quad    i[M, \, \dot{M}]_{i j} = \sum_\a^{N_f} \psi_{\a i} \psi^\dagger_{\a j} - \chi_{\a i} \chi^\dagger_{\a j}  \, ,
\ee
and
\be
\frac{\delta S_{total}}{\delta \tilde{A}_{\a \b}} = 0 \, \Rightarrow \quad i [\dot{u}, u^{-1}]_{\alpha \beta} = \sum_{i = 1}^N \psi_{\a i} \psi^\dagger_{\beta i} - \chi_{\a i} \chi^\dagger_{\beta i} \, ,
\ee
that mediate a representation theoretic interconnection between the two models and the groups $SU(N)$ and $SU(N_f)$, since the bifundamental excitations ``feed" the different representations that can appear in the Hamiltonians of the uncoupled models. It is clear from these constraints that we need to include also antifundamental fields as in~\cite{Betzios:2017yms}, so that the trace of the righthand side is zero\footnote{ Notice that we are now allowed to add Chern-Simons type of terms $k \int \tr A$ or $\tilde{k} \int \tr \tilde{A}$ for more generality to shift the overall $U(1)$ charge of each group~\cite{Betzios:2017yms}.}.

\subsection{Canonical partition function}

As we show in the previous section, our proposed ZZ/FZZT model is comprised of $SU(N)$ gauged MQM coupled to a (gauged) action of a particle on the group manifold $SU(N_f)$
via the presence of bi-fundamental fields. 

To compute the partition function of the system, we need to specify the topology of the
FZZT branes. Since they stretch to infinity, we shall momentarily assume that there is just one boundary there and hence their topology is of the disk type. From now on, we parametrise with $U = e^{i \theta}$ the $SU(N)$ holonomy zero modes and with $\Omega = e^{i \phi}$ the $SU(N_f)$ ones. The partition function of $2d$ YM on the disk is
\be
Z_{disk}(\Omega) = \sum_{\mathcal{R}: \, \ell(\mathcal{R}) \leq N_f} D_{\mathcal{R}} \chi_{\mathcal{R}}(\Omega) e^{- \frac{g_{YM}^2}{N} C_{\mathcal{R}}^2 }   \, ,
\ee
with $C_{\mathcal{R}}^2$ the quadratic Casimir.

The total result for the partition function then comes from coupling this disk piece to  the piece coming from MQM and the  $N \times N_f$ bi-fundamentals. After we integrate out all the matter fields, we get the following expression for the total finite temperature partition function as a holonomy integral (where $x = e^{- \beta m}$ with $m$ the masses of the bifundamentals)
\bea
{Z}_{system} &=& \sum_{\mathcal{R}: \, \ell(\mathcal{R}) \leq N_f, N} x^{|\mathcal{R}|} \int D U \, \chi_{\mathcal{R}} (U) Z^{MQM}_N (U) \, \int D \Omega \,  \chi_{\mathcal{R}} (\Omega^\dagger) Z_{disk}(\Omega) \nn \\
 &=& \sum_{\mathcal{R}: \, \ell(\mathcal{R}) \leq N_f, N} x^{|\mathcal{R}|}  \, D_{\mathcal{R}}    \,  e^{- \frac{g_{YM}^2}{N} C_{\mathcal{R}}^2 }  \,  Z^N_{\mathcal{R}} \, , 
\eea
with $Z^N_{\mathcal{R}} = \tr_{\mathcal{R}} e^{- \beta \hat{H}_{\mathcal{R}}}$ the MQM partition function in the arbitrary representation\footnote{The role of the bi-fundamentals was to relate the allowed representations of $SU(N)$ and $SU(N_f)$ in the partition function.}. For more details see appendix~\ref{Ungauged} and~\ref{holonomypf}. We observe that the result is similar to the ungauged matrix model partition function, with the difference that the sum over representations is weighted with the quadratic Casimir. The length of the allowed representations is capped by the minimum of $N, N_f$ and we naturally wish to study a limit where both of them are large and of the same order.

\subsection{Grand Canonical partition function}

We can also pass to the Grand-Canonical ensemble. The simplest idea is to introduce a single-chemical potential dual to $N$ as in the usual study of MQM. Labelling once more the representations with partitions ($\mathcal{R} \equiv \lambda$), the grand canonical partition function acquires a form similar to the $\tau$-functions we studied (see section~\ref{Tau}) 
\be
\mathcal{Z}(\mu, R ; g_{YM}) =  \sum_{\lambda : \, \ell(\lambda) \leq N_f } \dim (\lambda) \,  e^{- \frac{g_{YM}^2}{N}  \, C_\lambda^2 \, - \, \beta m  |\lambda|} (-1)^{|\lambda|}  {\bf G}_{\lambda \lambda}(\mu , R) \,   .
\ee
where ${\bf G}$ is the $GL(\infty)$ element for MQM, see eqn.~\eqref{BBB}. It would be interesting to study further the properties of this completely gauged model, since it also admits a saddle point analysis in the limit of large representations. The main difference with the examples we studied so far is that the quadratic Casimir creates an effective quadratic potential in the space of representations, that stabilises the logarithmic potential even for radii $R>R_{KT} = 2$, where it was unstable, see section~\ref{Saddles}. So perhaps this model has a better chance to capture the physics
of a semi-classical black hole for very large radii $R \gg 2$.

\section{Discussion}\label{Discussion}

In this work, we developed a new technique for analysing $\tau$-functions, by expanding them in terms of representations/partitions and considering the limit of continuous Young-diagrams. The resulting saddle point equations in the space of highest weights can then be studied with standard matrix model techniques.

Our physical motivation behind this analysis was to understand the origin of the microstates of a string theoretic version of a (non-supersymmetric) two dimensional black hole in asymptotically flat space and its thermodynamic properties. The later
are summarised in section~\ref{thermo} and in particular in fig.~\ref{fig:Phasediagram}. Using our techniques we not only managed to reproduce the results of~\cite{Kazakov:2000pm,Kazakov:2001pj} in the appropriate Sine Liouville limit for compactification radii $R< R_{KT} = 2$, but we also found the presence of new phases and extended the description of the model in some parametric regime above $R > R_{KT} = 2$. Along the way we clarified various subtleties in the physics of the various regimes and ensembles. We also encountered some unexpected features, such as the need to consider both types of Liouville dressing for the winding mode operators if we wish to go beyond the $R_{KT}$ barrier. There are still several salient features and details that remain to be explored and understood, such as the analysis of observables that could truly distinguish a black hole from a gravitating long string (winding mode) condensate and the complete description of the phase diagram in the three dimensional parameter space spanned by $(R, \mu, \xi$), perturbing away from the strict Sine Liouville plane. We now finish our discussion with some specialised comments and ideas.


\paragraph{Lorentzian version of the FZZ duality and regimes - }
In section~\ref{introduction} we discussed the Euclidean version of the FZZ duality, which is the one that is better understood. The Lorentzian version of the duality, is still somewhat mysterious.
The analytic continuation of the winding modes around the thermal circle can be argued to be related to the presence of long-strings that stretch from the weakly coupled region towards the strong coupling region (Liouville wall)~\cite{Maldacena:2005hi}. This is a perturbative picture around the linear dilaton background and uses the winding modes/long strings with the $(-)$ type of dressing, $\mathcal{T}^-_{\pm R}$ in eqn. \eqref{I10}, that have support in the weakly coupled region. If one wishes to describe the Lorentzian black hole background, it has been recently argued that a more refined version of the ER/EPR correspondence holds~\cite{Attali:2018goq,Jafferis:2021ywg}. In the ER side of the duality one finds a $2d$ black hole horizon and an Einstein-Rosen (ER) bridge connecting the two sides of the eternal black hole. In the EPR side one has a state that corresponds to the supersposition of two disconnected spacetimes with the presence of entangled pairs of long strings that emanate from the strong coupling region (see also~\cite{Yogendran:2018ikf,Giveon:2019gfk} for related ideas). Due to this peculiarity and the inherent strong coupling physics, it is much harder to understand their properties\footnote{ Nevertheless it was proposed in~\cite{Agia:2022srj} that a version of angular quantisation of the worldsheet CFT could shed light to the physics of such configurations.}.

In this work we propose that these long strings whose endpoints reside at strong coupling, are the Lorentzian continuation of the $\mathcal{T}^+_{\pm R}$ operators in eqn. \eqref{I10}, the main indication being that the $(+)$ type of Liouville dressing does in fact have support in the strongly coupled region of Liouville. An additional motivation for this correspondence is that these winding modes constitute normalisable deformations that change the state/background of the string theory. The picture that we advocate here seems to be consistent both with the presence of a black hole/string transition and the two types of long string condensates that have appeared in the literature\footnote{It is also noteworthy that the Lorentzian version of T-duality exchanges the singularity of the black hole with the horizon~\cite{Dijkgraaf:1991ba,Giveon:1994fu}, allowing perhaps to directly probe the singularity using these long strings.}. This again would mean that the importance of the two possible dressings of the winding mode operators/long strings is exchanged around the BH/string transition point, on the one side of which the black hole starts to exist in the spectrum of the theory. This is consistent with our findings of section~\ref{Saddles} in the regime $R> R_{KT}$. 

Going back to the canonical ensemble in Euclidean signature, our picture is that at very high temperatures (small radii) the background should be thought of as a condensate of winding modes that are explicitly sourced and emanate from the weakly coupled region (non normalisable deformation of the linear dilaton background). At somewhat lower temperatures\footnote{At very low temperatures one expects the linear dilaton background to dominate.} (larger radii) the system transitions into a strongly coupled winding mode condensate (having support on the opposite strong coupling region of the Liouville wall). Its dual weakly coupled (simple/perturbative) description according to the FZZ duality is in terms of a geometric black hole background and perturbative excitations around it. In this regime of parameter space the black hole starts to exist as an excited state in the string theory spectrum. This picture is also consistent with the remarks and findings of~\cite{Giveon:2005mi,Kutasov:2005rr}. 

In Lorentzian signature, there is no notion of canonical ensemble, but we can consider instead the scattering of long strings and tune their number $n$ and energy $E$. The Hamiltonian is that of $N$ interacting fermions in the inverted oscillator potential that can be found in appendix~\ref{Ungauged} (for the FZZT-brane models it is related to Spin-Calogero models~\cite{Betzios:2017yms}). The wavefunction transforms in a representation of $SU(N)$ with a fixed total number of boxes equal to $n$. Their number $n$ should be thought of as roughly the conjugate variable to the logarithm of the SL coupling $\log \xi$, see section~\ref{Measures} and~\ref{thermo} for the details. The question then is what are the various regimes and thresholds of scattering processes as we change these parameters\footnote{See~\cite{Ahmadain:2022gfw} for a more detailed analysis of these regimes.}. If the energy of the scattered constituents does not scale with $N$, then they do not backreact on the linear dilaton background. Based on general principles, we expect that the formation of a black hole or gravitating long string condensate could potentially happen only when the energy scales as $O(N^2)$. In this case the long strings penetrate very far in the strongly coupled region and interact very strongly, possibly forming a compact object. If their number is relatively small $1 < n \ll N $
it is not clear whether the interactions are enough to form such an object and even if it forms, its entropy will be small since the number of boxes in the allowed representations does not scale with $N$. If they are of order $1 \ll N \ll n \sim O(N^2)$, then they can potentially form a quasi-stable object, that has the interpretation either of a gravitating long string condensate or that of a black hole.
This is consistent with our findings in section~\ref{Saddles}, where we had to consider very large representations of $O(N^2)$ to define a non-trivial scaling limit. We expect that the precise fraction of $N^2$ might be important in distinguishing between these two cases.

\paragraph{Observables in a non-singlet background -} Our analysis and the definition of a non-singlet density of (energy) states $\rho_{n.s.}(\epsilon)$ from the density of boxes $\mathcal{N}(z)$ in the leading Young diagram, using the formula  eqn.~\eqref{densitiesrelations} in the appendix
\be
\rho_{n.s.}(\epsilon) =  \frac{1}{2 \pi} \int_{\text{supp.}} d z \, \mathcal{N}(z) \frac{z}{z^2+ \epsilon^2} \, ,
\ee
makes possible the computation of other types of observables in the black hole or long string condensate background. In particular one can use it to compute Tachyon or loop correlation functions on a background for which non-singlets have condensed (even if they originate from the presence of winding modes)\footnote{Mixed correlators of any kind are much harder to compute in any background~\cite{Alexandrov:2002fh,Kostov:2006dy,Bourgine:2007ua}.}, adjusting the formalism developed in~\cite{Moore:1991sf,Moore:1991zv,Betzios:2020nry}, by replacing the singlet $\rho_{singlet}(\epsilon)$ in the appropriate formulae. In essence, this is a form of mean field approximation that incorporates the effects of the non-singlets in a coarse-grained fashion. In addition, one could also determine an effective potential and a reflection amplitude that can reproduce the non-singlet density of states in our cases of interest. Such an analysis is extremely important, since it would allow us to clarify and elaborate on the properties of the object that we would like to identify as a black hole or long string condensate (for example whether it has or not effective absorption properties, a quasinormal spectrum and so on).

\paragraph{Integrability vs Chaos -}
All the (non-singlet) MQM models that have been proposed to describe a black hole in string theory~\cite{Kazakov:2000pm,Betzios:2017yms} satisfy certain properties of integrable systems, a fact which seems to be in clash with the expected chaotic behaviour of black holes. On the other hand the nature of the inverted oscillator\footnote{The inverted oscillator is related to a Dilation operator that has been discussed numerous times in relation to "quantum chaotic systems", according to the Berry-Keating proposal~\cite{BerryKeating1,BerryKeating2,Betzios:2020wcv}.}, gives rise to a continuous unbounded spectrum and hence it is not entirely clear if these models are integrable in a strict sense\footnote{Introducing appropriate effective boundary conditions could perhaps give rise to a discrete almost random (chaotic) spectrum, coinciding for example with the non-trivial Riemann zeros.}, especially when non-trivial representations and the resulting Spin-Calogero type of term are introduced into the model~\cite{Betzios:2017yms}. An answer to this question could perhaps be given upon studying the energy spectrum for non trivial representations of a typical large size\footnote{In particular one would need to analyse not only the density of energy states, but also the statistics of eigenvalues and whether the system exhibits any form of spectral rigidity/eigenvalue repulsion as expected from quantum chaotic systems.}. Another option is to analyse four point OTOC's in the non-singlet sector of MQM. Perhaps this is possible using our proposed non-singlet density of states.

\paragraph{Higher dimensional Black Holes -} In higher dimensional theories, when the string theory is defined in an asymptotically $AdS$ space, one can use the AdS/CFT correspondence to provide for a non-perturbative definition of the bulk quantum gravity path integral. The partition function of the gauge theory should then be able to capture
all the bulk black hole microstates. For weakly coupled gauge theories it is possible to directly cast the partition function in terms of a unitary integral that can be subsequently analysed using the technology of free fermions and partitions~\cite{Aharony:2003sx,AlvarezGaume:2005fv,AlvarezGaume:2006jg,Dutta:2007ws}. More generally, supersymmetry and the power of localisation allows a similar description of superconformal indices (even at strong coupling) in terms of matrix integrals with more complicated integrands, see for example~\cite{Cabo-Bizet:2018ehj,Choi:2018hmj,Benini:2018ywd} for some recent works clarifying how to extract the black hole entropy from these integrals. We believe that it is of great interest to  perform an analysis of such examples in the spirit of the present work (a relevant recent analysis can be found in~\cite{Murthy:2022ien}). 

Finally, if we wish to consider non-supersymmetric (asymptotically flat) black hole backgrounds, there is a useful expansion parameter that connects such solutions directly with our simple two dimensional example. As shown in~\cite{Emparan:2013xia}, in a large dimension limit, a wide class of non-extremal neutral black holes has a universal near horizon limit given by the two dimensional black hole of string theory that is the focus of the present work\footnote{See~\cite{Betzios:2018kwn} for a method to ``glue" UV AdS asymptotics on such solutions.}. A recent analysis of the black hole string transition in this setup/limit~\cite{Chen:2021emg,Chen:2021dsw}, indicates that these large dimension black holes can have temperatures a bit higher than the Hagedorn temperature, but it proved hard to analyse the transition to the string condensate. It would be extremely interesting to understand the implications and extensions of our work in this context.

We hope to report on some of these directions in the future.

\acknowledgments

We are especially grateful to Juan Maldacena for discussions at various stages of this work and for sharing with us his unpublished notes. Some special thanks also go to Gordon Semenoff for encouraging comments and discussions on matrix models. We wish to thank the string theory groups at UBC and Perimeter Institute as well as the participants of the online \href{https://danninos.wixsite.com/matrixmodels}{Workshop on Matrix Models and String Theory} (2020) for several stimulating discussions in relation to the topic of this work. We thank the authors of~\cite{Ahmadain:2022gfw} for useful exchanges and discussions on our respective works. We also acknowledge the hospitality of the Aspen Center for Theoretical Physics during the last stages of completion of this work.

\noindent The research of P.B. is supported in part by the Natural Sciences and Engineering Research Council of Canada. P.B. and O.P. acknowledge support by the Simons foundation. Research at Perimeter Institute is supported in part by the Government of Canada through the Department of Innovation, Science and Economic Development and by the Province of Ontario through the Ministry of Colleges and Universities.

\newpage

\appendix

\section{Partitions - Representations}\label{Partitionsreps}

We first provide some terminology and conventions on partitions~\cite{Macdonald}.  A part of this section also appears in~\cite{Betzios:2021fnm}, here we present a version describing in addition the Frobenius notation and composite representations that are useful for our purposes.

\subsection{Partitions}

\begin{itemize}
\item
A partition $\l$ is a sequence of non-increasing integers such that
\be
\l_1 \geq \l_2 \geq \, ... \, \geq \l_{\ell(\l)+1} = 0
\ee
The number of non-zero elements $\ell(\l)$ is called the length of the partition. The sum of all the elements $|\l|= \sum_{i \geq 1} \l_i$ is called the weight of the partition.
\item
The multiplicity $m_j(\l)$ of the positive integer j is how many times the number j appears in the partition $\l$ (such that $\l_i = j$).
\item
The partitions are represented graphically using Young diagrams. They are an array of boxes where the $i$'th row contains $\l_i$ boxes. This means that the number of rows is the length of the partition and the number of columns  is just $\l_1$. The total number of boxes is then equal to the weight $|\l|$.
\item
The conjugate or transpose of a partition $\l'$ or $\l^T$ is obtained by reflecting the Young diagram along the diagonal exchanging rows and columns. As an example one obtains $\l_1' = \ell(\l)$. 
\end{itemize}
As a simple example, the partition $(7,5,3^2,1^2)$ corresponds to the following Young diagram
\be \yng(7,5,3,3,1,1)\nn\ee
The irreducible representations of the symmetric group $S_n$ are in one-to-one correspondence with the Young-diagrams $\lambda$. The rows are symmetrizers of the elements while the columns anti-symmetrizers. Furthermore once appended with a totally ordered set (an ``alphabet") the Young Tableaux parametrise the irreps of $GL(N)$, where the $\lambda_i$ are a relabelling of the highest weights and the diagram has at most $N$ non-empty rows (this just means that the length of the partition is $\ell(\lambda) = N$). For the group $SL(N)$ the diagram has instead at most $N-1$ non-empty rows.

Using $\ell(\lambda)$ as the number of rows of the partition and by $d(\lambda)$ the number of diagonal elements, we can also adopt a Frobenius notation for the partition $\lambda \equiv (\alpha_1,..,\alpha_d | \beta_1, ... \beta_d)$, where $\alpha_i = \lambda_i - i, \, \beta_i = \lambda_i' - i$. This means that we count with $\alpha_i$ the number of rows above the main diagonal and with $\beta_i$ the number of columns below the main diagonal. If $\lambda = (\vec{\alpha} | \vec{\beta})$ then the transpose partition is $\lambda' = (\vec{\beta} | \vec{\alpha})$ and the $\alpha$'s, $\beta$'s are now strictly ordered. We also get the sum rule
\be
\sum_{i=1}^{d(\lambda)} (\alpha_i + \beta_i) + d = |\lambda|
\ee
As a simple example $\lambda=(5,4,1,1) \equiv (4,2|3,0)$, see below:
\\
\\
\yng(5,4,1,1)
\\
\\
One can connect the various partitions with the algebra of free fermions that we define in appendix~\ref{Fermionic}. In particular one can construct an arbitrary representation/partition by acting either with current or fermionic operators on a vacuum state (this can have an overall $U(1)$ charge-$s$) and create states corresponding to Young-diagrams/ $U(N)$ representations. The fermions we use carry half integers modes $p_i, q_i$ that are related to the Frobenius integers via $p_i = \beta_i + \half$, $q_i = \alpha_i + \half$. For more details see appendix~\ref{Frobeniusreps}.

\subsection{Composite representations}\label{UvsSU}

Composite representations naturally appear in the ungauged version of MQM, since the allowed $SU(N)$ representations need to admit a zero weight state and arise in the decomposition of tensor products of the adjoint~\cite{Gross:1990md,Boulatov:1991xz,Karczmarek:2008sc,Betzios:2021fnm}. This means that the diagram can be written in a boxes/anti-boxes form that we shall describe. More details on composite representations can be found in~\cite{Gross:1993hu,Marino:2009mw}.

Let us denote by $\mathcal{R}$ a $U(N)$ representation and decompose it into an $SU(N)$ representation labeled by $R$ and a $U(1)$ charge denoted by $q$. The Young diagrams are then related by
\be 
\mathcal{R}_i = R_i + r \, , \quad i=1,2,...,N-1 \, ,  \qquad \mathcal{R}_N = r
\ee
In other words we add $r$ columns of length $N$. The $U(1)$ charge that counts the number of boxes is
\be
|\mathcal{R}| = q = |R| + N r
\ee
The Casimirs are related as
\be
C_1(\mathcal{R}) = q \, , \qquad C_2(\mathcal{R}) = C_2(R) + \frac{q^2}{N}
\ee
with
\be
C_2(R) = \sum_{i=1}^{N-1} R_i \left(R_i - 2 i + 1 \right) + N |R| - \frac{|R|^2}{N}
\ee
\\
\\
One can subsequently factorise the representation Hilbert space at large $N$ (it fails non-perturbatively in $N$), by expressing the $SU(N)$ representation $R$ as $R \equiv \, R_+ \, \bar{R}_-$, or in other words as a coupled or composite representation. To label the composite representation is is most useful to transpose the Young-diagram and work with lengths of columns instead of rows. In particular if $R_\pm$ have a total number of $|R_\pm|$ boxes, and $r^\pm_i$ boxes in each column, we then form the composite $SU(N)$ representation $R$ as
\be
r_i = N - r^-_{L(R_-) + 1 - i}\, , \quad i \leq L(R_-) \, , \qquad r_i = r^+_{i - L(R_-)} \, , \quad i> L(R_-) \, .
\ee
In this expression we have denoted by $L(R_-)$ the number of boxes in the first row of the Young diagram for $R_-$.

We can understand this decomposition in a simple example (for $N=10$)
\be
R = \yng(7,6,4,4,4,4,4,3,2) \, , \qquad R+ = \yng(3,2) \, , \qquad \bar{R}_- =  \young(:::\hfil,::\hfil\hfil,\hfil\hfil\hfil\hfil) \, , \qquad  R_- = \yng(4,2,1)  \nn
\ee
Graphically this means that to construct the representation $R$ we start with a big parallelogram of $N \times L(R_-)$ boxes, and we remove the boxes from its bottom (forming the shape of $\bar{R}_-$) and we add boxes next to its first rows (forming the shape of $R_+$). As far as the weight is concerned the overall $N \times L(R_-)$ does not count in the $SU(N)$-rep (but it would count if it was viewed as a $U(N)$ rep). So we end up with a positive contribution from the boxes added on top and a negative from the boxes that are removed from the bottom (the $\bar{R}_-$ diagram is also said to contain \emph{anti-boxes}). For the ungauged matrix model that we describe in~\ref{Ungauged}, there is an additional constraint that the number of boxes is equal to the number of anti-boxes (only the representations that admit a zero weight state are allowed~\cite{Boulatov:1991xz,Betzios:2021fnm}).

As another useful example, the dimension of the composite representation at large N is
factorised as follows:
\bea
\dim R  = \dim R_+ \dim R_-  \, Q(R_+,R_-) \, , \qquad \text{with} \, , \nn \\
Q(R_+,R_-) = \prod_{i , j}^{rows} \frac{(N+1- i - j)(N+1-i - j + r_i^+ + r_j^-)}{(N+1-i - j + r_i^+ )(N+1-i - j + r_j^-)}  \sim 1+ \frac{r^+ r^-}{N^2} + ... \, , \nn \\
\eea
with $r^\pm = \sum_i r_i^\pm$ and the product is over the rows of $R^\pm$ for each index.



\section{Free fermions and Integrable Hierarchies}\label{hierarchies}

This appendix is dedicated to a brief introduction of free fermions, integrable hierarchies and how the general (vortex perturbed) grand canonical partition function of MQM is a $\tau$ function that can be expanded in a sum over representations/partitions. We follow the conventions in~\cite{Kazakov:2000pm,Alexandrov:2003ut}. A part of this appendix also appears in~\cite{Betzios:2017yms} and we repeat it here for the reader's convenience. Here our main focus is to analyse the connections between the $\tau$ functions and the representation theory of the infinite symmetric group. A useful review is~\cite{Alexandrov:2012tr}\footnote{The reader should be aware of the slightly different conventions used in this reference.}.

\subsection{Fermionic Algebra}\label{Fermionic}

To set up our conventions we define $\psi_n\, , n \in \mathbb{Z}+\half$ to be free fermionic operators satisfying
\be\label{C1}
[\psi_n\, , \psi^*_m]_+ = \delta_{n m} \,, \quad [\psi^*_n\, , \psi^*_m] = [\psi_n\, , \psi_m] = 0\, ,
\ee
and the fermion fields $\psi(z) = \sum_{n \in \mathbb{Z}+\half} \psi_n z^{-n-\half}$ and $\psi^*(z) = \sum_{n \in \mathbb{Z}+\half} \psi^*_{-n} z^{-n-\half}$. We then define the vacuum with charge $l$ as follows
\bea\label{C2}
\psi_m | l \rangle = \langle l | \psi^*_m = 0\, \quad m>l \nn \\
\psi^*_m | l \rangle = \langle l | \psi_m = 0\, \quad m<l\, ,
\eea
which in particular means that $\psi^*$ creates holes and $\psi$ creates particles (holes have a unit of charge $+1$ and particles $-1$) and the fermi sea has a total charge $l$. The vacuum is normalised $\langle l | l \rangle = 1$. Since
\be\label{C3}
\psi^*_n | n \rangle = | n+1 \rangle \, , \qquad \psi_n | n+1 \rangle = | n \rangle \, ,
\ee
we can create any excited state by acting with a repetition of these operators. If we define a vaccum, for example the Dirac vacuum $|0\rangle$, we can define a normal ordering of operators with respect to that vacuum. We denote this operation by $\normord ( ... )\normord$. By convention this is moving all the creation operators to the left and all the annihilation operators to the right (with respect to that vacuum) in the expression inside. For example $\normord \psi_1 \psi_1^* \normord = - \psi^*_1 \psi_1$ etc. We will then define the normal ordering with respect to any vacuum of charge $l$ by a subscript using $\normord ( ... )\normord_l$. The ordering is most simple with respect to the completely bare vacuum with no fermi sea (that contains only holes) $| \infty \rangle$, so that one finds $\normord ( \psi^*_n \psi_m  )\normord_\infty = -  \psi_m \psi^*_n \, , \, \forall m,n $. We will call this \emph{complete normal ordering}.
\\
\\
With these definitions and expanding into modes, one finds the fermionic correlators to take determinantal forms
\bea\label{C4}
\langle l | \psi(\z_1)...\psi(\z_m) \psi^*(z_m)...\psi^*(z_1) | l \rangle &=& \prod_{k=1}^m z_k^l \z_k^{-l}  \frac{\prod_{i<i'} (\z_i - \z_i') \prod_{j<j'} (z_j - z_{j'})}{\prod_{i j}(\z_i - z_j)} \nn \\
&=& \prod_{k=1}^m z_k^l \z_k^{-l} \det_{i j} \frac{1}{\z_i - z_j}\, .
\eea
Bilinear expressions of the fermions generate an infinite dimensional Lie algebra $GL(\infty)$ and a generic element of the group will take the form
\be\label{C5}
{\bf G} = \exp \sum_{m, n \in \mathbb{Z}+\half} b_{m n} \psi_m \psi^*_n\, .
\ee
One can also define the $GL(\infty)$ element using a completely normal ordered operator
\be\label{C6}
\normord \, {\bf G} \, \normord_\infty = \normord \, \exp \sum_{m, n \in \mathbb{Z}+\half} B_{m n} \psi_m \psi^*_n \, \normord_\infty \, = 1 + 	B_{i k} \psi_i \psi_k^* + \frac{1}{2!} B_{i k}B_{i' k'} \psi_i \psi_{i'}\psi_{k'}^* \psi^*_{k} + ...
\ee
Using the fermionic algebra, one can prove the following relations that hold for the action of $GL(\infty)$ elements onto the fermionic modes (see~\cite{Alexandrov:2012tr} for details)
\bea\label{glcommute}
{\bf G} \psi_n = \sum_l R_{l n} \psi_l {\bf G} \, , \quad \psi^*_n {\bf G} = \sum_{l} R_{n l} {\bf G} \psi^*_l \, \nn \\
 \psi_n {\bf G} = \sum_l \tilde{R}_{l n} {\bf G} \psi_l  \, , \quad {\bf G} \psi^*_n  = \sum_{l} \tilde{R}_{n l}  \psi^*_l {\bf G} \, .
\eea
One also finds $\tilde{R} R = I$ and $R = e^{b}$ with $b_{m n}$ the matrix element in~\ref{C5}. On the other hand for the completely normal ordered elements $\normord \, {\bf G} \, \normord_\infty$ one finds similar commutation relations but for which $R_{l n} = \delta_{l n} + B_{l n}$. In other words we find
\be\label{GLrelationsordering}
\exp \left( \sum_{m, n \in \mathbb{Z}+\half} b_{m n} \psi_m \psi^*_n \right) = \normord \, \exp \left( \sum_{m, n \in \mathbb{Z}+\half} \left(e^b - I \right)_{m n} \psi_m \psi^*_n \right) \, \normord_\infty
\ee
\\
\\
We then define the current operators 
that generate the Toda time flows as
\be\label{C9}
J_n = \sum_{r \in \mathbb{Z}+\half} \psi^*_{r-n} \psi_r \, ,   
\ee
which obey the following
\be\label{C10}
[J_n\, , J_m] = n \delta_{n+m, 0}\, ,  \quad J_n | l \rangle = \langle l | J_{-n} = 0 \, , \quad n>0 \, .
\ee
A special case of these operators is the total charge operator
\be\label{C11}
Q = \sum_{r \in \mathbb{Z}+\half} \psi^*_{r} \psi_r
\ee
One can pass to a bosonic description via the bosonization formulas 
\be\label{C12}
\psi(z) = :e^{-\phi(z)}:\, \quad \psi^*(z)= :e^{\phi(z)}:\, \quad \partial \phi(z) = :\psi^*(z) \psi(z): 
\ee
The bosonic field has the modes ($[\hat Q \, , \hat P] = 1$)
\be\label{C13}
\phi(z) = \hat P + \hat Q \log z + \sum_{n\neq 0} \frac{1}{n} J_n z^{-n}
\ee
and the normal ordering is defined by putting all $J_n\, , n>0$ to the right and $J_n\, , n<0$ to the left as well as $:\hat P \hat Q: = :\hat Q \hat P: = P Q$. The bosonic vacuum is defined as
\be\label{vacuum}
\hat Q | s \rangle = s | s \rangle\, , \quad J_n| s \rangle = 0 \, , \quad (n>0)\, ,
\ee
so consistently with our definitions the operator $\hat Q$ measures the total charge of the state\footnote{Notice that in string theory notation one labels this operator with $\hat p$, since the charge then would be the momentum of the string.}. The operator $\hat P$ is a shift operator that transitions between vacua of different charge $e^{\pm \hat P} | l \rangle = |l \pm 1 \rangle$.



\subsection{Free fermions, currents and Young diagrams}\label{Frobeniusreps}

Due to~\ref{vacuum}, one can split the Hilbert space on which the free fermions act into the direct sum of Hilbert spaces of definite charge (here $s$ is a real number)
\be\label{C19}
\mathcal{H} = \bigoplus_s \mathcal{H}^s \, , \qquad \hat{Q} | s \rangle = s | s \rangle \, .
\ee
The full Hilbert space is constructed by the repeated action of the currents
\be\label{C20}
J_{-\lambda_1} ... J_{-\lambda_k} | s \rangle \, \equiv | \lbrace \lambda_1, \lambda_2 , ... \lambda_k \rbrace ; s \rangle \, .
\ee
The states $| \lbrace \lambda_1, \lambda_2 , ... \lambda_k \rbrace ; s \rangle \equiv | \lambda ; s \rangle $ are into one to one correspondence with a Young diagram with the $\lambda_k$'s labeling the rows. We also have $\sum_k \lambda_k = |\lambda|$, the total number of boxes.  The representation states are orthonormal in the sense
\be\label{C23}
\langle \lambda , n| \mu , m \rangle = \delta_{\lambda \mu} \delta_{n m}\, .
\ee

Another point of view is to split further the Hilbert space into
\be\label{C25}
\mathcal{H}^s = \bigoplus_{n\in \mathbb{Z} + \half} \mathcal{H}^s_{n}
\ee
using the fermionic modes. Now $s \in \mathbb{R}/\mathbb{Z}$. 
We can now use the free fermion operators defined in~\ref{Fermionic} to act
on the vacuum and create states corresponding to Young-diagrams/representations as
\be\label{repstates}
|\lambda ; s \rangle = \psi_{s - \beta_1 - \half } ... \psi_{s - \beta_{d(\lambda)} - \half } \psi^*_{s + \alpha_{d(\lambda)} + \half} ... \psi^*_{s + a_1 + \half}   |s \rangle \, .
\ee
In this formula we denote by $d(\lambda)$ the number of diagonal elements in the partition. The Frobenius notation for the partition, that was introduced in appendix~\ref{Partitionsreps}, is  $\lambda \equiv (\alpha_1,..,\alpha_d | \beta_1, ... \beta_d)$, where $\alpha_i = \lambda_i - i, \, \beta_i = \lambda_i' - i$. We also notice that $\psi^*$'s create the part of the Young diagram above the diagonal, while $\psi$'s the one below the diagonal. They can also be considered as particles/anti-particles (holes). 

Since the fermions have half integer excitation levels which are all defined with respect to the vacuum charge $s$, this motivates to use the compact notation
\be\label{C27}
|\lambda ; s \rangle = \prod_{i=1}^{d(\lambda)} {\psi}_{- p_i} {\psi^*}_{q_i} |  s \rangle
\ee
In this description $p_i, q_i$ are half integers corresponding to the fermionic pseudomomentum excitations $p_i = \beta_i + \half$, $q_i = \alpha_i + \half$, that are defined with respect to the vacuum  of charge s. 
These pseudomomenta also correspond to the excitations of the lower/upper fermi sea defined for the particle on a group action that could be used to label the different representations. 


\subsection{The $\tau$-function as a sum of partitions}\label{Tau}

Given a $GL(\infty)$ operator ${\bf G}$ one can express the $\tau$ function of the 2D Toda Hierarchy as a transition amplitude
\be\label{C15}
\tau_s(t_+, t_-) \, = \, \langle s | e^{J_+ [t]} {\bf G} e^{- J_-[t]} | s \rangle \, =  \langle t_+ ;  s |  {\bf G}  | t_- ;  s \rangle  \, 
\ee
with
\be\label{C16}
J_+[t] = \sum_{r>0} t_r J_r\, , \quad J_-[t] = \sum_{r<0} t_r J_r \, ,
\ee
and where the $t$'s are Miwa ``time" parameters. One could instead use for the definition of the $\tau$-function a completely normal ordered operator $\normord \, {\bf G} \, \normord_\infty $, the two descriptions being related via eqn.~\ref{GLrelationsordering}. As we mentioned before the parameter-s measures the charge of the vacuum but can in general be a complex parameter (for example when incorporating the chemical potential $\mu$ of the double scaled MQM). 

This form of the $\tau$-function hiddenly contains all the various representations in $GL(\infty)$. To show this, we can use the following relations in order to expand the current deformed states in terms of representations/partitions (for a proof see~\cite{Alexandrov:2012tr})
\bea\label{C17}
|t_- ;  s \rangle  = e^{J_-(t_-)} | s\rangle &=& \sum_\lambda (-1)^{b(\lambda)} s_\lambda(t_-) | \lambda ; s \rangle \, , \nn \\
\langle t_+ ; s | = \langle s | e^{J_+(t_+)}   &=& \sum_\lambda (-1)^{b(\lambda)} s_\lambda(t_+) \langle \lambda ; s | \, , \nn \\
b(\lambda) = \sum_{i=1}^{d(\lambda)} (\beta_i +1) \, .
\eea
For the definition and properties of states $|\lambda ; s \rangle$ that label an arbitrary representation/partition see section~\ref{Frobeniusreps} and in particular~\ref{repstates}.
Using these expansions, one finds that the $\tau$ function can be expressed equivalently as
\be\label{C18}
\tau_s(t_+ , t_- ) = \sum_{\lambda , \kappa} {\bf G}_{\lambda \kappa} s_\lambda(t_+) s_\kappa(- t_-)\, , \qquad  {\bf G}_{\lambda {\kappa}}^{(s)} = (-1)^{b(\lambda)} (-1)^{b(\kappa)} \langle \lambda, s | {\bf G} | \kappa , s \rangle \, .
\ee
The elements $ {\bf G}_{\lambda \kappa}$ can be thought of as transition amplitudes between different representations where the dynamics is encapsulated in the specific form of the $GL(\infty)$ element ${\bf G}$. The notation is such that the first index of $ {\bf G}_{\lambda \kappa}$ corresponds to the ket and the second to the bra state used to form the inner product. Some comments on this form of the $\tau$ function are in order: 

\begin{itemize}
\item  The transition amplitude/partition function splits into a combinatorial sum of Schur polynomials with the objects  ${\bf G}_{\lambda \kappa}$ that are transition amplitudes between fixed representations. From a mathematical perspective the objects ${\bf G}_{\lambda \kappa}$ correspond to Pl\"ucker coordinates on the infinite dimensional Sato Grassmannian. The Schur polynomials define the in/out states to be scattered. In particular they can be thought of as ``wavefunctions in the space of Miwa times".
\be
s_\lambda(t_+) = \langle t_+ | \lambda \rangle \, .
\ee

\item For MQM the $GL(\infty)$ element is explicitly diagonal ${\bf G}_{\lambda \kappa} = \delta_{\kappa \lambda} {\bf G}_{\lambda \lambda} $ (either in the winding mode or the scattering basis). This is to be expected since the MQM dynamics is diagonal in the representation basis and does not mix different representations. This nevertheless does not mean that the representations themselves are self conjugate, since each $\lambda$ is generic, or in other words the $p_i$ and $q_i$ of the $|\lambda \rangle$ state in Frobenius notation~\ref{C27} can be different.

\end{itemize}

\subsubsection{Schur polynomials}\label{Schur}

There exist various useful formulae for the Schur polynomials, see~\cite{Macdonald} for many examples. We follow and extend appendix C. of~\cite{Betzios:2017yms}. 

Schur's polynomials $s_\l(X)$ are symmetric polynomials in $N$ variables that form a linear basis for the space of all symmetric polynomials. Seen from a representation theory point of view they are characters for the irreducible representations of the general linear groups.

To be more concrete the representations $R$ of $GL(\infty)$ are labeled by Young diagrams, or ordered integer partitions $R \equiv \lambda \, : \l_1 \geq \l_2 ....\geq 0$ with $|\lambda|$ the number of boxes in the diagram. We will thus use the representation index $R$ or the partition index $\l$ interchangeably in this case.
For example can represent the characters either through the use of the eigenvalues $x_i$ of the matrix $X$, or through auxiliary Miwa (``time") variables of a matrix $X$ as $\chi_R(X) \equiv s_\lambda(X)$ or $\chi_R(t) \equiv s_\lambda(t)$. The definition of Miwa variables is $t_k = \tr X^k / k = \sum_i x_i^k/k$ with $x_i$ the eigenvalues of the matrix $X$.
We then define the characters using Weyl's formula (the denominator is the Vandermonde determinant)
\be
\chi_R (X) \equiv s_\lambda(x_1,x_2,...,x_N) = \frac{\det x_i^{\lambda_j + N -j}}{\det x_i^{N - j}}
\ee
We define the generating functional of time variables via
\be
\xi(t, z) = \sum_{k \geq 1} t_k z^k \, , \quad e^{\xi(t,z)} = \sum_{k \geq 0} h_k(t) z^k \, ,
\ee
which contains the complete symmetric functions $h_k(x_1,x_2,...)$ in a time variable formalism. The Schur polynomials are then expressed in terms of time variables as (Jacobi-Trudi formula)
\be\label{JTformula}
s_\lambda(t) = \det_{i,j = 1, ... \ell(\lambda)} h_{\lambda_i - i + j}(t) \, .
\ee
There is also a Frobenius version of the Schur polynomials (also called Giambelli's formula)
\be
s_\lambda(t) = \det_{i, j = 1, ... d(\lambda)} s_{(\a_i | \b_j)} (t) \,  \quad s_{(\a_i | \b_j)} (t) = (-1)^{\b_j} \sum_{m=0}^{\b_j} h_{\b_j-m}(-t) h_{\a_i + m + 1}(t) \, .
\ee
Finally there exists a Tableaux representation of them
\be
s_\lambda(x_1,x_2,...,x_N) = \sum_{\text{T}} x_1^{\mu_1} x_2^{\mu_2} ... x_N^{\mu_N} \, .
\ee
In this expression, the collection $\mu_1,\mu_2,...,\mu_N$ is the collection that gives the weight of the Tableaux, or in other words the number $1$ appears $\mu_1$ times in the Tableaux and so on. The sum is over all semistandard Young Tableaux $T$ of shape $\lambda$ and weight $\mu = (\mu_1,\mu_2,...,\mu_N)$. Depending on the application some of these formulae can be more useful.

Some trivial examples of Schur polynomials are the dimensions of $GL(N)$ ($\lambda$ is the associated partition)
\be
D_R = \dim_R \left( GL(N) \right) = \chi_R \left(t_k = \frac{N}{k} \right) = \chi_R (I) = s_\lambda \left((1)^N \right) =\prod_{j<k}^{N=|\lambda|} \frac{\lambda_j - \lambda_k + k - j}{k-j} \, ,
\ee
and the dimensions of $S_N$ 
\bea
d_R \equiv d_\lambda &=& \frac{1}{N!} \dim_R (S_N) = \chi_R (t_k = \delta_{k 1}) = \prod_{j<k}^{\infty} \frac{\lambda_j - \lambda_k + k - j}{k-j}   \,  \nn \\
&=& \prod_{k=1}^N \frac{(N-k)!}{(N + \lambda_k - k )!} \prod_{ 0< i < j}^N  \frac{\lambda_j - \lambda_k + k - j}{k-j}  \, , \quad N = |\lambda|
\eea

Let us finally list two important specialisations of Schur polynomials that are discussed in the main text and in particular in section~\ref{Measures}:

In the model of~\cite{Kazakov:2000pm} we simply have the first times turned on, so that
\be
s_\lambda(t_1) = \frac{t_1^{|\lambda|}}{H_\lambda} = \frac{t_1^{|\lambda|} \dim \lambda }{|\lambda| !}
\ee
while in the model of~\cite{Betzios:2017yms} we use the expression ($x=e^{-\beta m}$)
\be
s_\lambda(t_k = N_f x^k/k) = \frac{ x^{|\lambda|} (N_f)_\lambda}{H_\lambda} = \frac{ x^{|\lambda|} (N_f)_\lambda  \dim \lambda}{|\lambda| !} \, ,
\ee
with $H_\lambda = \prod_{x \in \lambda} H(x)$ the product over hook lengths and $(u)_\lambda$ the product over the content of the boxes (product of Pochhammers)
\be
(u)_\lambda = (u)_{\lambda_1} (u-1)_{\lambda_2} ... (u-\ell(\lambda) +1)_{\lambda_\ell} \, . 
\ee
The product over Hook lengths can also be written using the shifted highest weights $h_i = \lambda_i - i + \ell(\lambda) $ as
\be
H_\lambda = \frac{\prod_{i=1}^\ell  h_i !}{\prod_{i < j} (h_i - h_j)}  = \frac{\prod_{i=1}^\ell \Gamma(h_i + 1) }{\Delta(h_i)}\, .
\ee

\subsection{$\tau$-function from the reflection amplitude}\label{reflctau}

An S-matrix element can serve as a $GL(\infty)$ element ${\bf G}$ rotating the basis elements. In this case the matrix element $b$ in~\eqref{C5} has the interpretation of a scattering phase $b = i \Phi$. Generically a scattering spectrum is continuous and hence the free fermions in~\eqref{C1} and~\eqref{C5} should have a continuous index.
Nevertheless, we shall consider the case of compactified Euclidean time on $S^1$.
In the particular example of bosonic MQM at finite temperature the specific reflection amplitude is provided by eqn. \eqref{reflnp} with the properties 
\be\label{C30}
\mathcal{R}_{n m}= \delta_{n m} \mathcal{R}\left(E = i \frac{n}{R} + \mu \right)\, , \qquad \mathcal{R}_{n}^{-1} =  \mathcal{R}_{-n}^{*}  \, .
\ee
Since it is diagonal in the discrete energy basis, this means that we can simply define the $GL(\infty)$ element ${\bf G}$  as
\be\label{reflectionGLelement}
{\bf G} = \exp \left( \sum_{n \in \mathbb{Z} + 1/2} \log \mathcal{R}_{n} \, {\psi}_n  \,{\psi^*}_n \right) \, , \quad  \log \mathcal{R}_{n} = i \Phi_n \, .
\ee
Moreover we work directly using the double scaled model (grand-canonical ensemble), so there is no parameter $N$ (of the original MQM), only its dual chemical potential $\mu$. This chemical potential dictates the location of the fermi sea and the reflection  amplitude depends on it, see eqn. \eqref{reflnp} for the explicit expression.

The fermionic fields in the scattering case can be expressed as
\be\label{C31}
\psi (z) = \sum_{n \in \mathbb{Z}+\half} \psi_n z^{-n-\half } \, , \quad {\psi^*}(z) = \sum_{n \in \mathbb{Z}+\half} {\psi^*}_{-n} z^{-n-\half }\, , \quad z = e^{\rho - i \theta} \, .
\ee
In this realisation they are propagating modes in a Euclidean space of coordinates $\rho, \theta$ where $\theta$ is an angular coordinate parametrising the thermal circle. This sets the fermionic charge to be equivalent to a Euclidean energy (we have mapped the $S^1$ modes to the compact $U(1)$ modes). The coordinate $\rho \in [0,\infty)$.

The reflection amplitude encapsulated in the $GL(\infty)$ element \eqref{reflectionGLelement} can be equivalently be thought of as arising from a boundary state $| B \rangle$ at $\rho = 0$ that reflects the fermions. In particular the boundary state is determined from the reflection amplitude for particles and holes\footnote{A similar expression can be found in~\cite{Maldanotes}. The factor of $i$ is related to the fact that in Euclidean signature we rotate the boundary by $\pi/2$, but will not play any role in the following, since it is an overall constant phase.}
\bea\label{boundarystate}
\langle B | \left( \psi_n^{in} - i \mathcal{R}_{n}^{-1} \psi_n^{out}  \right) = 0 \, , \nn \\
\langle B | \left( \psi_n^{in \, *} - i \mathcal{R}_{n}^{\dagger} \psi_n^{out \, *}  \right) = 0 \, , \nn \\
\langle B | \, \sim \, \langle 0 |  \exp \left(i \sum_{n \in \mathbb{Z}+\half} \, \psi_n^{out} \,  \psi^{in \, *}_n \, \mathcal{R}_{n}^{-1} \, +  \,  \psi_n^{out \, *} \psi_n^{in} \,  \, \mathcal{R}_{n}^{\dagger} \, \right) \, = \, \langle 0 | \normord \, {\bf B} \, \normord_\infty \, .
\eea
Any amplitude is then computed as an expectation value using states that belong in the doubled Hilbert space $\mathcal{H}_{in} \otimes \mathcal{H}_{out}$
\be
\langle B | \left(|\Psi_{in } \rangle \otimes |\Psi_{out } \rangle  \right) \, .
\ee 
The advantage of this second description is that it dispenses with the need to define a Euclidean reflection amplitude, since all the processes are just overlaps (form factors) with the boundary state. See also for a recent discussion of this formalism~\cite{Jiang:2019xdz} in the context of $\mathcal{N}=4$ SYM. 

Using an explicit expression for the $GL(\infty)$ element, either eqn. \eqref{reflectionGLelement}, or the completely normal ordered boundary state eqn. \eqref{boundarystate} we can prove the following:

\begin{itemize}

\item Moving the $GL(\infty)$ elements past the fermions using~\ref{glcommute} we can obtain the T-dual of the partition function for a generic representation eigenstate $|\lambda \rangle \equiv |(\vec{q} \, | \, \vec{p} \, ) \rangle \, , q_i = \lambda_i - i + \half \, , p_i = \lambda'_i - i + \half $, using the definitions \eqref{C27} and \eqref{C18}
\be\label{PartitionRep}
{\bf G}_{\lambda \lambda}^{(s)}(\mu, R) =  \langle s | {\bf G} |  s \rangle \prod_{i=1}^{d(\lambda)} \mathcal{R}_{s - p_i} \prod_{j=1}^{d(\lambda)} \mathcal{R}^{-1}_{s + q_j}   \,  = \,  Z_{singlet}^{(s)}(\mu, R) \prod_{i=1}^{d(\lambda)} \mathcal{R}_{s - p_i} \prod_{j=1}^{d(\lambda)} \mathcal{R}^{-1}_{s + q_j} \, , 
\ee
where the singlet piece is
\be\label{C36}
 \langle s | {\bf G} |  s \rangle =  Z^{(s)}_{singlet}(\mu, R) = \prod_{n \in \mathbb{Z}+\half \, , n>s}  \mathcal{R} \left(E = i \frac{n}{R} + \mu \right)  = \prod_{m \geq 0}  \mathcal{R}_{m + \half +s} \, .
\ee
The first important thing to notice is that by construction of the $GL(\infty)$ element
the only non trivial expectation values are the ones for which the in/out representations are the same. This is also consistent with the fact that MQM is diagonal in the representation basis. The condition $n>s$ comes from the fact that the generic vacuum with $U(1)$ charge $| s \rangle$ is built out of the no particle vacuum $|0 \rangle$, with the action of particle modes that raise the value of the total charge/energy (so that there are less states available to excite for the particles). We then find that in the Frobenius formalism we can construct and scatter in any representation by removing particles and adding holes with respect to the singlet result for which the original vacuum is defined. For example the adjoint has $p_1 = q_1 = \half$, so that we have removed the lowest energy particle and added a hole.

\item The T-dual of the partition function in the Jacobi-Trudi coordinates (the original highest weight labels of the Young diagram) for which the representation eigenstate is created by
\be
|\lambda ; s \rangle = \psi^*_{\lambda_1 - \half + s} \, ... \, \psi^*_{\lambda_{\ell} - \ell(\lambda) + \half + s} | s - \ell(\lambda) \rangle \, ,
\ee
takes the form
\be\label{PartitionRep2}
{\bf G}_{\lambda \lambda}^{(s)}(\mu, R) \, = \, \langle s - \ell(\lambda) | {\bf G} |  s - \ell(\lambda) \rangle \,  \prod_{i=1}^{\ell(\lambda)} \mathcal{R}^{-1}_{\lambda_i - i  + \half + s}  \, .
\ee
By rearranging the products one can check that \eqref{PartitionRep2} can be exactly mapped to eqn. \eqref{PartitionRep}, so the two equations merely represent the same partition function in a slightly different fashion. It is also convenient to define the (strictly decreasing) shifted highest weights 
\be
h_i = \lambda_i - i + \ell(\lambda)  \, , \quad h_{i+1} < h_i \, ,
\ee
with respect to which the T-dual partition function in an arbitrary representation (of length $\ell$) is written
\be\label{PartitionRep3}
{\bf G}_{\lambda \lambda}^{(s)}(\mu, R) \, = \, \langle s - \ell(\lambda) | {\bf G} |  s - \ell(\lambda) \rangle \,  \prod_{i=1}^{\ell(\lambda)} \mathcal{R}^{-1}_{h_i  + \half - \ell(\lambda)  + s }  \, .
\ee
In this parametrisation both factors involve a product of reflection amplitudes computed in the same $s-\ell$ charge vacuum. These variables are the natural variables that describe a rotated Young diagram by $\pi/4$, see fig. \eqref{fig:grid}.

\end{itemize}

\subsection{$\tau$-function from the gauge field zero modes}\label{windtau}

We could also define an appropriate $GL(\infty)$ element (now for the direct partition function of MQM) using winding modes. The free fermions $\psi(z)$ now depend on the gauge field zero modes/angles $z = e^{i \theta} \, , \theta \in [0, 2 \pi)$. We first define a vertex operator 
\be\label{C39}
V_q(z) =\psi(q^{-\half} z) \psi^*(q^{\half} z)\, \, , \quad q= e^{- \omega \beta} .
\ee
Using this, we define a specific $GL(\infty)$ operator in a completely normal ordered form (see Appendix~\ref{Fermionic} and eqn. \eqref{GLrelationsordering} for the transition between normal and non-normal ordered operators) 
\be
\label{gloperator}
\normord \, {\bf G} \, \normord_\infty =  \normord \, \exp \left( e^{\beta \mu} \oint \frac{dz}{2 \pi i}  V_q(z) \right) \,  \normord_\infty  \, .
\ee
This can also be expressed in terms of fermionic modes as
\be \label{C41}
\normord \, {\bf G}^{(\hat Q)} \, \normord_\infty  = \normord \,  \exp \left( e^{\beta \mu} \sum_{n \in \mathbb{Z} + 1/2} q^{n}  \, {\psi}_n  \,{\psi^*}_n \right) \, \normord_\infty
\ee
with $\omega = 1 \, , \omega = i$ for the normal/inverted oscillator respectively.
\\
\\
Expanding the $GL(\infty)$ operator~\ref{gloperator} in a series and commuting the fermions past the currents $J_\pm$ using the definitions eqn.~\ref{C15} and \eqref{C9}, the $\tau$ function becomes (up to an overall normalisation)
\bea\label{Taufunctionwinding}
\tau_s[t]&=&  \sum_{N=0}^\infty \frac{ e^{\beta \mu N}}{N!} \oint  \prod_{n=1}^N \frac{d z_n \, }{2 \pi i} e^{\sum_{m \neq 0} (q^{-m/2} - q^{m/2}) t_m z_n^m}\, \times \,  \nn \\
&\times& \langle s | \normord \prod_{n=1}^N \psi(q^{\half} z_n) \psi^*(q^{-\half} z_n) \normord_\infty \, | s \rangle \nn \\
&=&  \sum_{N=0}^\infty \frac{q^{s N} e^{\beta \mu N}}{N!} \oint  \prod_{n=1}^N \frac{d z_n }{2 \pi i} e^{\sum_{m \neq 0} \tilde{t}_m z_n^m} \det_{i j} \frac{1}{q^{\half} z_i - q^{-\half} z_j} =  \mathcal{Z}_G (\tilde{t}, \mu+i s ) \, ,  \nn \\
\eea
where in the second equality we used the fermionic correlator for normal ordered insertions~\ref{C4}. This is the explicit integral in terms of matrix model gauge field zero modes that has appeared in~~\cite{Kazakov:2000pm} and~~\cite{Betzios:2017yms}.
This expression allows to determine the rescaling between the ``physical Toda time parameters"  $t_m = \frac{\tilde{t}_m}{q^{-m/2}- q^{m/2}}\,$ in terms of the apparent $\tilde{t}$ that appear in the integral expression for the partition function.  The chemical potential is related to a shift to the charge $s$, that is real for the usual oscillator but imaginary for the case of the inverted oscillator ($q= e^{i \beta}$). It is also possible to prove that the Chern-Simons level $k$ can be identified with the real charge $s$ of the $\tau$ function and therefore allows to define the complex string coupling $g_{str}^{-1} = \mu + i k$~\cite{Betzios:2017yms}.

If we expand the partition function in representations labelled by Young diagrams, we find again that since the $GL(\infty)$ element is diagonal and bilinear in the fermions, the only non-trivial elements are diagonal in the representation basis
\be\label{C43}
 {\bf G}^{(s)}_{\lambda \lambda}(\mu) =  \langle \lambda, s | \normord  \, {\bf G} \normord_\infty \, | \lambda , s \rangle \, =  \langle s | \normord {\bf G} \normord_{\infty} \,	 |  s \rangle \prod_{i=1}^{d(\lambda)} R_{s - p_i} \prod_{j=1}^{d(\lambda)} R^{-1}_{s + q_j}   \,   
\ee
where we chose again to adopt a Frobenius notation for the representations using the strictly ordered $p_i, q_i$'s. The matrices $R_{i j}$ are now diagonal and found using eqn.~\ref{GLrelationsordering} (and $R = I + B$)
\be\label{C44}
R_{m n} = \delta_{m n} + e^{\beta \mu} q^{n} \delta_{m n} \, , \quad R_n = 1 +  e^{\beta \mu} q^{n} \, . 
\ee
In particular the singlet partition function is expressed as
\be\label{singletgauge}
\langle s | \normord {\bf G} \normord_{\infty} \,	 |  s \rangle = \prod_{n \in \mathbb{Z} + 1/2,\, n > s} R_n =  \prod_{n \in \mathbb{Z} + 1/2,\, n > s} \left(1 + e^{\beta (\mu + i n)} \right) \, ,
\ee
that is the grand-canonical partition function of free fermions in the harmonic/inverted harmonic potential. We observe that the chemical potential indeed defines the fermi vacuum since $n > s = - i \mu$ (remember that excitation energies are imaginary for the inverted oscillator). Once more the generic representation is built by removing particles and adding holes to the singlet result.

In the next  appendix we shall analyse further the expressions of MQM in a fixed representation eqns. \eqref{C43} and \eqref{PartitionRep} and show the precise T-duality relation between them.

\section{MQM in a fixed representation}\label{MQMnonsingletsapp}

\subsection{Free energy in a fixed representation}

In the previous appendix we found that the $\tau$ function can be written as a combinatorial sum involving Schur-polynomials/characters and the fixed representation partition function ${\bf G}_{\lambda \lambda}(\mu, R)$, see eqn.~\ref{C34}. In this section we consider an arbitrary but fixed representation and analyse the properties of the partition function and free energy in the grand canonical ensemble with fixed chemical potential $\mu$. We want also to analyse the effects of T-duality, expressed as~\footnote{Notice that we work in units where $\alpha' = 1$, which can be reinstated if we wish to keep track of the dimensions of various parameters.}
\be\label{aa6}
R \rightarrow 1/R \, , \qquad \mu \rightarrow \mu R \, .
\ee

\subsubsection{From the reflection amplitude}\label{fixedrepreflect}

Let us first consider the free energy in a fixed representation computed using the reflection amplitude formalism. 
Since the T-duality properties of the partition function in a fixed representation are not a priori evident\footnote{For example there could exist non-trivial leg pole factors that need to be introduced when T-dualising the result.}, we will therefore also compute the free energy using the expression for the gauge field zero modes in the next section in order to contrast and compare the two approaches.

Let us first define the scattering phase $\Phi(E)$ via the perturbative scattering amplitude in the inverted oscillator potential\footnote{Notice that for consistency with the rest of the sections and appendices, we followed conventions similar to~\cite{Kazakov:2000pm,Boulatov:1991xz,Alexandrov:2002fh}, that are different compared to some other references, such as~\cite{Moore:1991zv} (for example their reflection amplitudes are mutual inverses). One can change conventions but it should be done consistently in all the various formulae.} 
\be\label{bosrefl}
\mathcal{R}^{-1}(E) = e^{- i \Phi(E)} = \sqrt{\frac{\Gamma(1/2 - i E)}{\Gamma(1/2 + i E)}}  \, , \qquad \, E = i \frac{n}{R} + \mu  ,
\ee
where in the second equation we denote the discrete modes that are relevant in the case of a compactified Euclidean time, when the free fermions in the inverted oscillator are at chemical potential $\mu$. 

The non-perturbative (in  $1/\mu$) bosonic reflection amplitude was derived in~\cite{Moore:1991zv}
\bea\label{reflnp}
\mathcal{R}^{-1} \left(E = i \frac{n}{R} + \mu \right) =   \sqrt{\frac{1+ i e^{- \pi(\mu + i n/R)}}{1- i e^{-\pi(\mu + i n/R)}}} \sqrt{\frac{\Gamma\left(\half - i \mu  + \frac{n}{R} \right)}{\Gamma\left(\half + i \mu - \frac{n}{R} \right)}} \, \nn \\
= \sqrt{\frac{2}{\pi}} \cos \left(\frac{\pi}{2}\left(\half + i \mu  - \frac{n}{R} \right) \right) \Gamma\left(\half - i \mu  + \frac{n}{R} \right)  \, .
\eea
We observe that up to non-perturbative effects in $1/\mu$ it can be simplified into either (this is the form of the scattering phase that is obtained using the large distance WKB asymptotics of the parabolic cylinder functions neglecting tunneling)
\be\label{reflpert}
\mathcal{R}^{-1} \left(E = i \frac{n}{R} + \mu \right) =    \sqrt{\frac{\Gamma\left(\half - i \mu  + \frac{n}{R} \right)}{\Gamma\left(\half + i \mu - \frac{n}{R} \right)}} \, ,
\ee
or
\be
\mathcal{R}^{-1} \left(E = i \frac{n}{R} + \mu \right) =  e^{i \pi n/2 R} \frac{\Gamma\left(\half - i \mu  + \frac{n}{R} \right)}{\Gamma\left(\half - i \mu  \right)} \, .
\ee
In this work we shall not delve deeper into a study of non-perturbative effects and we can hence use these expressions interchangeably. We can also fix a notion of positive energy for the reflection amplitude by arguing as follows: The Gamma functions have poles in the negative axis. This means that the numerator in the reflection amplitude eqn. \eqref{bosrefl} correctly contains poles at negative imaginary energies which are the analogues of QNM's/decaying states. 

Using the scattering phase, the density of states is defined as in~\cite{Boulatov:1991xz} (we neglect any cutoff-dependence $C \sim \log \Lambda /2\pi$)
\be\label{DOS}
\rho(E) = - \frac{1}{2 \pi} \frac{d \Phi(E)}{d E} \, , \qquad \rho_{singlet}(E) = - \frac{1}{2 \pi} \Re \psi \left(\half + i E \right) = - \frac{1}{2 \pi} \sum_{k=0}^\infty \frac{k + \half}{\left(k +\half \right)^2 + E^2} + C \, .
\ee
We can then express the singlet free energy using the scattering phase and the density of states as follows (the singlet vacuum does not carry any $U(1)$ charge $s$ and $\beta = 2 \pi R$)
\bea\label{C48}
\mathcal{F}_{singlet} &=&  i \sum_{n \in \mathbb{Z} +\half , \, n > 0} \Phi \left(i \frac{n}{R}  + \mu  \right)  \nn \\
 &=& - \frac{\beta}{2 \pi} \int_{- \infty}^\infty d \epsilon \frac{\Phi(\epsilon)}{1 + e^{ \beta ( \epsilon - \mu)}} = \int_{- \infty}^\infty d \epsilon \rho_{singlet}(\epsilon) \log(1 + e^{\beta (\mu - \epsilon)})
\eea
After performing T-duality this expression becomes
\bea\label{C49}
\mathcal{F}_{singlet} &=&  i \sum_{n \in \mathbb{Z} +\half , \, n > 0} \Phi \left(i n R + \mu R \right)  \nn \\
 &=& - \frac{1}{R} \int_{- \infty}^\infty d \epsilon \frac{\Phi(\epsilon)}{1 + e^{ 2 \pi (-\mu + \epsilon /R)}} = \int_{- \infty}^\infty d \epsilon \rho_{singlet}(\epsilon) \log(1 + e^{ 2 \pi (\mu - \epsilon /R)})
\eea
The singlet free energy is known to be T-duality invariant~\cite{Klebanov:1991qa} so these last expressions are in fact equivalent. In these two expressions we can pick the poles in the energy integral by closing the contour in the upper half plane, where the integrand is convergent. It is convenient to use the integral expression for the Digamma function in eqn. \eqref{DOS} (dropping an irrelevant $\mu$ independent infinite term) to obtain (assuming $\Im \e > 0$)
\bea\label{intexpressions}
\rho_{singlet}(\epsilon) = \frac{1}{2 \pi}  \Re \, \int_0^\infty \frac{d t}{ 2  \sinh(t/2)} e^{ i \epsilon t} \, , \quad \Phi(\epsilon) =  \Im \, \int_0^\infty \frac{ d t}{ 2  t \sinh(t/2)} e^{ i \epsilon t} \, , \nn \\
\mathcal{F}_{singlet} =  \Im  \int_0^\infty \frac{- i dt}{4 t \sinh (t/2)} \frac{e^{ i \mu R t}}{\sinh(R t/2)} \, .
\eea
This is the form of the singlet free energy used in the literature with manifest T-duality properties. This last expression is most convenient for performing a perturbative expansion in genera that reads (this is prior to a Legendre like transform that changes the genus zero sign)
\be\label{C50}
\mathcal{F}_{singlet} = - \frac{R}{2} \mu^2 \log \mu - \frac{1}{24}\left(R + \frac{1}{R} \right) \log \mu + ...  \, .
\ee
We then observe that the genus-0 entropy vanishes (while for higher genera there is entropy)
\be\label{C51}
S_{singlet} = -R \frac{ \partial}{\partial R} \mathcal{F}_s + \mathcal{F}_s = - \frac{1}{48} \frac{1}{R} \log \mu + ... 
\ee
This entropy should be associated to the presence of quantum (one-loop and higher) perturbative modes on the linear-dilaton background. Therefore the singlet states have no-classical entropy, as expected. 

Let us now move on to the adjoint representation excitation on top of the vacuum. Using $p_1 = q_1 = \half$ in~\ref{PartitionRep} (moving a single box from the bottom of the singlet so that the first row contains now two boxes), after T-dualising we find
\be\label{PFadjoint}
\mathcal{F}_{adj} - \mathcal{F}_{singlet} =  i \left[ \Phi\left(-i \frac{R}{2} + \mu R  \right) - \Phi\left(i \frac{R}{2} + \mu R  \right) \right] \, ,
\ee
which using the explicit bosonic (perturbative) reflection amplitude \eqref{reflpert}, reads
\be\label{C52}
\mathcal{F}_{adj} - \mathcal{F}_{singlet} = \half \left( \log \frac{\Gamma(\half + \frac{R}{2} + i \mu R)}{\Gamma(\half - \frac{R}{2} - i \mu R)} + \mu \rightarrow - \mu \right)
\ee
This expression is invariant under $\mu \rightarrow - \mu$ (which holds for self-conjugate reps and is expected in the presence of only closed string states more generically). It is easy to see from these equations that in order to form the adjoint, one has to remove the lowest mode of the singlet spectrum and to add the dual mode of the hole (or equivalently lifting a particle from the fermi sea to fill the first excited state). Due to this fact, the difference in the free energies can be expressed in terms of a density of states having a single pole 
\be\label{C53}
\mathcal{F}_{adj} - \mathcal{F}_{singlet} =  - \frac{1}{2 \pi} \int_{- \infty}^\infty d \epsilon  \frac{\half}{\frac{1}{4} + \epsilon^2} \left( \log(1 + e^{ 2 \pi (\mu - \epsilon/R)}) +  \log(1 + e^{ - 2 \pi (\mu + \epsilon/R)}) \right)
\ee
This matches the expression given by Boulatov and Kazakov in~\cite{Boulatov:1991xz}\footnote{In this reference a factor of $2$ seems to be wrong in subsequent calculations.}.

For an arbitrary representation indexed by the strictly ordered positive half integers $p_i, q_j$ (Frobenius coordinates, see~\ref{Partitionsreps} and~\ref{Frobeniusreps}), we find the general expression
\be\label{Freeenergygeneralrep}
\mathcal{F}(\lbrace p_i, \, q_j \rbrace) - \mathcal{F}_{singlet} =  i \sum_{i,j =1}^d   \left( \Phi\left(-i p_i R + \mu R  \right) - \Phi\left(i q_j R + \mu R  \right) \right) \, .
\ee
Using the integral expression for the perturbative scattering phase in eqn. \eqref{intexpressions}, we obtain the following result for the difference in free energies
$\Delta \mathcal{F} = \mathcal{F}(\lbrace p_i, \, q_j \rbrace) - \mathcal{F}_{singlet} $
\be\label{generalrepdualexpansion2}
\Delta \mathcal{F}    =  \sum_{i,j =1}^d   \Im \, \int_0^\infty \frac{ i d t}{ 2  t \sinh(t/2)} \left( e^{ - p_i R t -  i \mu R t} + e^{ - q_j R t +  i \mu R t} \right)
\ee
An equivalent result is obtained by using a density of states for isolated resonances 
\be\label{isolresonfree}
\Delta \mathcal{F}  =  - \frac{1}{2 \pi} \int_{- \infty}^\infty d \epsilon \sum_{i,j =1}^d   \left( \frac{q_j}{q_j^2 + \epsilon^2} \log(1 + e^{ 2 \pi (\mu - \epsilon /R)}) +  \frac{p_i}{p_i^2 + \epsilon^2}    \log(1 + e^{- 2 \pi (\mu + \epsilon/R)}) \right) \, .
\ee
The equivalence of eqns. \eqref{generalrepdualexpansion2} and \eqref{isolresonfree} can be proven using the integral representation for the resonances
\be
\frac{z}{z^2 +\epsilon^2} = \Re \int_0^\infty d x e^{i x \epsilon - z x} \, ,
\ee
to get for the difference in the free energies 
\be\label{Freeenergygeneralrep2}
\frac{\partial \Delta \mathcal{F}}{\partial \mu} = -  \frac{R}{2} \Im \, \sum_{i,j =1}^d    \int_0^\infty  \frac{ d x }{\sinh \frac{R x}{2}} \left(e^{- p_i x - i \mu R x} - e^{-q_j x + i \mu R x} \right) \, .
\ee
If we integrate this we obtain \eqref{generalrepdualexpansion2}.

From eqn. \eqref{generalrepdualexpansion2}, we observe that for self-conjugate representations ($q_i = p_i$) the expression is real and the odd in $\mu$ terms vanish.
For reps with a finite number of $p_i, q_j$ the leading term in the $1/\mu$ expansion is
\be
\mathcal{F}(\lbrace p_i, \, q_j \rbrace) - \mathcal{F}_{singlet} =  R \log \mu \sum_{i,j =1}^d    (p_i + q_j)  \, ,
\ee
that exhibits the well known logarithmic gap between the singlet and higher representations~\cite{Gross:1990md}.
We conclude again that there is no classical entropy for a fixed finite size representation (fixed total number of boxes). Nevertheless it is still interesting to study the limit of infinite size reps, since also for the singlet the infinite summation over the modes did provide a genus zero contribution to the free energy (but not to the entropy).

In the main text we shall find that in explicit models, a genus zero entropy arises due to appropriate weights in the space of representations. These weights naturally come from the Schur polynomials (Schur measure) in the expression for the grand free energy (see the discussion around eqn. \eqref{C18}), that also need to be taken into account. The microscopic dynamical Spin-Calogero type of models~\cite{Betzios:2017yms} and the simpler model of~\cite{Kazakov:2000pm}, both contain such weights. The properties of the Schur measure are discussed in section~\ref{Measures} of the main text.

\subsubsection{From the gauge field zero modes}

One can also compute the free energy in a fixed representation using the gauge field zero modes and fermions that depend on them. This takes into account the representation~\ref{gloperator} of the $GL(\infty)$ element. Proceeding as in the previous section we can again compute the singlet grand canonical partition function~\ref{singletgauge}
\be
\log \left( \langle 0 | \normord {\bf G} \normord_\infty \, |  0 \rangle \right) = \log Z_{singlet}(\mu) =  \sum_{n \in \mathbb{Z} + 1/2,\, n > 0}  \log(1 + e^{\beta \mu} q^n ) \, .
\ee
Repeating the same steps we observe again that higher representations will have different particle/hole resonances missing/added so that for a rep labelled by the ordered half-integers $p_i, q_j$ we find
\bea
\Delta \mathcal{F} &=&  \sum_{i,j =1}^d   \left(   \log(1 + e^{ \beta \mu } q^{- p_i}) - \log(1 + e^{\beta \mu  } q^{q_j}) \right) \, , \nn \\
&=& -  \frac{1}{2 \pi} \int_{- \infty}^\infty d \epsilon \sum_{\lbrace p_i, \, q_j \rbrace} \left( \frac{q_j}{q_j^2 + \epsilon^2}    \log(1 + e^{ \beta (\mu + \epsilon )}) +   \frac{p_i}{p_i^2 + \epsilon^2} \log(1 + e^{- \beta (\mu - \epsilon)}) \right) \, , \nn \\
\eea
with $q = e^{i \beta}$ for the inverted oscillator. The result is again to remove a number of particles from the vacuum and add instead a number of holes. The asymptotic expansion is the T-dual of the one coming from the reflection amplitude. 

From these expressions we finally conclude that even though the free energy in a fixed representation is not T-duality invariant, we can either use the expression from the reflection amplitude or the winding modes to perform any computation we like, the two results being directly related by the action of T-duality (without any need to introduce additional leg pole factors to relate the two expressions for the free energy).

\subsection{Limit of continuous representations}\label{ContRep}

Since we found that representations/partitions with a finite number of boxes do not have any classical genus zero free energy or entropy contribution associated with them, we now turn on to the study of the limit of large representations for the grand canonical ensemble. As we observed, even the grand canonical singlet partition function can be thought of as containing an infinite number of boxes associated with all the negative energy harmonic oscillator resonances (that are filled to form the vacuum). The infinite summation over these fermionic excitations contributes to a genus zero classical free energy (but zero classical entropy).

\begin{figure}
\begin{center}
\includegraphics{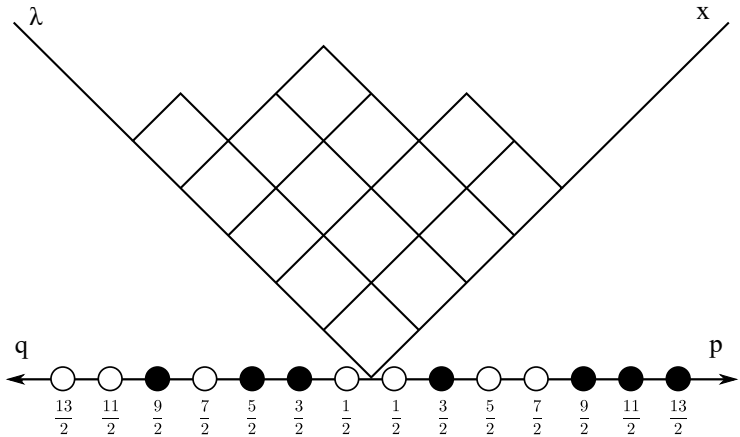}
\end{center}
\caption{A rotated Young diagram corresponding to the partition $\lambda$. The lower axis is labelled in terms of the half integer Frobenius coordinates $q_j, p_i$ (both positive). There is a piecewise-linear curve starting with the diagonal axis on the left, running along the top border of the diagram of the partition and ending with the diagonal axis on the right. It is clear that the contour of $\lambda$ has slope $-1$ wherever there is a white pebble, and has slope $+1$ in the intervals where  there is a black pebble. The association of sequences of pebbles to a Young diagram is denoted as a \emph{Maya diagram}. The collection of all the Maya diagrams is into 1-1 correspondence with the fermionic Fock space. We are especially interested to understand the limiting shape of the contour as the number of boxes tends to infinity, while their size tends to zero (keeping the area of the diagram fixed). See fig.~\ref{fig:Plancherel} for the limiting shape corresponding to the Plancherel measure.}
\label{fig:grid}
\end{figure}

In order to study the limit of continuous Young diagrams, we momentarily review our Frobenius notation for partitions: $\lambda \equiv \lbrace \lambda_i \rbrace \equiv (\vec{q} \, | \, \vec{p} \, )\, , q_i = \lambda_i - i + \half \, , p_i = \lambda'_i - i + \half $. The idea is then to find the leading representation for Young diagrams with a large number of boxes. The infinite harmonic resonance boxes for the overall singlet can be also thought as providing an overall negative shift in the density of boxes (since the singlet involves inverse reflection amplitude elements with respect to the non-singlets).

Let us define a number $N$ that will describe the size of the partition $\lambda$\footnote{This number $N$ need not be the $N$ of the gauge symmetry $SU(N)$ of MQM. For example it could be $N_f$ i.e. a number of flavors, or the number of diagonal elements in the partition $d(\lambda)$, for the Frobenius coordinates below.}. In order to study the limit of continuous representations, we define in analogy to~\cite{Douglas:1993iia}
\be
\lambda(x) = \frac{\lambda_i}{N} \, , \quad x = \frac{i}{N}\, , \qquad \lambda(x) \geq \lambda(y) \, , \quad \text{if} \, \, x \leq y
\ee
and then we go to strictly decreasing adapted coordinates
\be
h(x) = \lambda(x) - x + \half \, , \qquad \frac{d h}{d x} \leq  - 1
\ee
Equivalently, we can define continuous analogues for the Frobenius coordinates
\be
q(x) = \frac{q_i}{N} \, , \quad q(x) = \lambda(x) - x + \half \, , \qquad p(x) = \frac{p_i}{N} \, , \quad p(\lambda) = x(\lambda) - \lambda + \half \, .
\ee
In the last equation we are using the inverse function, since we had to flip the diagram along the diagonal.
It is easy to see that for $ 0 \leq x \leq \lambda^* = \lambda(x^*) = x^*$ we find $h=q$, and for $x \geq x^* = \lambda^*$ we find $h = 1 - p$  .

We can think of the diagram as being rotated  so that it is defined inside an infinite wedge, see fig.~\ref{fig:grid}. The horizontal axis is now labelled in terms of $q_j,  p_i$, or in other words in terms of strictly ordered half integers. We can now define a density of boxes $\mathcal{N}$ via
\be
\mathcal{N}(h) \, = \, - \frac{d x(h)}{d h} \, \leq 1
\ee
or sectionally via
\be
 \mathcal{N}(q) = -\frac{d x}{d q}\, , \quad \text{for} \quad q \geq 0 \, , \lambda \geq x \, , \qquad  \tilde{\mathcal{N}}(p) = - \frac{d \lambda}{d p} \, , \quad \text{for} \quad 0 \leq p \, , \lambda \leq x \, .
\ee
We shall also assume that the density is continuous at $p=q=0$. The coordinates $h$ or $p, q$ are depicted in fig.~\ref{fig:grid}.

We can now express the difference in the free energy~\ref{Freeenergygeneralrep} using continuous variables
\bea\label{conti}
\Delta{\mathcal{F}} &=&  i  \int_0^\infty d p \, \tilde{\mathcal{N}}(p) \, \Phi\left(  -i p R + \mu R \right) - i \int_0^\infty d q \, \mathcal{N}(q) \, \Phi\left(  i q R + \mu R \right)  \, \nn \\
\Delta{\mathcal{F}_{s.c.}} &=&  \Re   \int_{0}^\infty d y \, 2 \mathcal{N}(y) \, i \Phi\left( - i y R + \mu R \right)  \, ,
\eea
where we find that in the case of self conjugate (reflection symmetric) reps ($\mathcal{N}(q) = \tilde{\mathcal{N}}(p)$), we can naturally combine the two sections into a single distribution as expected. We should emphasize at this point that this expression is similar to the one describing the singlet free energy, but now the density of boxes is not provided by the dynamics of the inverted harmonic oscillator, but by a measure in the space of Young diagrams - that is equivalently an occupation density of the various energy states.
For example the total energy of $q_j$ excitations is
\be
E_{q} = \sum_{\lbrace q_j \rbrace} q_j \, \rightarrow \int_0^\infty d q  q \mathcal{N}(q) \, .
\ee
From eqns. \eqref{conti} and \eqref{C48}, we observe that one can also incorporate the contribution of the singlet states by a constant negative shift in the density of boxes, (since for the singlet all the energy states are uniformly occupied). Finally we can write the weighting factor in terms of
\be
\frac{\mathbf{\Omega}'(y)}{2} = \half -  \mathcal{N}(y) \, ,
\ee
with the function $\mathbf{\Omega}(y)$ providing the limiting shape of the Young diagram in fig.~\ref{fig:grid}, see fig.~\ref{fig:Plancherel} for the example of the Plancherel measure. 

The open question left is to find a rule that determines the appropriate measure $\mathcal{N}(y)$. Generically this measure depends on the degrees of freedom that source non-singlets. For example in the model of~\cite{Betzios:2017yms} this arises from the presence of the extra bi-fundamental fields. To calculate this measure, one should study the associated Schur measure in eqn.~\ref{C34}, for more details see section~\ref{Measures} and~\ref{Saddles}.

\paragraph{Further analysis and a non-singlet density of states -} There are some further manipulations that one can perform on eqn.~\eqref{conti}, using the perturbative bosonic scattering phase. Using the expression \eqref{generalrepdualexpansion2} we find
\be
 \Delta{\mathcal{F}} =  \Im    \, \int_0^\infty \frac{ i d t}{ 2  t \sinh(t/2)} \left( \int_{0}^\infty d q \, \mathcal{N}(q)  e^{ - q R t +  i \mu R t}  + \int_0^\infty d p \, \tilde{\mathcal{N}}(p) e^{ - p R t - i \mu R t} \right) \, .
\ee
We can then compute a Laplace-like transform of the density of boxes (in general this depends on the region of support of the density of boxes that could be finite)
\be
\mathcal{N}^L ( R t) = \int_{0}^\infty d q \, \mathcal{N}(q)  \,  e^{-  q R t}  \, ,
\ee
and evaluate
\be\label{continuousboxesfree}
\Delta{\mathcal{F}} =  \Im    \, \int_0^\infty \frac{ i d t}{ 2  t \sinh(t/2)} \left( \mathcal{N}^L ( R t)  e^{   i \mu R t}  + \tilde{\mathcal{N}}^L ( R t)  e^{ -  i \mu R t} \right) \, .
\ee
Once more for symmetric reps one can simply combine the two terms above into a single expression.

If we assume that the support of the density of boxes does not depend on $\mu, R$, then we find from \eqref{continuousboxesfree} that upon expanding in powers of $1/\mu$, the leading genus zero term is linear in $R$ and hence it cannot contain any entropy.
On the other hand in explicit models, we shall observe in section~\ref{Measures} that the support for the density of boxes actually depends on such parameters (as well as additional ones). Hence the models of interest in general do contain a genus zero free energy and entropy contribution, but we need an additional principle in order to determine the density of boxes. This principle is analysed in the main text in section~\ref{Saddles}.

We close this appendix with the important definition of a \emph{non-singlet density of energy states}\footnote{One can also define through them an associated coarse-grained scattering phase and reflection amplitude.}. The density of energy states can be defined via a convolution with resonances (or as  a fourier cosine transform of the Laplace transform of the density of boxes/occupation number)
\bea\label{densitiesrelations}
\mathcal{\rho}(\epsilon) &=&  \frac{1}{2 \pi} \int_{\text{supp.}} d y \, \mathcal{N}(y) \frac{y}{y^2+ \epsilon^2} =  \frac{1}{2 \pi} \Re \int_{\text{supp.}} d y \, \mathcal{N}(y) \frac{1}{y - i \epsilon} \, \nn \\
&=& - \frac{1}{2 \pi} \Re \int_0^\infty \, d t \, e^{i t \epsilon} \, \mathcal{N}^L(t) \, .
\eea
For the singlet, eqn. \eqref{densitiesrelations} together with eqn. \eqref{intexpressions} define the occupation density and its Laplace transform for singlets
\be
\mathcal{N}^L_{\text{singlet}}(t) = \frac{1}{2 \sinh (t/2)} \, , \qquad \mathcal{N}_{\text{singlet}}(y) =  \sum_{n=0}^\infty \delta ( y -  n - 1/2) \, .
\ee
For the non-singlets, the simplest application of \eqref{densitiesrelations} is to determine the non-singlet free energy by the simple formula
\be
\Delta{\mathcal{F}} = - \int_{- \infty}^\infty \, \left[ d \epsilon \mathcal{\rho}_{n.s.}(\epsilon)  \log(1 + e^{ \beta (\mu - \epsilon )}) + \tilde{\mathcal{\rho}}_{n.s.}(\epsilon)   \log(1 + e^{- \beta (\mu + \epsilon)}) \right] \, .
\ee
More generally, the non-singlet density of states is useful if we wish to compute Tachyon or loop correlation functions on a background for which non-singlets have condensed\footnote{See also~\cite{Betzios:2016lne} for another application of the non-singlet density of states.} (even if they originate from the presence of winding modes), adjusting the formalism of~\cite{Moore:1991sf,Moore:1991zv}, by replacing the singlet $\rho_{singlet}(\epsilon)$ in the appropriate formulae. In essence, this is
a form of mean field approximation that incorporates the effects of the non-singlets in a coarse-grained fashion.

\subsection{Determining the density of boxes from a coherent state}\label{Coherent}

Another way to determine the density of boxes, is to exploit the fact that it also describes the occupation number (density) of fermionic states\footnote{A similar analysis appears in~\cite{Maldanotes}.}. Since it is a density, it corresponds to a bilinear operator of fermions or in other words in a bosonic scalar field $\phi$ (see also the bosonisation formulae of appendix~\ref{Fermionic}). The profile/coherent state of the bosonic field $\phi$ is most simply determined by the measure of the unitary integral that describes the zero modes of the gauge field ($\theta$), see appendix~\ref{windtau}. In practice it is determined by the following equation
\be
{\phi'(\theta)} =  \sum_{n>0} t_n^+ e^{i n \theta} +  \sum_{n>0} t_n^- e^{- i n \theta} \, ,
\ee
and the density of fermions/occupation density is then given by
\be\label{occupationdensity}
\mathcal{N}(y) = \int_0^{2 \pi} \frac{d \theta}{2 \pi} \Theta(\phi' - y) \,. 
\ee
We observe that symmetric deformations $t_n^+ = t_n^-$, will result in reflection symmetric Young diagrams.

Let us now analyse our two cases of interest:

\begin{itemize}

\item For the model of~\cite{Kazakov:2000pm} $\phi' = \xi \cos \theta$. The profile is then found to be
\be
\mathcal{N}(y/\xi) = \frac{1}{\pi} \arccos \left( \frac{y}{\xi} \right) \, , \quad y \leq \xi \, .
\ee
The parameter $\xi$ also governs the area of the Young diagram, which is also to be expected since from eqn. \eqref{partsize} we find
\be
\text{Area} = \langle |\lambda| \rangle_{\text{Planch.}} = \xi^2 \, . 
\ee

\item For the model of~\cite{Betzios:2017yms}. In this case we expect an ansatz of the form (for bi-fundamental bosons) - $N_f$ governs the number of boxes and hence the Area of the Tableaux
\be\label{zmeasurecoherent}
\mathcal{N}(p/N_f) = \int_0^{2 \pi}  \frac{d \theta}{2 \pi} \Theta(\phi' - p) \, , \quad \phi' = - N_f \log \left[ (1 - e^{- \beta m + i \theta})  (1 - e^{- \beta m - i \theta}) \right]
\ee
so that the solution is given by 
\be
\cos \theta =  \frac{1 + e^{- 2 \beta m} - e^{ -p/N_f}}{2 e^{- \beta m}} \, ,
\ee
resulting into
\be
\mathcal{N}(p/N_f) = \frac{1}{\pi}  \arccos \frac{1 + e^{- 2 \beta m} - e^{ -p/N_f}}{2 e^{- \beta m}} \, , \quad  p/N_f \leq - 2 \log (1-e^{- \beta m})
\ee
If we perform the double scaling limit $N_f \rightarrow \infty, \,e^{- \beta m } \rightarrow 0 , \, N_f e^{- \beta m } = fixed $ in \eqref{zmeasurecoherent}, we obtain the previous result for the Plancherel measure. We could also consider instead $p/N_f = y$ as a new scaling variable for the continuous Tableaux. In this case we also find
\be
\mathcal{N}'(y) = - \frac{1}{ \pi}  \frac{e^{\beta m - y}}{ \sqrt{4 - e^{2 \beta m}(1 + e^{- 2 \beta m} - e^{- y} )^2}} \, ,
\ee
which exhibits singular behaviour (that determines the edges of the support) at 
\be
y_1 = - 2  \log (1 + e^{- \beta m}) \, , \quad y_2 =  - 2 \log (1-e^{- \beta m}) \, .
\ee
This density of boxes describes the scaling limit of the more general $z$-measures~\cite{Borodin2,Okounkov2}, when $z = z' = N_f$.

\end{itemize}

We close by mentioning two (apparent) shortcomings of this method. First the limiting shape does not seem to exhibit any phase transition in contrast with the Gross-Witten-Wadia transition of the model with the first winding mode~\cite{Gross:1980he,Wadia:1980cp} (or its dual Douglas Kazakov transition~\cite{Douglas:1993iia}). This can be remedied if we translate the condition $\mathcal{N}(y) \leq 1 \, , \, \forall y$ in the occupation density formalism. The second and most important deficiency of this method is that we can only find the leading density of boxes related to the Schur measure alone (equivalently taking into account only the diagonal measure piece in the integral over the zero modes of the gauge field). The saddle point though should be determined more consistently, taking also into account the contribution from the adjoint MQM matrix field $M$ in the integrand (equivalently the inverted oscillator reflection amplitude). It is not clear how to achieve this with the coherent state approach, and that is the reason why we resort to the techniques of section~\ref{Saddles} of the main text to determine the density of boxes and the shape of the leading Young diagram in a completely consistent manner.

\section{Saddle point equations for the model with dynamical bifundamentals}\label{bifundeffective}

It is interesting to repeat the large representation analysis for the more general bifundamental model of~\cite{Betzios:2017yms} that includes higher winding modes and can potentially reach different phases. The relevant Schur measure is the so called $z$-measure and reads (see also eqn. \eqref{z-measure})
\be
s_{(p_i|q_j)}\left(t_k = N_f \frac{\xi^{k}}{k} \right) = \xi^{|\lambda|} \prod_{i=1}^d  \frac{\Gamma(N_f+p_i + 1/2) \Gamma(1/2- N_f + q_i)}{\Gamma(p_i+1/2)\Gamma(q_i+1/2)\Gamma(N_f) \Gamma(1-N_f)} \det_{i, j} \frac{1}{p_i + q_j} \, .
\ee
To obtain the large $d$ limit we define $N_f = d n_f$ with $n_f \sim O(1)$. One finds a more general effective action (up to constant terms irrelevant for the saddle point analysis)
\bea\label{effectiveaction2}
S_{eff} &=&  - \int_0^1 d x \int_0^1 d y  \, \log \frac{|(p(x) - p(y))(q(x) - q(y))|}{(p(x)+q(y))^2}   \, + \tilde{V}(p) + V(q)  \, ,  \nn \\
\tilde{V}(p) &=& \int_0^1 d x \, 2  p(x) ( \log p(x) - 1 -  \log \xi)   -R\left(p(x)-i \mu\right)\left[\log\left[R\left(p(x)- i  \mu \right)\right]-1\right]  \, \nn \\
&-& \int_0^1 d x \, 2  (n_f + p(x)) \left( \log(n_f + p(x)) - 1  \right) \,  , \nn \\
V(q) &=& \int_0^1 d x \, 2 q(x) ( \log q(x) - 1 -  \log \xi)   - R\left(q(x)+i \mu\right)\left[\log\left[R\left(q(x)+ i  \mu \right)\right]-1\right] \, \nn \\
&+& \int_0^1 d x \, 2 (q(x) -  n_f ) \left( \log( q(x) -  n_f) - 1 \right) \, . \nn \\
\eea
The resulting saddle point equations are now
\bea\label{saddlepointbif1}
 \log \left(\frac{q^2}{\xi^2} \right) + 2 \log(q - n_f) -R \log\left[R\left(q + i \mu \right) \right]\,  = \, 2  \dashint d s \frac{\mathcal{N}(s)}{q - s} \, - 2  \dashint d s \frac{\tilde{\mathcal{N}}(s)}{q + s}\, , \nn \\
 \log \left(\frac{p^2}{\xi^2} \right)  - 2 \log(n_f + p) - R \log\left[R\left(p -i\mu \right)\right] \,  = \, 2 \dashint d s \frac{\tilde{\mathcal{N}}(s)}{p - s} \, - 2  \dashint d s \frac{{\mathcal{N}}(s)}{p + s}\, .
\eea
By adding and subtracting the equations in~\eqref{saddlepointbif1} we find
\bea\label{saddlepointbif2}
 2\log \left(\frac{p^2}{\xi^2} \right) + 2 \log\frac{(p+ n_f)}{(p-n_f)} - R \log\left[R^2 \left(p^2 + \mu^2 \right) \right]\,  = \, 4  \dashint d s  \frac{s(\mathcal{N}(s)+\tilde{\mathcal{N}}(s))}{p^2 - s^2}\, , \nn \\
  2 \log(p^2 - n_f^2) - R \log\left[\frac{p +i\mu}{p-i\mu}\right] \,  = \, 4p \dashint d s \frac{(\mathcal{N}(s)-\tilde{\mathcal{N}}(s))}{p^2 - s^2}
\eea
We observe that for finite $n_f$ the saddle point equations are not anymore equivalent as in the case of the Plancherel measure. The spectral asymmetry now contains a real part.

We therefore expect the presence of phases in which the leading Young diagrams are not reflection symmetric (even for $\mu = 0$). These are novel phases that cannot be reached using the model of~\cite{Kazakov:2000pm}. Nevertheless, once more for $R>2$ the potential becomes unstable.  We leave the analysis of this more general model for the future.

\section{A method to determine the general resolvent}\label{methodforgenresolvent}

In this appendix we shall describe how it is possible to determine the (real part) of resolvent and the density of states in the general case with non-zero $\mu$, extending slightly the method of~\cite{Dutta:2015noa}. The general form of the resolvent is obtained without even doing a single integral. The most difficult part of the analysis is determining the edges of the support in terms of the physical variables $\mu, \xi, R$. Our starting point will be eqn. \eqref{saddlerealpart} 
\be\label{saddleappendix}
 \log \left(\frac{u}{\xi^2} \right) - \frac{R}{2} \log\left[R^2 \left(u + \mu^2 \right) \right]\,  = \, 2  \dashint d v  \frac{\Re \mathcal{N}(u)}{u - v} = 2 \slashed{\Omega}(u) \, .
\ee

Let us use an ansatze for the resolvent of \eqref{saddleappendix} of the form
\be\label{ansatz1}
{\Omega}(u) =  \lambda_1 \log \left(\frac{g_1(u) - \sqrt{g_1(u)^2 - f_1(u)^2}}{ h_1} \right) + \lambda_2 \log \left(\frac{g_2(u) - \sqrt{g_2(u)^2 - f_2(u)^2}}{ h_2} \right) \, ,
\ee
which is expected to be valid for a single resolvent cut $u \in [a^2,b^2]$, leading to the fact that $g_{1,2},f_{1,2}$ are functions of degree up to one in powers of $u$, such that $g_1^2 - f_1^2 = (u-a^2)(u-b^2) = g_2^2 - f_2^2$. Using this ansatze we find
\be
 F(u) \, = \,  \log \left(\frac{u}{R^{- R}\xi^2} \right) -  \frac{R}{2} \log \left(\mu^2+u \right) \, = \lambda_1 \log \frac{f_1^2(u)}{ h^2_1} + \lambda_2  \log \frac{f^2_2(u)}{ h^2_2} \, ,
\ee
so that a natural identification is $\lambda_1=1, \lambda_2 = -R/2 ,$ with
\be
f_1^2(u) = f_1^2 u  \, , \quad h^2_1 = f_1^2 R^{- R} \xi^2 \qquad f^2_2(u) = f_2^2( u + \mu^2)  \, , \quad h^2_2 = f_2^2 \, .
\ee
This means that we are left with the task to determine $g_{1,2}^2 = (u + c_{g_{1,2}})^2$, that are order $u^2$. Consistency of the ansatze under the roots also demands
\be
2 c_{g_{1}} - f_1^2 = 2 c_{g_{2}} - f_2^2 \, , \qquad c_{g_{1}}^2 = c_{g_{2}}^2 - f_2^2 \mu^2  \, .
\ee
This determines
\be
c_{g_{1}} = \frac{1}{4} \left( f_1^2 - f_2^2 \right) +  \frac{ f_2^2 \mu^2}{f_2^2 - f_1^2} \, , \quad c_{g_{2}} = \frac{1}{4} \left( -f_1^2 + f_2^2 \right) +  \frac{ f_2^2 \mu^2}{f_2^2 - f_1^2} \, .
\ee
Finally, from the leading asymptotics of the resolvent at $u=\infty$ we find
\be
\log \frac{f_2^R}{2^R} + \log \frac{f_1^2}{4 R^{-2 R} \xi^4} = 0 \, .
\ee 
So far, with these conditions, we have managed to determine all but one parameter.
Usually this is determined from the subleading asymptotic at $u = \infty$, so that
$\Omega \sim 1/u$. Unfortunately our resolvent is not normalised in the $u$ variable. 
In order to fix all of the parameters we shall need to use period integrals of the resolvent expressed in the $z^2=u$ variable. These involve elliptic integrals and, while it is possible to write a general expression determining the normalisation it is rather long and opaque.

Things simplify considerably when $\mu=0$. In this case we find the resolvent quoted
in eqns.\eqref{ressingle1} and \eqref{ressingle3}, with the parameters obeying \eqref{identifications1}. Again there is a single undetermined parameter that should be fixed in terms of a period integral. In the case where there is a saturated region in the cut, one can derive an effective equation that needs to be solved with an additional term in the potential as in \eqref{sat1}. 

Having an explicit resolvent being the sum of several terms of the form of \eqref{ansatz1}, the density of boxes is then computed from its discontinuity along the common cut(s)
\be\label{densitygeneralformula}
\mathcal{N}(u) = - \sum_i \frac{\lambda_i}{ \pi i } \log \left( \frac{g_i - i \sqrt{f_i^2 - g_i^2}}{g_i + i \sqrt{f_i^2 - g_i^2}} \right) =  \sum_i \frac{\lambda_i}{ \pi  } \cos^{-1} \left( \frac{|g_i|}{f_i} \right) \, .
\ee

\section{Determinantal formulae for the $\tau$ function}

\subsection{A determinantal formula in the space of shifted weights}

Using the general formula for the $\tau$ function~\ref{C34}
\be
\tau(s, t_+ , t_- ;\mu) = \sum_{\lambda } s_\lambda(t_+) s_\lambda(- t_-) \langle \lambda, s | {\bf G} | \lambda , s \rangle \, ,
\ee
and the determinantal formula for the Schur polynomials eqn. \eqref{JTformula}
\be
s_\lambda(t) = \det_{i,j = 1, ... \ell(\lambda)} h_{\lambda_i - i + j}(t) \, ,
\ee
we find
\bea
\tau(s, t_+ , t_- ;\mu) = \sum_{\ell = 1}^\infty \sum_{ \lbrace \lambda_i \rbrace } \det_{i,j = 1, ... \ell} h_{\lambda_i - i + j}(t_+) \det_{i,j = 1, ... \ell} h_{\lambda_i - i + j}(- t_-) \,  \langle s - \ell | {\bf G} |  s - \ell \rangle \, \prod_{i =1}^\ell  \mathcal{R}^{-1}_{\lambda_i - i + \half + s}   \nn \\
= \sum_{\ell = 1}^\infty   \langle s - \ell | {\bf G} |  s - \ell \rangle \det_{j,k =1...\ell} \left( \sum^\infty_{p = - \ell}   \mathcal{R}^{-1}_{p + \half + s}  h_{p + j}(t_+)   h_{p + k}(- t_-)   \right)  \, , \nn \\
\eea

\subsection{Determinantal formulae in Frobenius coordinates}\label{determinantal}

One could try to find an exact expression for the free energy using the space of representations (taking into account both the Schur measure and the transition amplitude ), using some determinantal formulae in Frobenius coordinates $p_i,q_j \in \mathbb{Z}^{+} + \half$, defined in appendices~\ref{Partitionsreps} and \ref{Frobeniusreps}.

The first such formula is (Cauchy-Binet formula)
\be
\sum_{l_1=1}^M ... \sum_{l_N=1}^M \det_{j,k \leq N}(f_{j-1}(y_{l_k}) \det_{j,k \leq N}(g_{j-1}(y_{l_k}) = N! \det_{j,k \leq N} \left( \sum_{l=1}^M f_{j-1}(y_l) g_{k-1} (y_l) \right) \, .
\ee
If the $l_k$ are strictly ordered then one can simply drop the $N!$ on the right hand side.

The second formula is by Giambelli. This expresses the Schur functions of an arbitrary diagram as determinants of Hook ones in Frobenius coordinates
\be
s_\lambda(t) = \det_{i, j = 1, ... d(\lambda)} s_{(p_i | q_j)} (t) \,  \quad s_{(p_i | q_j)} (t) = (-1)^{q_j - \half} \sum_{m=0}^{q_j - \half} h_{q_j - \half -m}(-t) h_{p_i + m + \half}(t) \, ,
\ee
the functions $h_i(t)$ being complete symmetric functions.
This means that the tau function can be expressed as a sum of determinants
\bea
\tau(s, t_+ , t_- ;\mu) = \sum_{\lambda } s_\lambda(t_+) s_\lambda(- t_-) (-1)^{|\lambda|} \langle \lambda, s | {\bf G} | \lambda , s \rangle    \nn \\
= \sum_{d=1}^\infty \sum_{\lambda: \lbrace p_i, q_j \rbrace } \det_{i, j} s_{(p_i | q_j)}(t_+)   \det_{i,j} s_{(p_i | q_j)}(t_-)     \,  \prod_{i=1}^d (-1)^{p_i + q_i} \mathcal{R}( \mu R  - i p_i R) \mathcal{R}^{-1}(\mu R  + i q_i R)   \nn \\ 
=  \sum_{d=1}^\infty \sum_{p_i} \prod_{i=1}^d \mathcal{R}( \mu R  - i p_i R)  d! \det_{i, j} \left( \sum_{q=1}^\infty  \mathcal{R}^{-1}( \mu R  + i q R)   s_{(p_i | q)}(t_+) s_{(p_j | q)}(t_-)  \right) = \nn \\
= \sum_{d=1}^\infty   d!  d! \det_{i, j} \left( \sum_{p=1}^\infty \mathcal{R}( \mu R  - i p R)  \mathcal{R}^{-1}( \mu R  + p R)   s_{(p | i)}(t_+) s_{(p | j)}(t_-)  \right) \,.
\eea
Let us now specialise to the model of~\cite{Kazakov:2000pm}, with only the first winding mode turned on. The Poissonised Plancherel measure is 
\bea
\mathfrak{M}^\xi(\lambda) = e^{- \xi^2} \xi^{2 |\lambda|} \left(\frac{\dim \lambda}{|\lambda|!}\right)^2  = e^{- \xi} \xi^{|\lambda|} \det \left[ \frac{1}{p_i + q_j} \right]^2 \prod_{i=1}^d  \left[\frac{1}{(p_i-\half)! (q_i-\half)!}\right]^2  \, .
\eea
We can then multiply this with the partition function/reflection amplitude in a fixed representation to obtain the total formula for the Kazakov-Kostov-Kutasov partition function $\mathcal{Z}_{KKK} = \mathcal{Z}_{singlet} \mathcal{Z}_{imp}$
\bea
\mathcal{Z}_{imp} &=& \sum_{d=0}^\infty \sum_{\lambda: \lbrace p_i, q_i \rbrace} \, \mathfrak{M}^\xi(\lambda) \, \prod_{i=1}^d \mathcal{R}( \mu R  - i p_i R) \mathcal{R}^{-1}( \mu R  + i q_i R)    \,  \nn \\
&=& \sum_{d=0}^\infty \sum_{\lambda: \lbrace p_i, q_i \rbrace } \,  \det_{d \times d} \left[\frac{( i {\xi})^{p_i + q_j}}{p_i + q_j} \frac{ e^{i \delta_+(-p_i) -i \delta_+(q_j)}}{\Gamma(p_i + \half) \Gamma(q_j + \half)} \right] \det_{d \times d} \left[\frac{ (i {\xi})^{p_i + q_j}}{p_i + q_j} \frac{ e^{i \delta_-(-q_j) -i \delta_-(p_i)}  }{\Gamma(p_i + \half) \Gamma(q_j + \half)} \right] \, \nn \\
&=& \sum_{d=0}^\infty \sum_{\lambda: \lbrace p_i, q_i \rbrace } \det_{2 d \times 2 d} \begin{pmatrix}
0 & e^{i \delta_+(-p_i)} L(p_i, q_j ; \xi) e^{ -i \delta_+(q_j)} \\
e^{i \delta_-(-q_j)}  L(q_j, p_i ; \xi)e^{-i \delta_-(p_i)}  & 0
\end{pmatrix}
\eea
with $e^{i \delta_\pm(x)} = \Gamma(\half \pm i \mu R + x R)$ ( this phase is for the case of $0B$, for the bosonic case one should replace it with its square root). 

Next it is useful to employ Gram's identity
\be
\sum_{n=0}^{n=N} \sum_{A \subset N\, , \, |A|=n}  z^n \det_{i,j \in A} M_{i j} = \det_{N \times N} (1+ z M_{i j})
\ee
in order to sum over all reps with $d \in [0, \infty)$ to obtain
\be
\mathcal{Z}_{imp} = \det \left( \hat{I} - \xi^2 \hat{M} \right)
\ee
where the $2 \times 2$ kernel $\hat{M}$ acts on functions as (using that $L(x,y)$ is hermitean symmetric matrix)
\be
\hat{M} \left[\begin{pmatrix}
F_1 \\ F_2
\end{pmatrix} \right](x) = \int_0^\infty d y \begin{pmatrix}
0 & e^{i \delta_+(-x)} L(x, y) e^{-i \delta_+(y)} \\
e^{i \delta_-(-x)} L(x, y) e^{-i \delta_-(y)} & 0 
\end{pmatrix} 
\begin{pmatrix}
F_1(y) \\ F_2(y)
\end{pmatrix} 
\ee
(we chose to remove $\xi$ from the definition of $\hat{M}$, since it is a prefactor in the expression for the Fredholm determinant.). The interpetation is the following: 

When we sum over arbitrary sized representations, the spectrum of this integral equation gives the exact solution for the free energy. For reps with a cutoff in size, it would involve the diagonalisation of a big matrix.

\subsection{The $\tau$ function in terms of two coupled normal matrix models}\label{couplednormal}

It is also possible to transform the expressions that involve sums over ordered (half) integers in a coupled normal matrix model following the analysis of~\cite{Eynard:2008mt}.

Using the function
\be\label{residue1}
f(x) = - x \Gamma(-x) \Gamma(x) e^{- i \pi x} = \frac{\pi e^{- i \pi x}}{\sin (\pi x)}
\ee
that has poles with residue one at all integers and shifting by a half-integer, we can rewrite the partition function in terms of two coupled $d \times d $ normal matrices $P,Q$
\be
\mathcal{Z}_{imp} \, = \, \sum_{d=0}^\infty \,  \int_{M_d(\mathcal{C})} d P  \, \int_{M_d(\mathcal{C})} d Q \,  e^{-  \tr V(Q) - \tr V(P) - V_{int}(P,Q)}
\ee
The matrices $P,Q$ are normal $[Q, Q^\dagger] = [P, P^\dagger] = 0$, and the contour $\mathcal{C}$ in the particular example is a hairpin contour surrounding the positive semi-axis. We can then perform a usual saddle point analysis (large - $d$) of the coupled matrix integral.

In particular for the measure corresponding to the Kazakov-Kostov-Kutasov model we find
the explicit potentials
\bea\label{KKKcoupled}
V(p) = \log \Gamma \left( \half +  p \right) - \log \Gamma \left( \half - p \right)  - 2 p \log \xi - i \Phi\left( -i p R + \mu R  \right) \nn \\
V(q) = \log \Gamma \left( \half +  q \right) - \log \Gamma \left( \half - q \right)  -   2 q \log \xi   + i \Phi\left( i q R + \mu R  \right) \nn \\
V_{int}(P,Q) = \log \det (P + Q) \, ,
\eea
with $\Phi(E)$ the scattering phase of the inverted oscillator, see appendix~\ref{fixedrepreflect}.
Universality arises in the large-$d$ limit of continuous representations, that should be taken along the lines of section~\ref{Saddles}. We observe that the large-$d$ effective action is the same whether we use the discrete model or the continuous model. The saddle point equations of this coupled normal two-matrix model then coincide with the ones analysed in section~\ref{Saddles} of the main text.

\section{Ungauged Matrix Quantum Mechanics}\label{Ungauged}

In this appendix, we consider the dynamics of ungauged Matrix Quantum Mechanics described by the following action ( $\omega \rightarrow i \omega$ one passes from the normal to the inverted oscillator)
\be
S_{MQM} = \half \int dt \tr \left((\partial_t M)^2 - \omega^2 M^2 \right) \, .
\ee
It has the classical solution between two configurations $M_1(t_1)= M_1$ and $M_2(t_2) = M_2$
\be
M^c(t) = \frac{1}{\sin\omega(t_2 - t_1)}  \left(M_2 \sin \omega (t - t_1) - M_1 \sin \omega(t-t_2) \right) \, ,
\ee
that leads to the on-shell action
\be
S[M^c] = \frac{\omega}{\sin \omega(t_2-t_1)} \left((\tr M_1^2 + \tr M_2^2) \cos \omega(t_2 - t_1) - 2 \tr M_1 M_2 \right)
\ee
To pass to the Euclidean one rotates $t = - i \tau$.
The states are specified in terms of the $N^2$ entries of the matrix
\be
\Phi(M) = \langle M_{i j} | \Phi \rangle \, .
\ee
Using the global $SU(N)$ to diagonalise the matrices $M = U^\dagger \Lambda U$, the inner product can be expressed as ($\Phi(\lambda ; U) = \Phi(U^\dagger \Lambda U)$)
\be
\langle \Phi | \Phi' \rangle = \int \mathcal{D} M \, \overline{\Phi(M)} \Phi'(M) = \int \mathcal{D} U \int \prod_i d \lambda_i \Delta^2(\lambda) \overline{\Phi(\lambda ; U)} \Phi'(\lambda ; U) \, .
\ee
Since the action is invariant under global $SU(N)$, the dynamics of the $N^2$ free matrix degrees of freedom can be reorganised by splitting the total Hilbert space $\mathcal{H} = \oplus \mathcal{H}_R$ into a direct sum using appropriate $SU(N)$ representations. 
The wavefunctions can also be represented as matrices (depending on the rep) in the diagonal eigenvalue basis using
\be
\Psi^R_{a b}(\lambda) = \langle \lambda ; R , a, b | \Psi \rangle \, , \quad \langle \Psi | \Psi' \rangle = \sum_R \frac{1}{D_R} \int \prod_{i=1}^N d \lambda_i \sum_{a, b} \overline{\Psi^R_{a b}(\lambda)} {\Psi'}^R_{a b}(\lambda)   \, ,
\ee
where we used the resolution of the identity
\be
\hat{I} = \sum_R \frac{1}{D_R} \int \prod_{i=1}^N d \lambda_i \Delta^2(\lambda) \sum_{a, b} | \lambda ; R , a, b  \rangle \langle  \lambda ; R , a, b | \, ,
\ee
and chose to absorb the vandermonde factor in the definition of the wavefunctions i.e. $\Psi^R_{a b}(\lambda) = \Delta(\lambda) \Phi^R_{a b}(\lambda)$. We can also define
\be
\langle \lambda ; U | \lambda' ; R, a, b \rangle = \sqrt{D_R} R_{a b}(U) \frac{\delta(\lambda - \lambda')}{\Delta(\lambda)} \, , \qquad \Psi^R_{a b}( U^\dagger \lambda U) = R_{a c}(U) \Psi^R_{c b}( \lambda ) \, .
\ee
and then
\be
\langle R, a, b | U \rangle = \overline{\langle U| R, a, b \rangle}  \, \Rightarrow R_{b a}(U^{\dagger}) = \overline{R_{a b}(U)}\, ,
\ee
This expression guarantees that the inner product of states is invariant under unitary rotations in each irrep sector.

The wavefunctions evolve with the appropriate hamiltonian for each irrep
\be
i \hbar \frac{\partial \Psi^R_{a b}}{\partial t} = \sum_c^{D_R} \hat{H}^R_{a c} \Psi^R_{c b} \, .
\ee
Notice also that $D_R$ is a degeneracy in energies for each rep $R$ that should be taken into account when one works in the energy basis. This is evident since the last index $b$ in the equation above is just a spectator index.

The Hamiltonian can be expressed in the eigenvalue basis as 
\be
(\hat{H}_{R})_{a b} {\Psi}^R_{b d}(\lambda)= \hat{P}_{a c} \left[\left(\sum_i^N -\half \frac{\partial^2}{\partial \lambda_i^2} + V(\lambda_i) \right) \delta_{c e} + \half \sum_{i \neq j} \frac{\left(T_{i j}^R T_{j i}^R\right)_{c e}}{(\lambda_i - \lambda_j)^2} \right]\hat{P}_{e f} { \Psi}^R_{f d}(\lambda) 
\ee
with $\hat{P}$ a projector to zero weight\footnote{This is the subset of representations that admit a zero weight state and arise in the decomposition of tensor products of the adjoint~\cite{Gross:1990md,Boulatov:1991xz,Karczmarek:2008sc,Betzios:2021fnm}. In the Young diagram notation of appendix~\ref{UvsSU}, this means that the number of boxes should be the same as the number of anti-boxes.} states. Practically this means that
\be
\hat{P} = \int_0^{2 \pi} \prod_i \frac{d \theta_i}{2 \pi} e^{i \sum_m \theta_m T_{m m}^R } \, , \quad T_{i i}^R {\Psi}^R(\lambda) =0 \, .
\ee

Transition amplitudes between generic states are given using the matrix propagator
\be
\langle \Phi_2  ; t_2 | \Phi_1 ; t_1  \rangle = \int \mathcal{D} M_2 \int  \mathcal{D} M_1 \, \Phi_2^\dagger(M_2) \langle M_2, t_2 | e^{- i \hat{H} \Delta t} | M_1 , t_1 \rangle \Phi_1(M_1) \,  .
\ee
For vacuum to vacuum amplitudes it is also useful to introduce the vacuum wavefunction (for the oscillator potential) 
\be
\langle M | 0 \rangle = \Phi_0(M) = \left(\frac{\omega}{\pi} \right)^{N^2/4} \exp \left(- \frac{\omega}{2} \tr M^2 \right) \, .
\ee
The Euclidean propagator ($t = - i s$) solves the Heat kernel equation
\be
\left( \frac{\partial}{\partial s}  - \half \left( \nabla^2_M - \omega^2 M^2 \right) \right) \langle M_2, s | M_1 , 0 \rangle = 0 \, , \quad  \langle M_2, s | M_1 , 0 \rangle_{ s \rightarrow 0} \rightarrow \delta^{N^2}(M_1 - M_2)
\ee
This is a matrix generalisation of the harmonic oscillator Euclidean Schroendinger equation with solution
\be
\langle M_2, s | M_1 , 0 \rangle  = \left(\frac{\omega}{2 \pi \sinh \omega s } \right)^{N^2/2} \exp \left(- \frac{\omega}{2 \sinh(\omega s)} \left[ (\Tr M_1^2 + \Tr M_2^2 ) \cosh \omega s \, - \, 2 \Tr M_1 M_2 \right] \right)
\ee
It is also useful to pass in the representation/energy eigenbasis using  a resolution of the identity and energy eigenstates for each rep
\bea
\langle M_2, t_2 | e^{- i \hat{H} \Delta t} | M_1 , t_1 \rangle = \sum_R  \sum^{D_R}_{a b ; c d} \, \int d E_R e^{- i E_R  \Delta t} \, R_{a b}(U_2)  {\Psi}^{R}_{a b}(E_R ; \lambda_2) \overline{{\Psi}_{c d}^{R}(E_R ; \lambda_1)} \overline{R_{c d}(U_1)}   \nn \\
\eea
A simple application and check of this formalism is the computation of the partition function for the $N^2$ oscillators, expanded in the representation basis
\bea
Z_{N^2} &=& \tr e^{- \beta \hat{H}_{N^2}} = \int \mathcal{D} M \, \langle M, \beta | e^{-\beta \hat{H}_{N^2}} | M , 0 \rangle = \int D U \int \prod_i d \lambda_i \Delta^2(\lambda) \langle U ; \lambda | e^{-\beta \hat{H}_{N^2}} | U ; \lambda \rangle \nn \\
&=& \int D U \int \prod_i d \lambda_i   \sum_R \sum^{D_R}_{a b ; c d} \, \int d E_R e^{- \beta E_R } \, R_{a b}(U)  \overline{R_{c d}(U)} {\Psi}^{R}_{a b}(E_R ; \lambda) \overline{{\Psi}_{c d}^{R}(E_R ; \lambda)}  \nn \\
&=& \sum_R \frac{1}{D_R} \sum^{D_R}_{a b} \int d E_R \langle \Psi^R ; E_R , a, b | e^{- \beta E_R} | \Psi^R ; E_R, a, b \rangle = \sum_R D_R \tr_{\mathcal{H}_R} \left( e^{- \beta \hat{H}_R}\right) 
\eea

\subsection{The canonical partition function in the holonomy basis}\label{holonomypf}

The canonical partition function for matrix quantum mechanics after one decomposes into $SU(N)$ representations can be written as
\be
Z_N  = \sum_R D_R Z_N^R = \sum_R D_R \tr_R e^{- \beta \hat{H}_R}\, .
\ee
An equivalent way of computing such a partition function involves to first compute the twisted partition function using the quadratic oscillator hamiltonian of the singlet with the inclusion of a rotation operator
\be
Z_N(U) = \tr \left(e^{- \beta \hat{H}_{MQM}} \hat{\Theta}(U) \right)\, , \quad Z_N^R = \int D U \chi_R (U) Z_N (U)
\ee
the twisted partition function takes the form ($q = e^{- \beta \omega} / e^{i \beta \omega} $ for normal/inverted oscillator)
\be
Z_N(e^{i \theta}) = q^{\half N^2} \prod_{i, j = 1}^N \frac{1}{1 - q e^{i (\theta_i - \theta_j)}}
\ee
this then results into
\bea
Z_N^R &=& \frac{1}{N!} \oint \prod_{i=1}^N \frac{d \theta_i}{2 \pi} |\Delta(e^{i\theta})|^2 \chi_R(e^{i \theta}) Z_N(e^{i \theta}) = \sum_{R'} q^{\half N^2} \int D U  q^{|R'|} \chi_R(U) \chi_{R'}(U) \chi_{R'}(U^{-1}) \nn \\
\eea

\end{document}